\documentclass{article}

\usepackage{microtype}
\usepackage{graphicx}
\usepackage{subcaption}
\usepackage{booktabs} 

\usepackage{hyperref}

\usepackage[accepted]{icml2026}

\usepackage{amsmath}
\usepackage{amssymb}
\usepackage{mathtools}
\usepackage{amsthm}
\usepackage{enumitem} 
\usepackage{placeins} 
\usepackage{xspace} 

\usepackage{listings}
\usepackage{xcolor}
\definecolor{codebg}{RGB}{245,245,250}
\definecolor{codeframe}{RGB}{200,200,210}
\definecolor{codekeyword}{RGB}{0,100,180}
\definecolor{codecomment}{RGB}{100,100,100}
\definecolor{codestring}{RGB}{180,80,0}
\lstdefinestyle{codestyle}{
    backgroundcolor=\color{codebg},
    frame=single,
    rulecolor=\color{codeframe},
    basicstyle=\ttfamily\small,
    keywordstyle=\color{codekeyword}\bfseries,
    commentstyle=\color{codecomment}\itshape,
    stringstyle=\color{codestring},
    breaklines=true,
    xleftmargin=1em,
    xrightmargin=1em,
    aboveskip=1em,
    belowskip=1em,
}

\theoremstyle{plain}

\theoremstyle{definition}

\theoremstyle{remark}

\definecolor{taskCDTcolor}{HTML}{1F77B4} 
\definecolor{taskNTcolor}{HTML}{2CA02C}  
\definecolor{taskCLTcolor}{HTML}{9467BD} 
\definecolor{taskABTcolor}{HTML}{D62728} 
\definecolor{taskALTcolor}{HTML}{FF7F0E} 
\definecolor{taskMIAcolor}{HTML}{17BECF} 
\definecolor{taskTDEcolor}{HTML}{8C564B} 
\definecolor{taskRNCcolor}{HTML}{2C3E50} 
\definecolor{taskSPFcolor}{HTML}{E377C2} 
\definecolor{taskSMcolor}{HTML}{BCBD22}  

\newcommand{\taskCDT}{\texorpdfstring{\textcolor{taskCDTcolor}{\textsc{Colored Dot Tracking}}}{Colored Dot Tracking}\xspace}
\newcommand{\taskNT}{\texorpdfstring{\textcolor{taskNTcolor}{\textsc{Neuron Tracking}}}{Neuron Tracking}\xspace}
\newcommand{\taskCLT}{\texorpdfstring{\textcolor{taskCLTcolor}{\textsc{Cell Lineage Tracking}}}{Cell Lineage Tracking}\xspace}
\newcommand{\taskABT}{\texorpdfstring{\textcolor{taskABTcolor}{\textsc{Animal Behavioral Tracking}}}{Animal Behavioral Tracking}\xspace}
\newcommand{\taskALT}{\texorpdfstring{\textcolor{taskALTcolor}{\textsc{Animal Limb Tracking}}}{Animal Limb Tracking}\xspace}
\newcommand{\taskMIA}{\texorpdfstring{\textcolor{taskMIAcolor}{\textsc{Multichannel Image Alignment}}}{Multichannel Image Alignment}\xspace}
\newcommand{\taskTDE}{\texorpdfstring{\textcolor{taskTDEcolor}{\textsc{3D Exploration}}}{3D Exploration}\xspace}
\newcommand{\taskRNC}{\texorpdfstring{\textcolor{taskRNCcolor}{\textsc{Road Network Construction}}}{Road Network Construction}\xspace}
\newcommand{\taskSPF}{\texorpdfstring{\textcolor{taskSPFcolor}{\textsc{Spectral Plume Finding}}}{Spectral Plume Finding}\xspace}
\newcommand{\taskSM}{\texorpdfstring{\textcolor{taskSMcolor}{\textsc{Shape Matching}}}{Shape Matching}\xspace}

\icmltitlerunning{A Systematic Study of Behavioral Cloning for Scientific Data Annotation}

\begin{document}

\twocolumn[
\icmltitle{A Systematic Study of Behavioral Cloning for Scientific Data Annotation}



  \icmlsetsymbol{equal}{*}

  \begin{icmlauthorlist}
    \icmlauthor{Ishaan Singh Chandok}{equal,harvard,vi}
    \icmlauthor{Core Francisco Park$^\dagger$}{equal,harvard,vi,comp}
  \end{icmlauthorlist}

  \icmlaffiliation{harvard}{Department of Physics, Harvard University, Cambridge, MA, USA}
  \icmlaffiliation{vi}{Visual Intuition, Cambridge, MA, USA}
  \icmlaffiliation{comp}{Prior Computers, Cambridge, MA, USA}

  \icmlcorrespondingauthor{Ishaan Singh Chandok}{ichandok@g.harvard.edu}
  \icmlcorrespondingauthor{Core Francisco Park}{corefranciscopark@g.harvard.edu}

  \icmlkeywords{Machine Learning, ICML}

  \vskip 0.3in
]



\printAffiliationsAndNotice{}  

\begin{abstract}
Scientific data annotation, such as tracking animals in video or proofreading neural reconstructions, remains bottlenecked by the ``last mile'' problem: even with strong automation, verification and correction consume substantial human effort. Standard approaches train models to directly predict annotations, discarding the rich supervision in how experts navigate, click, verify, and correct. We introduce a framework for studying behavioral cloning on scientific annotation: 9 synthetic tasks paired with synthetic annotations that simulate realistic human strategies including exploration, mistake correction, and strategic decision-making. Our experiments reveal several findings. First, skills emerge hierarchically: models learn GUI mechanics before task-critical decisions, and commit fewer mistakes than the training data while retaining the ability to correct errors when they occur. Second, scaling models on multi-task behavioral cloning shows that larger models are more data efficient within our scale range. Third, multi-task pretraining enables efficient fine-tuning to new tasks, while training from scratch fails entirely. Fourth, linear probes reveal that models internally represent latent variables of the annotation process such as task phase and data position; interestingly, we find a shared mistake representation that generalizes across different annotation tasks. Overall, our framework establishes systematic benchmarks and identifies key bottlenecks, providing a foundation for scaling behavioral cloning to real-world scientific data annotation.
\end{abstract}

\section{Introduction}

Scientific data analysis is bottlenecked by manual annotation, from tracking animals in behavioral videos~\citep{mathis2018deeplabcut,pereira2022sleap,lauer2022multi} to reconstructing wiring diagrams of brains~\citep{scheffer2020connectome,microns2025functional}. Deep learning has reduced the bulk of annotation, but the ``last mile'' remains expensive: even when automated systems are mostly correct, humans must still find and fix the remaining errors. In connectomics, for example, proofreading a single fly brain required 33 person-years of effort despite state-of-the-art automated segmentation~\citep{dorkenwald2022flywire}.

Most ML approaches to automate annotation treat the process as a direct mapping from data to labels, optimizing only for final output accuracy. However, human annotation is an interactive workflow: an expert navigates through data, clicks or edits elements in the interface, verifies uncertain cases, and revises their mistakes. These action sequences provide rich supervision about the \emph{process} by which high-quality annotations are produced, not just the final labels. Behavioral cloning (BC)~\citep{pomerleau1988alvinn,bain1995framework}, where models learn to imitate human actions, offers a natural way to exploit this supervision by training directly on annotation traces, an approach that has proven effective for autonomous driving~\citep{bojarski2016end} and game-playing agents~\citep{baker2022video}. 

Despite BC being a standard approach in robotics and embodied AI~\citep{mandlekar2021matters,chi2025diffusion}, its potential for scientific annotation remains poorly characterized. The barriers are largely practical: collecting behavioral data requires instrumented interfaces, no standardized benchmarks exist for long-horizon annotation sessions, and uncertain expectations discourage scientists from diverting effort from their core research. As a result, neither ML researchers nor domain scientists have the tools to answer basic questions: How hard is it for a model to learn an annotation workflow? What challenges remain even with substantial data? What factors make some tasks harder than others?

\begin{figure*}[ht]
\vskip 0.2in
\begin{center}
\includegraphics[width=0.9\textwidth]{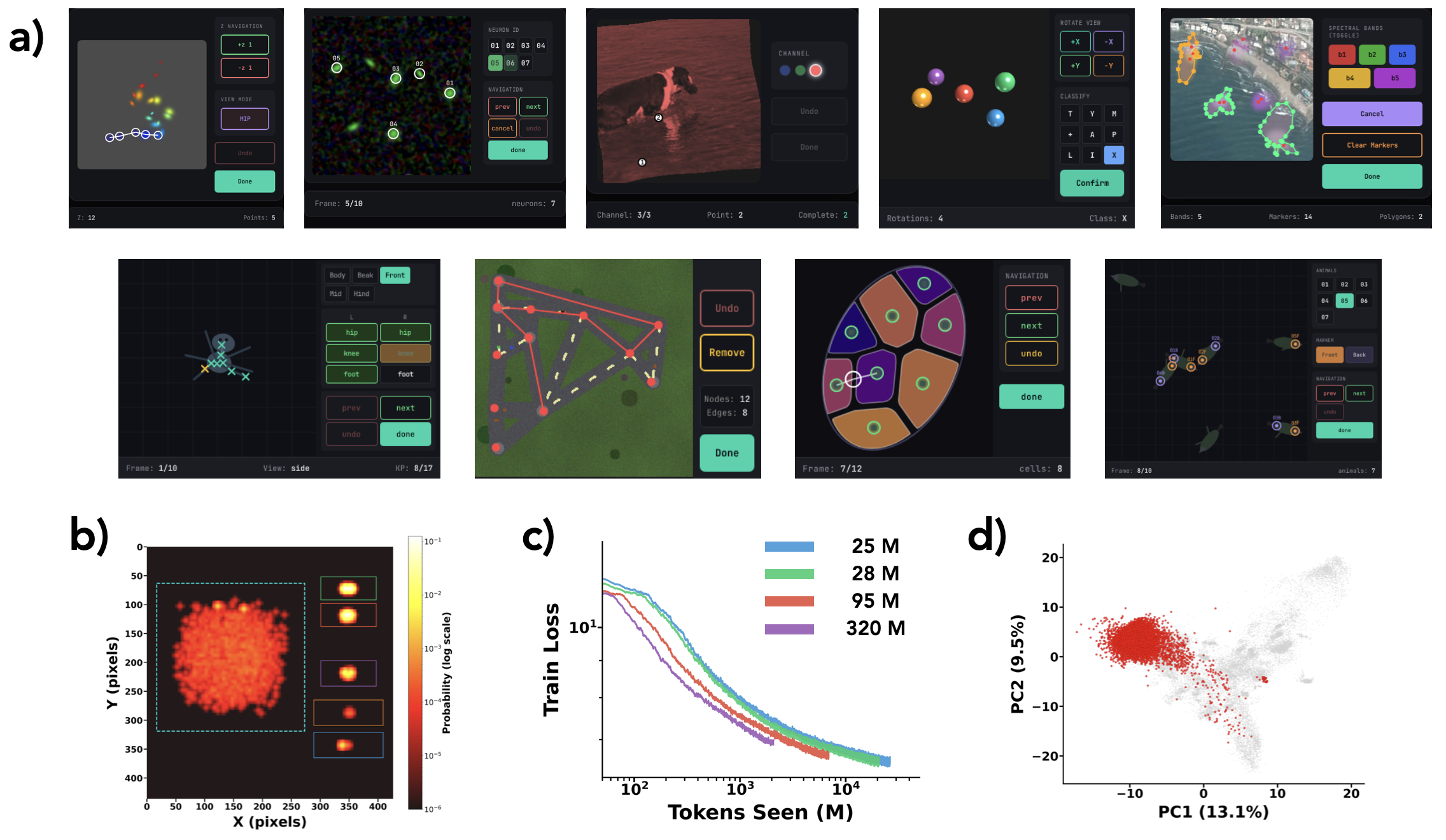}
\vspace{1.5em}
\caption{\textbf{Framework Overview.} (a) The 9 synthetic annotation tasks: \taskCDT, \taskNT, \taskMIA, \taskTDE, \taskSPF, \taskALT, \taskRNC, \taskCLT, and \taskABT. (b) Click heatmap from a trained model. The model predicts $(x, y)$ probability distributions directly and places clicks in semantically reasonable locations. (c) Scaling behavior: training loss vs.\ tokens seen for multiple model sizes. Larger models are more data-efficient learners. (d) Internal representations across tasks. Each dot is an activation from a different task; mistakes (red) cluster together regardless of task, indicating a shared mistake representation learned across the multi-task model.}
\label{fig:framework-main}
\end{center}
\vskip -0.2in
\end{figure*}

\textbf{This paper} provides a framework for answering these questions. We construct 9 synthetic annotation tasks with procedurally generated data and simulated human behavior, enabling controlled experiments that would otherwise require years of data collection. The synthetic setting lets us isolate variables: we can generate arbitrary amounts of training data, vary task difficulty systematically, and study failure modes without confounds from real-world data collection. Our framework serves as a meeting ground: ML researchers gain access to well-defined scientific annotation challenges, while domain scientists see standard approaches applied to tasks resembling their own, providing baselines and realistic expectations.


\textbf{Our contributions are:}
\begin{enumerate}[leftmargin=*,nosep]
    \item \textbf{A multi-task procedural framework for studying behavioral cloning on scientific annotation} (\S\ref{sec:framework}, Figure~\ref{fig:framework-main}). We introduce 9 synthetic tasks inspired by scientific data annotation, paired with virtual annotators that simulate realistic human strategies including information gathering, exploration, mistake-and-correction sequences, and strategic decision-making. A GUI simulator enables both data generation and closed-loop evaluation.

    \item \textbf{A detailed analysis of training dynamics and skill emergence in single-task learning} (\S\ref{sec:single-task}, Figure~\ref{fig:single-task}). We find hierarchical emergence: GUI mechanics (clicking buttons, placing markers) are learned before task-critical decisions (when to undo, when to finalize). Notably, models commit fewer mistakes than present in the training data, yet when teacher-forced into error states, they still correctly trigger undo actions, a beneficial effect of majority-class learning.

    \item \textbf{Scaling laws and transfer properties for multi-task annotation models} (\S\ref{sec:multi-task}--\ref{sec:downstream}, Figures~\ref{fig:multi-task},~\ref{fig:downstream}). Larger models are more data-efficient learners: 10$\times$ more parameters yields roughly 3$\times$ better sample efficiency. Multi-task pretrained models transfer efficiently via fine-tuning, while in-context learning and training from scratch both fail entirely.

    \item \textbf{Interpretability analysis revealing shared mistake representations across tasks} (\S\ref{sec:internals}, Figure~\ref{fig:internals}). We find that mistake representations show cross-task transfer: a pooled probe achieves 0.87 ROC AUC, and probes trained on 8 tasks transfer to held-out tasks with mean 0.71 ROC AUC. One exception is \taskTDE (0.29), where errors are classification mistakes rather than spatial placement errors, suggesting the shared representation is specific to placement-type errors rather than a fully task-general ``mistake'' concept. Models also linearly encode annotation phase, data-relative time (position within the data, not sequence position), and task state.
\end{enumerate}

\paragraph{Conflict of Interest Disclosure.} The authors are affiliated with Visual Intuition, a company that develops data annotation models.

\section{Related Work}

\paragraph{Behavioral Cloning and GUI Agents.}
Behavioral cloning~\citep{pomerleau1988alvinn,bain1995framework} learns policies via supervised learning on expert demonstrations. Classic challenges include covariate shift, addressed by DAgger~\citep{ross2011reduction} and DART~\citep{laskey2017dartnoiseinjectionrobust}, and multimodal action distributions, handled by diffusion policies~\citep{chi2025diffusion}. Sequence modeling approaches like Decision Transformer~\citep{chen2021decisiontransformerreinforcementlearning} and Trajectory Transformer~\citep{janner2021offlinereinforcementlearningbig} frame action prediction as next-token prediction, while multi-task agents like Gato~\citep{reed2022generalistagent} extend this to diverse domains. Deep learning revived interest in end-to-end BC for driving~\citep{bojarski2016end} and manipulation~\citep{mandlekar2021matters}, with vision-language-action models like RT-2~\citep{brohan2023rt2visionlanguageactionmodelstransfer} and OpenVLA~\citep{kim2024openvlaopensourcevisionlanguageactionmodel} transferring web-scale knowledge to robotic control. In parallel, GUI agents learn to navigate computer interfaces through clicks and keystrokes; systems like WebArena~\citep{zhou2024webarenarealisticwebenvironment} and SeeClick~\citep{cheng2024seeclickharnessingguigrounding} train models to complete web tasks from screenshots. Our setting resembles GUI agents but focuses on scientific annotation rather than general web navigation, providing cleaner evaluation (ground truth is known) and exposing distinct challenges like long-horizon credit assignment and rare critical actions.

\paragraph{Scientific Annotation Automation.}
Automation in scientific imaging has primarily pursued direct prediction: flood-filling networks for connectomics~\citep{januszewski2018high}, tracking algorithms for cells and animals~\citep{lauer2022multi}, pose estimators for keypoints~\citep{mathis2018deeplabcut,pereira2022sleap}. These approaches discard the rich supervision in how humans actually annotate: navigation patterns, verification habits, and error correction strategies. Interactive methods like SAM~\citep{kirillov2023segment} incorporate user clicks but model single refinement interactions rather than full annotation sessions. PseudoClick~\citep{liu2022pseudoclickinteractiveimagesegmentation} imitates human click placement for interactive segmentation but operates in a Markovian single-image setting without modeling sequential annotation workflows. RLCorrector~\citep{nguyen2022rlcorrectorreinforcedproofreadingcelllevel} uses reinforcement learning to proofread cell segmentations, learning correction strategies through reward optimization rather than behavioral cloning from annotation traces. Active learning~\citep{settles2009active} reduces annotation burden by selecting informative samples but still treats annotation as atomic labeling events.

\paragraph{Synthetic Data and Positioning.}
Synthetic data enables controlled study of learning dynamics~\citep{chan2022datadistributionalpropertiesdrive,raventos2023pretrainingtaskdiversityemergence,allen2023physics1,park2024competition}. We use synthetic annotation data to isolate factors confounded in real studies: task difficulty, behavior patterns, error rates. The closest methodological analog is Video PreTraining (VPT)~\citep{baker2022video}, which applies behavioral cloning to Minecraft gameplay. We adapt this approach to scientific annotation, using synthetic data to overcome the absence of large-scale behavioral datasets. Prior work has rarely focused on training from full annotation sessions, the sequences of navigation, placement, verification, and correction that constitute expert behavior.

\section{Framework}
\label{sec:framework}

%

Training behavioral cloning models requires interleaved sequences of GUI screenshots and user actions: $(\text{img}_0, \text{click}_0, \text{img}_1, \text{click}_1, \ldots)$. We generate this data by separating three concerns: \emph{what exists in the world} (task instances with ground truth), \emph{how a human would annotate it} (action sequences), and \emph{how to render it} (GUI screenshots). This separation enables controlled experiments: we can vary task difficulty, annotation behavior, and error rates independently. Concretely, each task is implemented as a 5-step pipeline: data preparation, instance creation, virtual-annotation generation, ML dataset rendering, and visualization (see Appendix~\ref{app:tasks:framework}).

\paragraph{Virtual Human Annotation Model.}
The key component is a procedural model of human annotation behavior. Given a task instance with ground truth, it generates realistic action sequences including:
\begin{itemize}[leftmargin=*,nosep]
    \item \textbf{Navigation}: moving through z-slices, rotating 3D views, switching spectral channels
    \item \textbf{Placement}: clicking to place markers, draw polygons, or select objects
    \item \textbf{Verification}: checking work by toggling views or revisiting previous locations
    \item \textbf{Mistakes and corrections}: simulated errors (10--30\% of placements, varying by task) followed by undo actions
\end{itemize}
This captures supervision about the annotation \emph{process}, the navigation patterns, verification habits, and error recovery strategies that constitute expert behavior, rather than just the final output. We validate this procedural model against four real human annotators on \taskCDT; the virtual annotator's action distributions (navigation, placement, correction rates) fall within the human range (Appendix~\ref{app:human_annotation}).

\paragraph{Task Suite.}
We construct 9 tasks spanning diverse annotation challenges (Table~\ref{tab:task_overview}). \emph{Tracking tasks} require maintaining object identity across frames: colored dots through z-slices (connectomics-inspired), neurons through elastic deformation, cells through division events, and animals through independent motion. \taskALT requires inferring keypoint positions from rendered morphology without explicit markers. \taskMIA tests cross-channel correspondence under geometric distortion. \taskTDE requires active information gathering, rotating an occluded object before classification. \taskRNC involves graph annotation (nodes at intersections, edges along roads). \taskSPF combines multi-band reasoning with polygon drawing. Full task details appear in Appendix~\ref{app:tasks}.

\begin{table*}[h]
\centering
\small
\vspace{1.5em}
\begin{tabular}{llc}
\toprule
\textbf{Task} & \textbf{Key Challenge} & \textbf{Seq. Len.} \\
\midrule
\taskCDT & 3D navigation, order constraints & 100--170 \\
\taskNT & Identity from spatial continuity & 155--275 \\
\taskCLT & Division events, tree structure & 50--90 \\
\taskABT & Multi-object, orientation & 255--540 \\
\taskALT & Implicit keypoints from shape & 200--660 \\
\taskMIA & Cross-channel correspondence & 20--65 \\
\taskTDE & Active exploration, occlusion & 5--10 \\
\taskRNC & Graph topology (nodes + edges) & 30--70 \\
\taskSPF & Multi-band reasoning, polygons & 55--125 \\
\bottomrule
\end{tabular}
\vspace{1.5em}
\caption{\textbf{Task Overview.} Sequence lengths indicate typical annotation episode duration (number of actions).}
\label{tab:task_overview}
\end{table*}

\paragraph{Model Architecture.}
We use a vision-language model~\citep{alayrac2022flamingovisuallanguagemodel,liu2023visualinstructiontuning} that processes interleaved screenshots and click coordinates. A DINOv2 vision encoder~\citep{oquab2024dinov2learningrobustvisual} extracts patch tokens from each frame, spatially pooled to $12 \times 9 = 108$ tokens. A transformer head processes the sequence with \emph{block-causal attention} (a standard pattern for interleaved image-token sequences): bidirectional within each frame's patches, but causal across frames. The model predicts the next click as independent distributions over $x$ and $y$ coordinates. We train four model sizes (25M to 320M parameters) to study scaling. Notably, long context annotations are hard to handle with off-the-shelf vision-language models; for example Qwen3-VL-4B-Instruct~\citep{bai2025qwen3vltechnicalreport} exceeds $80\:\mathrm{GB}$ GPU memory at 20 frames. Our architecture retains spatial structure (required for pixel-precise prediction) while remaining trainable. Full details are in Appendix~\ref{app:methodology}.

\paragraph{Action space.}
An action is a single click at canvas coordinates $(x, y)$; the simulator interprets it by hit-testing against the GUI, firing the corresponding button (e.g., \texttt{+z}, \texttt{undo}, \texttt{done}, an object-class selector) or recording a canvas placement. The model has a single output head (independent distributions over $x$ and $y$) that subsumes all action types. This canvas-click parametrization is intentional: any tool with clickable elements can be wrapped without modifying the model or its training objective.

\paragraph{Evaluation.}
We evaluate in two modes. \emph{Teacher-forced evaluation} measures next-action prediction accuracy given ground-truth history, enabling fine-grained analysis of per-action-type learning without compounding errors. \emph{Autoregressive evaluation} runs the model in closed loop with a GUI simulator: the model predicts clicks from screenshots, the simulator executes them through the GUI's JavaScript interface, and the process repeats until task completion or timeout. Critically, the simulator determines action type from coordinates alone; the model cannot specify action types directly, ensuring evaluation measures whether it has learned \emph{where} to click.

\paragraph{Key terminology.}
\textbf{Motor actions} place markers, polygons, or other content at specific pixel coordinates on the canvas; \textbf{task-critical decisions} are actions that affect workflow correctness rather than spatial precision (e.g., \texttt{undo}, \texttt{done}, object/class selection). \textbf{Action validity} checks whether a predicted click lands on a valid GUI element instead of inactive areas.

\section{Results}

\subsection{Single Task Model}
\label{sec:single-task}

\begin{figure*}[ht]
\vskip 0.2in
\begin{center}
\includegraphics[width=0.9\textwidth]{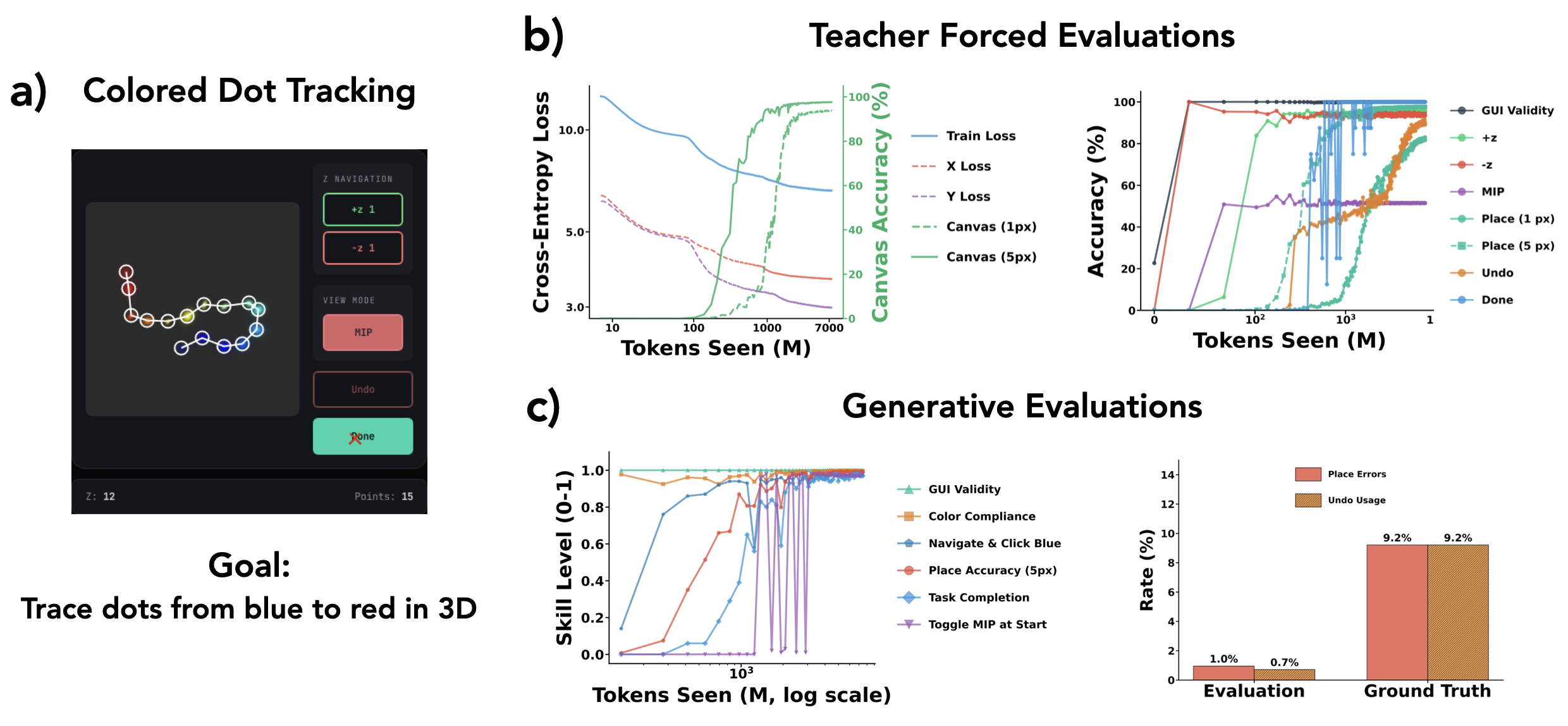}
\vspace{1.5em}
\caption{\textbf{Single Task Model Analysis.} (a) The \taskCDT task: dots are scattered in 3D from blue (start) to red (end); the annotator uses $+$z/$-$z buttons to navigate in depth and clicks dots in color order, then clicks Done. (b) Left: training loss (total, $x$, $y$) and canvas click accuracy at 1px and 5px precision; fine placement accuracy emerges sharply after sufficient loss reduction. Right: teacher-forced evaluation shows when capabilities emerge. GUI understanding and button usage emerge quickly, but placement accuracy and undo accuracy emerge slower; notably, undo accuracy eventually reaches near 100\%, indicating the model learns to perfectly correct mistakes. (c) Left: generative evaluation of skill acquisition; GUI validity is always high, navigation emerges fast, dot placement improves gradually, but MIP button usage (showing maximum intensity projection) emerges very late despite having a sharp transition. Right: models naturally make fewer mistakes than the training data distribution, a known phenomenon where models under-represent minorities; yet teacher-forced evaluation (b) confirms the model can still correct errors when they occur.}
\label{fig:single-task}
\end{center}
\vskip -0.2in
\end{figure*}

We first analyze how models learn annotation behavior on a single task (\taskCDT, depicted in Figure~\ref{fig:single-task}a) to understand the dynamics of skill acquisition. Unlike in natural language modeling, GUI annotation tasks decompose into clearly identifiable sub-skills: clicking specific buttons, navigating a dataset, placing markers at precise locations, and correcting errors. Training a base model (95M parameters) on \taskCDT provides insight into how these skills are acquired.

Figure~\ref{fig:single-task}b (left) shows training dynamics: total loss decomposes into $x$ and $y$ coordinate losses, while canvas click accuracy is measured at both coarse (5px) and fine (1px) precision. Fine placement accuracy emerges sharply only after sufficient loss reduction, suggesting a phase transition in spatial precision.

\textbf{Skills emerge in a hierarchy.} Figure~\ref{fig:single-task}b (right) shows teacher-forced accuracy per action type across training. General GUI understanding and button usage emerge quickly, as the model quickly learns to click valid portions of the interface and use navigation buttons ($+$z, $-$z). In contrast, capabilities requiring complex visual reasoning emerge more slowly: placement accuracy improves gradually, and undo accuracy (which requires recognizing that a mistake was made) emerges even later. Notably, undo accuracy eventually reaches near 100\%, indicating the model learns to perfectly identify and correct errors when landing into mistake states. Overall, simpler visual tasks are learned before those requiring more abstract reasoning.

We also studied the order in which skills are learned via generative evaluation (Figure~\ref{fig:single-task}c, left). GUI validity is consistently high throughout training. Navigation emerges quickly, as the model quickly learns to use $+$z to find and click the first blue dot. Dot placement accuracy improves more gradually through training. Interestingly, MIP button usage (which toggles a maximum intensity projection view) emerges very late despite exhibiting a sharp transition when it does appear. This stereotyped behavior from the training data (toggling MIP at episode start) requires the model to learn a specific sequential pattern rather than respond to an immediate visual cue, explaining its delayed acquisition.

\textbf{Some skills exhibit non-monotonic learning.} Done accuracy in teacher-forced evaluations (Figure~\ref{fig:single-task}b (right)) and the model's ability to toggle MIP at the start of generative evaluation (Figure~\ref{fig:single-task}c (left)) oscillate through learn-unlearn cycles before stabilizing. See Figure~\ref{fig:single_task_cdt_1} for individual plots with error bars.

\textbf{Models under-represent mistakes, but still know how to correct them.} Figure~\ref{fig:single-task}c (right) reveals a striking pattern: models naturally make far fewer mistakes than present in the training data. Training data includes mistakes followed by undo clicks (9.2\% of place actions), but trained models in generative evaluation make errors only 1.0\% of the time and rarely use undo (0.7\%). This is a known phenomenon where models amplify dataset biases by under-representing minority classes~\citep{zhao2017menlikeshoppingreducing}, effectively learning to ``skip'' the mistake-then-correct pattern. One might conclude the model simply cannot handle errors. However, teacher-forced evaluation (Figure~\ref{fig:single-task}b, right) shows otherwise: when forced into a mistake state, the model predicts undo with near-perfect accuracy. The model has learned error correction but chooses not to exercise it during generation: it prefers to avoid mistakes entirely rather than make and correct them.

\subsection{Multi Task Model}
\label{sec:multi-task}

\begin{figure*}[ht]
\vskip 0.2in
\begin{center}
\includegraphics[width=0.75\textwidth]{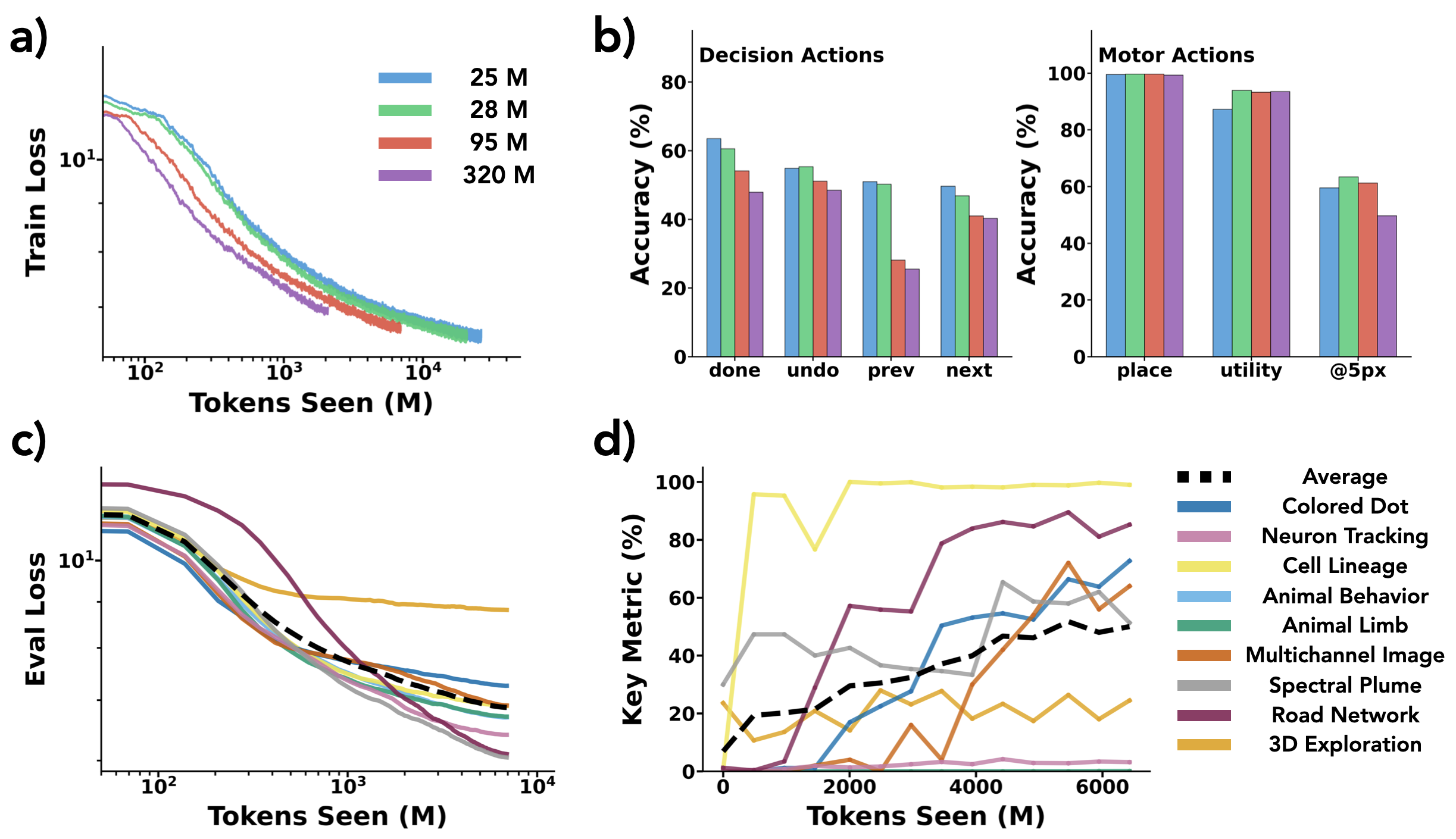}
\vspace{1.5em}
\caption{\textbf{Multi-task Model Analysis.} (a) Training loss vs.\ tokens seen for four model sizes (25M--320M parameters). Larger models are more data-efficient, achieving lower loss with fewer tokens. (b) Left: Decision action accuracy (done, undo, prev, next) at equal loss: surprisingly, smaller models outperform larger models. Right: Motor action accuracy (placing markers, using navigation) shows no model size dependence. (c) Per-task loss decomposition during training. Different tasks are learned at different times; \taskTDE dominates late-training loss despite containing learnable features (Section~\ref{sec:internals}). (d) Per-task evaluation metrics during training (see Appendix for metric definitions). The smooth aggregate improvement masks discrete task-level breakthroughs occurring at different times.}
\label{fig:multi-task}
\end{center}
\vskip -0.2in
\end{figure*}

We trained four model sizes (25M, 28M, 95M, 320M parameters) on all 9 tasks jointly. Each model consists of a frozen DINOv2 vision encoder plus a transformer head that processes interleaved image-action sequences. Larger models are more data-efficient, achieving lower loss with fewer tokens seen (Figure~\ref{fig:multi-task}a), consistent with language model scaling laws. However, at our compute budget, we did not reach the regime where larger models become compute-efficient; the 25M model trained longer outperforms the 320M model at equal FLOPs (Figure~\ref{fig:app_scaling_flops}). The main multi-task run took ${\sim}2$ days on 4$\times$A100; all scaling and ablation runs combined fit in ${\sim}1$ week of 4$\times$A100 compute.

Comparing models at \textit{equal training loss} reveals that within our scale range, smaller models are slightly \emph{better} at decision actions (Figure~\ref{fig:multi-task}b, left). When choosing whether to click done, undo a mistake, or navigate between frames, smaller models consistently outperform larger ones. In contrast, motor actions (placing markers at correct pixel locations, using navigation utilities) show no model size dependence (Figure~\ref{fig:multi-task}b, right). We hypothesize that smaller models, forced to compress information due to limited capacity, learn more abstract decision-making factors that transfer across contexts, while larger models can afford to memorize surface patterns.

Decomposing loss by task during training (Figure~\ref{fig:multi-task}c) reveals that different tasks are learned at different times, reminiscent of sudden breakthroughs in language model training~\citep{michaud2024quantizationmodelneuralscaling, kangaslahti2025hiddenbreakthroughslanguagemodel}. Early in training, simpler tasks like \taskNT improve rapidly. Later, \taskTDE dominates the loss and, notably, stops improving despite not being solved. In Section~\ref{sec:internals}, we show that the model \emph{does} learn the correct classification features internally, yet fails to use them for prediction, consistent with recent findings that VLMs can encode visual information they fail to act upon~\citep{fu2025hiddenplainsightvlms}.

Per-task evaluation metrics (Figure~\ref{fig:multi-task}d; see Appendix for metric definitions) show that the smooth aggregate improvement curve masks discrete task-level breakthroughs occurring at different training stages. This staged acquisition suggests that multi-task behavioral cloning, like language modeling, learns skills in a structured order determined by task complexity and data statistics. 






\subsection{Downstream Adaptation}
\label{sec:downstream}

\begin{figure}[ht]
\vskip 0.2in
\begin{center}
\includegraphics[width=1.0\linewidth]{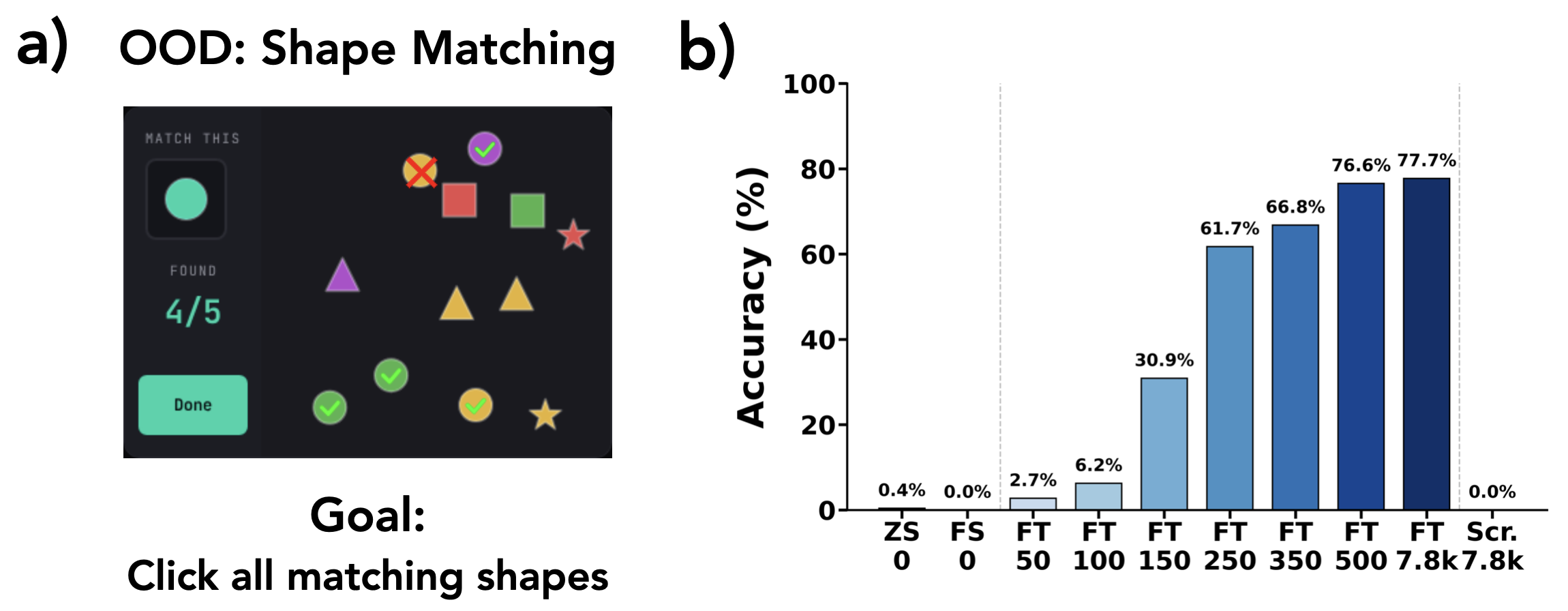}
\vspace{1em}
\caption{\textbf{Downstream Adaptation.} (a) The held-out \taskSM task: click all objects matching the template in the sidebar. This task was not seen during training and has a different GUI layout. (b) Evaluation across adaptation methods. Zero-shot (ZS) and few-shot (FS) in-context learning achieve negligible accuracy. Fine-tuning saturates at 500 sequences (76.6\%); 7,800 sequences yields similar performance. Training from scratch fails entirely (0\%) even with 7,800 sequences.}
\label{fig:downstream}
\end{center}
\vskip -0.2in
\end{figure}

To test whether multi-task pretraining produces models that adapt efficiently to new tasks, we evaluate on a held-out \taskSM task not seen during training (Figure~\ref{fig:downstream}a). This task requires clicking all objects matching a template shown in a sidebar. Unlike all training tasks, which have button elements on the right side of the GUI, this task places them on the left.

Fine-tuning the pretrained multi-task model on 500 sequences from the new task achieves strong performance (76.6\% accuracy, Figure~\ref{fig:downstream}b). In contrast, training the same architecture from scratch on 7800 sequences, over 15 times more data, achieves 0\% accuracy. The from-scratch model fails completely despite having access to far more task-specific data\footnote{It might be confusing why a task would fail to train on this seemingly simple task. We suspect that the main reason is that the action requires a choice between many objects (see Fig.~\ref{fig:downstream}a.) which dilutes the learning signal.}.




This gap demonstrates that multi-task pretraining builds transferable representations for annotation behavior. The pretrained model has learned general patterns of GUI interaction, visual attention, and action sequencing that transfer to novel tasks. The from-scratch model must learn all of this simultaneously with the task-specific behavior, and fails.

This suggests synthetic pretraining followed by fine-tuning on limited real data as a practical path forward.

However, in-context learning fails entirely (shown in  Figure~\ref{fig:downstream}b). We tested whether the pretrained model could adapt to the new task through examples provided in context, either as demonstration sequences or as prefixes of ground-truth actions. In all cases, accuracy remained at 0\%. The model can transfer via fine-tuning but cannot learn from in-context examples. This suggests that while pretraining builds useful representations, it does not induce meta-learning capabilities for annotation tasks, possibly due to limited task diversity (9 tasks).

\textbf{VLM Baselines.} We evaluated Gemini 3 Flash Preview and Qwen3-VL-32B-Instruct on all 9 tasks with a fully descriptive scaffold (text state plus the 3 most recent screenshots per step). Our 95M BC model outperforms both on teacher-forced placement accuracy @5px across all 5 placement tasks (e.g., \taskCDT 97.4\% vs.\ 80.0\% / 25.0\%), and in autoregressive evaluation both VLMs fail on most tasks (Gemini succeeds on 3/9, Qwen on 0/9). See Appendix~\ref{app:vlm_baselines}.

\textbf{UI Layout Adaptation.} To test whether the model has learned the underlying annotation task rather than memorizing pixel positions, we re-rendered \taskCDT with 8 alternative GUI variants (varying theme, panel position, button style, and size) plus 3 held-out OOD variants combining visual axes never seen during fine-tuning. The pretrained model degrades sharply on rearranged layouts; brief fine-tuning on the 8 ID variants restores performance across all 11 variants, including the held-out OOD ones, without harming performance on the original layout. See Appendix~\ref{app:ui_adaptation}.

\textbf{DAgger.} On the two tasks where autoregressive BC fails (animal\_behavioral\_tracking and animal\_limb\_tracking), $\beta$-DAgger~\citep{ross2011reduction} also fails to bootstrap performance, suggesting fundamental task difficulty rather than compounding error. See Appendix~\ref{app:dagger}.


\subsection{Model Internal Analysis}
\label{sec:internals}

Having trained models on annotation behavior, we examine what representations they learn using linear probing~\citep{li2022emergent,gurnee2023language,nanda2023emergent}.

\begin{figure*}[!ht]
\vskip 0.2in
\begin{center}
\includegraphics[width=0.74\textwidth]{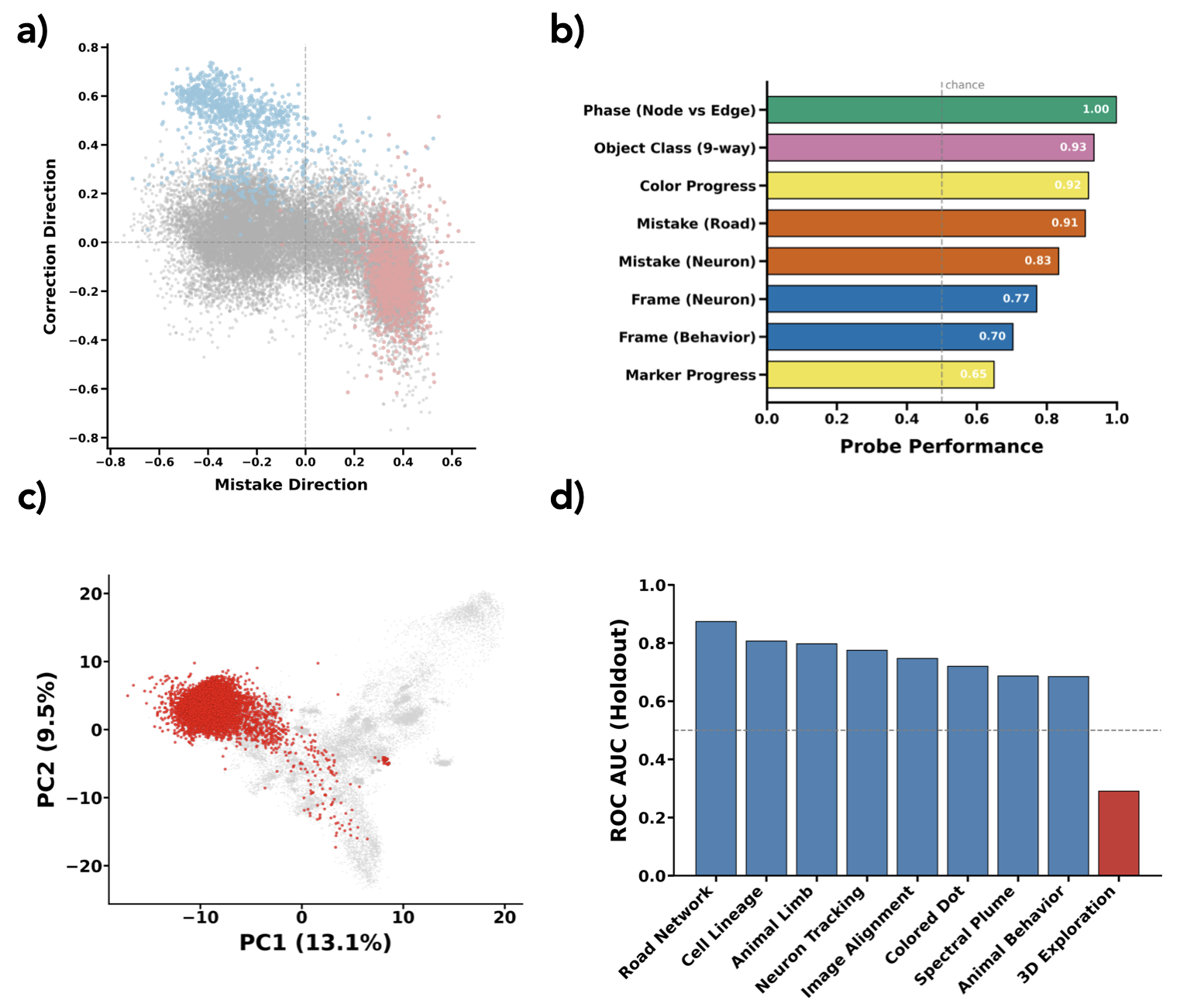}
\vspace{1.5em}
\caption{\textbf{Model Internals.} (a) Projection of activations onto learned mistake and correction directions. Mistakes (red) cluster with high mistake scores; corrections (blue) cluster with high correction scores; correct actions (gray) are distributed throughout. The orthogonal clustering demonstrates distinct representations for error states versus recovery actions (mistake probe ROC AUC = 0.87, correction probe ROC AUC = 0.99). (b) Linear probes decode diverse information from model activations: task phase (node vs.\ edge placement, AUC = 1.00), object identity (9-way classification, AUC = 0.93), task progress (R$^2$ = 0.65--0.92), error states (AUC = 0.83--0.91), and temporal position (AUC = 0.70--0.77). All probes perform well above chance (dashed line). (c) PCA of activations pooled across all 9 tasks. Mistakes (red) form a distinct cluster separate from correct actions (gray), suggesting a partially universal mistake representation. (d) Leave-one-task-out cross-validation: probes trained on 8 tasks and tested on the held-out task. 8/9 tasks show above-chance transfer (blue bars, mean AUC = 0.71), while \taskTDE shows poor transfer (red bar), indicating both generalizable and task-specific components of mistake detection.}
\label{fig:internals}
\end{center}
\vskip -0.2in
\end{figure*}

We first examine whether models represent mistakes and corrections. We train linear probes on the residual stream activations at the [cls] token (the click-generating position) to predict whether the upcoming action is a mistake or a correction. Due to the severe class imbalance ($<$5\% of actions are mistakes), we use ROC AUC rather than accuracy as our metric, since a classifier which predicts ``no mistake'' for all inputs would achieve $>$95\% accuracy. We find that both mistake and correction states are linearly separable: the mistake probe achieves 0.92 ROC AUC on single-task data and 0.87 when pooled across all 9 tasks; the correction probe achieves 0.99 ROC AUC. As shown in Figure~\ref{fig:internals}a, projecting activations onto the learned mistake and correction directions reveals clear clustering: mistakes (red) separate along the mistake axis while corrections (blue) separate along the correction axis, with correct actions (gray) distributed throughout.

Beyond mistakes, we probe for other latent variables relevant to annotation (Figure~\ref{fig:internals}b). These include: annotation phase in \taskRNC (node placement vs.\ edge construction; ROC AUC = 1.00), object class in \taskTDE (9-way classification; ROC AUC = 0.93), color progress in \taskCDT ($R^2 = 0.92$), frame position in \taskNT and \taskABT (ROC AUC = 0.70--0.89), and marker progress in \taskMIA. Importantly, variables like ``frame position'' are not simply the number of frames observed; annotators navigate back and forth with prev/next buttons, so the model must track position within the data rather than just count inputs. Similarly, color progress reflects semantic progress through the annotation task, not merely temporal sequence position. All probes perform well above chance, indicating that models develop internal representations of task-relevant state.

\textbf{A Shared Mistake Representation} A natural question is whether these representations are task-specific or shared across tasks. Mistake detection, for instance, could in principle be computed from visual features alone (e.g., a marker placed on empty space). To test for shared structure, we pool activations from all 9 tasks and visualize via PCA (Figure~\ref{fig:internals}c). Strikingly, mistakes cluster together across different tasks, suggesting a partially universal ``something is wrong'' representation that transcends task-specific error patterns.

To quantify cross-task generalization, we train mistake probes on 8 tasks and evaluate on the held-out task (Figure~\ref{fig:internals}d). Mean transfer ROC AUC = 0.71, with 8 of 9 tasks showing above-chance transfer. The exception is \taskTDE (ROC AUC = 0.29), which is below the 0.5 chance level, indicating the probe has learned a direction that is \emph{anti-correlated} with mistakes in this task. This makes sense: \taskTDE mistakes are classification errors (clicking the wrong class button), fundamentally different from spatial placement errors in other tasks. The model appears to encode these distinct error types in opposing directions.

\subsection{Connectomics Tracing}
\label{sec:connectomics_tracing}

\begin{figure}[!ht]
\centering
\begin{minipage}[c]{0.4\linewidth}
\centering
\includegraphics[width=0.9\linewidth]{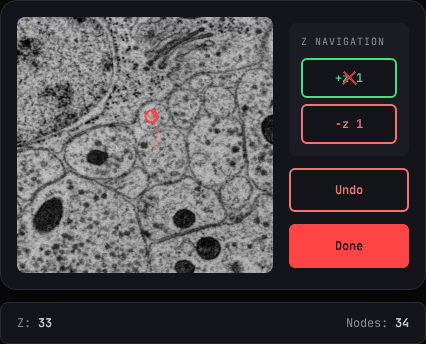}
\end{minipage}\hfill
\begin{minipage}[c]{0.59\linewidth}
\centering
\small
\setlength{\tabcolsep}{3pt}
\renewcommand{\arraystretch}{1.15}
\begin{tabular}{lcc}
\toprule
\textbf{Metric} & \textbf{Human} & \textbf{\textit{C. elegans}} \\
\midrule
Skeleton acc. & \textbf{95.1\%} & \textbf{89.4\%} \\
Placement @5px & 95.1\% & 78.0\% \\
Placement @10px & 97.5\% & 83.9\% \\
Done rate & 100\% & 100\% \\
\bottomrule
\end{tabular}
\end{minipage}
\vspace{0.5em}
\caption{\textbf{Connectomics Tracing.} Left: the neuron tracing GUI shared by both tasks. The annotator clicks the target neuron's cross-section at each z-slice and uses $\pm$z to navigate, Undo to correct, and Done to finish. Right: autoregressive evaluation after fine-tuning, on 28 held-out axons from H01~\citep{shapson2024petavoxel} and 13 held-out neurons from the \textit{C. elegans} nerve ring~\citep{witvliet2021connectomes}. See Appendix~\ref{app:connectomics_tracing} for full metrics.}
\label{fig:connectomics}
\end{figure}

We fine-tune the multi-task pretrained 95M model on real EM neuron tracing from two volumes: H01 human cortex~\citep{shapson2024petavoxel} and the \textit{C. elegans} nerve ring~\citep{witvliet2021connectomes}, using raw EM with no segmentation overlay. Autoregressive evaluation of tracing reaches 95.1\% (H01) and 89.4\% (\textit{C. elegans}) skeleton accuracy with 100\% task completion (Figure~\ref{fig:connectomics}); fine-tuning peaks well before one epoch in both runs. See Appendix~\ref{app:connectomics_tracing} for full metrics.

\section{Discussion}

How models learn annotation behavior is central to scaling behavioral cloning for scientific applications. Our analysis reveals that standard maximum likelihood training creates systematic blind spots: rare but critical actions like error correction are underweighted, leading to models that can correct mistakes when prompted but choose not to during generation. Understanding when and how behaviors emerge during training (specifically, the hierarchical acquisition we observe) could guide better data curation strategies. Oversampling critical decision points, or using loss weighting that reflects action importance rather than frequency, may address these bottlenecks.

Our findings also have implications for interface design. Current annotation GUIs are designed for humans alone. But if we understand what makes behaviors hard for models to learn (sparse critical actions, long-horizon dependencies, ambiguous decision points), we can design AI-native interfaces that make these patterns more frequent or more learnable. For instance, an explicit ``verify'' button might be easier to learn than an implicit toggle-check-toggle pattern. The synthetic framework we provide enables evaluating such designs before deployment.

More broadly, we hope this work helps resolve the collective action problem that has hindered progress on behavioral cloning for scientific annotation. The uncertainty around scaling (how much data is needed, what performance to expect, which tasks are tractable) has discouraged both ML researchers and domain scientists from investing in this direction. By providing systematic baselines across controlled tasks, we offer a foundation for making informed decisions about when behavioral cloning is viable and how to scale it effectively.

\subsection{Limitations}

Our scaling claims hold within the 25M--320M range we test and were not pushed to frontier scales; the decision-action finding in particular is scope-limited (Sec.~\ref{sec:multi-task}). Our action space is restricted to canvas clicks (intentional for generality, Sec.~\ref{sec:framework}) but excludes modalities like keyboard shortcuts, scrolling, or freehand strokes that some annotation tools rely on. Aside from DAgger (Appendix~\ref{app:dagger}), which left the failing tasks unchanged, we do not explore reward-based RL; reward shaping may address rare action underweighting. In-context learning failed entirely; inducing ICL for procedural tasks remains open. Our architecture is memory-intensive, limiting context length; efficient alternatives merit exploration.

\section*{Impact Statement}

This paper advances machine learning for scientific data annotation. Such tools should augment rather than replace the trained experts who perform this work, freeing them for higher-level judgement and edge cases. Behavioral cloning inherits the biases of training annotators, and deployment at scale can systematize a single lab's idiosyncrasies; multi-annotator collection and audits against ground truth help mitigate this. Models that learn GUI interaction patterns are general-purpose and could in principle be repurposed for surveillance-style monitoring; we encourage care when releasing models trained on real, identifiable annotation sessions.

\vspace{-0.5em}
\section*{Acknowledgements}

We thank Aravi Samuel, Jeff Lichtman, Hidenori Tanaka, Venkatesh Murthy, Puneet Bhutta, Xu Pan, Pranav Misra, Jorin Overwiening, Fuming Yang, Helena Casademunt, and Felix Sosa.

\bibliography{main}
\bibliographystyle{icml2026}

\newpage
\appendix
\onecolumn

\begin{center}
\LARGE\textbf{Appendix}
\end{center}
\vspace{1em}

\section{Tasks}
\label{app:tasks}

\subsection{Framework}
\label{app:tasks:framework}

Training behavioral cloning models requires interleaved sequences of GUI screenshots and user actions: $(\text{img}_0, \text{click}_0, \text{img}_1, \text{click}_1, \ldots)$. We generate this data by separating three concerns: \emph{what exists in the world} (task instances), \emph{how a human would annotate it} (action sequences), and \emph{how to render it} (GUI screenshots). This section describes the pipeline.

\paragraph{Architecture Overview.}
Our framework consists of three core components:

\begin{enumerate}
    \item \textbf{Task Instance Generator}: Produces the object to be annotated and its ground truth. This may be purely synthetic (e.g., 3D point clouds, procedural road networks) or derived from real images (e.g., natural images with synthetic deformations).

    \item \textbf{Virtual Human Annotation Model}: Simulates human annotation behavior given a task instance. Because we have access to ground truth, we can efficiently generate realistic action sequences that mimic human patterns: exploration before committing, verification after placement, occasional mistakes followed by corrections. We can also perturb parameters (error rates, verification frequency, navigation strategies) to study how behavioral variations affect learning.

    \item \textbf{GUI Renderer}: Takes both task instances and action sequences, then renders the interleaved image-action training data. The GUI is defined in HTML/CSS/JavaScript; a headless browser (Playwright + Chromium) executes the rendering.
\end{enumerate}

\paragraph{The 5-Step Pipeline.}
Each task is implemented as a 5-step pipeline:

\begin{table}[h]
\centering
\small
\begin{tabular}{clll}
\toprule
\textbf{Step} & \textbf{Name} & \textbf{Output} \\
\midrule
1 & Download Data & External datasets (optional) \\
2 & Prepare Instances & Task instance JSONs \\
3 & Make Human Annotations & Action sequence JSONs \\
4 & Make ML Dataset & PNG images + metadata CSV \\
5 & Visualize & Diagnostic figures \\
\bottomrule
\end{tabular}
\vspace{1.5em}
\caption{\textbf{Pipeline Steps.} Steps for each task.}
\label{tab:pipeline_steps}
\end{table}

\paragraph{Virtual Human Annotation Model.}
Step 3 implements a procedural model of human annotation behavior. Given a task instance, it generates the complete sequence of actions a human would perform, including:

\begin{itemize}
    \item \textbf{Navigation}: Moving through z-slices, rotating 3D views, switching channels
    \item \textbf{Placement}: Clicking to place markers, draw polygons, or select objects
    \item \textbf{Verification}: Checking work by toggling views or revisiting previous locations
    \item \textbf{Mistakes and corrections}: Simulated errors (10--30\% of placements, varying by task) followed by undo actions
\end{itemize}

The annotation model operates entirely on abstract task data. For example, in the \taskCDT task, it receives a list of 3D points with colors and decides: ``navigate to slice 7, place marker at (0.45, 0.62), toggle MIP view to verify, navigate to slice 9\ldots'' This produces action sequences like:

\begin{lstlisting}[style=codestyle]
[{z: 0, action: "+z_1"}, {z: 1, action: "+z_1"}, ...,
 {z: 7, action: "place", x: 0.45, y: 0.62},
 {z: 7, action: "mip"}, ...]
\end{lstlisting}

This separation also means task instances and annotation sequences can be reused with different GUI designs.

\paragraph{HTML-Based GUI Rendering.}
Each task's GUI is defined as a self-contained HTML file with embedded CSS and JavaScript. We use Playwright with headless Chromium to render screenshots. The same HTML template serves both synthetic data generation and model evaluation. During evaluation, the model's predicted clicks are executed through the GUI's JavaScript interface.

The HTML template exposes a JavaScript API for external control:

\begin{lstlisting}[style=codestyle]
// Inject task data (3D points, colors, etc.)
setTaskData(sample);

// Set GUI state (current z-slice, placed markers, etc.)
setState({z: 7, mip: false, markers: [[0.45, 0.62, 7]]});

// For evaluation: execute action and get result
executeAction("+z_1");  // Returns {success: true/false}
\end{lstlisting}

\paragraph{Standardized Output Format.}
All tasks produce training data in a unified format:

\begin{itemize}
    \item \textbf{Images}: PNG files organized as \texttt{images/\{entry\_idx\}/\{step\}.png}
    \item \textbf{Metadata}: CSV with columns \texttt{file\_name}, \texttt{action\_x}, \texttt{action\_y}, \texttt{action\_type}, plus task-specific fields
    \item \textbf{HDF5 mirror}: For efficient loading, metadata is also saved as HDF5
\end{itemize}

This standardization enables a single dataset loader to train on all 9 tasks jointly.

\paragraph{GUISimulator for Evaluation.}
The same HTML-based architecture supports model evaluation. We provide a \texttt{GUISimulator} class that wraps the Playwright browser and exposes methods for interactive evaluation:

\begin{lstlisting}[style=codestyle,language=Python]
with GUISimulator() as sim:
    sim.load_task(sample_data)
    while not sim.is_complete():
        screenshot = sim.screenshot()
        x, y = model.predict(screenshot)
        result = sim.execute_click(x, y)  # Determines action from coords
    correctness = sim.check_correctness()
\end{lstlisting}

Critically, \texttt{execute\_click(x, y)} determines the action type from coordinates alone, so the model cannot ``cheat'' by specifying action types directly. This ensures evaluation measures whether the model has learned to click the right locations, not just output valid action labels.

\subsection{Main Tasks}
\label{app:tasks:main}

\begin{figure*}[ht]
\centering
\includegraphics[width=0.8\textwidth]{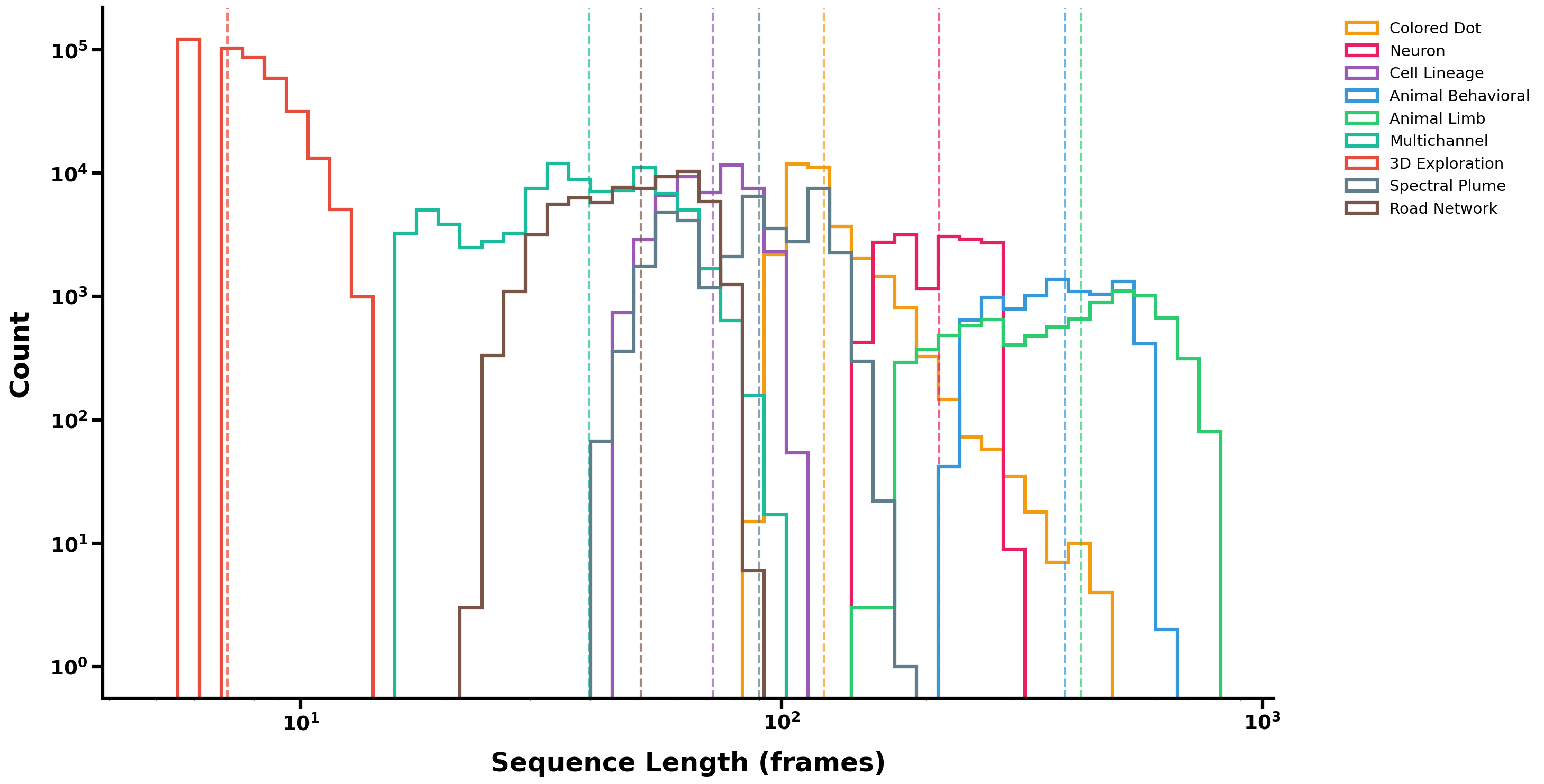}
\caption{\textbf{Distribution of sequence lengths across the multi-task dataset.} Sequence length varies substantially across tasks, from short episodes (\taskTDE, mean 7 actions) to long sessions (\taskALT, mean 420 actions; \taskABT, mean 389 actions). This variation reflects inherent differences in task complexity: \taskTDE requires only a few rotations before classification, while limb tracking demands placing many keypoints across multiple frames. The multi-task model must handle this 60$\times$ range in episode length.}
\label{fig:seq_l_per_task}
\end{figure*}

The sequence length distribution for all tasks is depicted in Figure~\ref{fig:seq_l_per_task}. 

\subsubsection{\taskCDT}
\label{app:tasks:colored_dot_tracking}

The annotator traces a trajectory of colored 3D dots through z-slices. Dots are colored by position along the path (jet colormap: blue$\rightarrow$red), derived from Omniglot handwriting strokes extruded into 3D. This task models the core challenge of connectomics proofreading: navigating a volumetric GUI to place markers in the correct order.

\paragraph{GUI.}
A 256$\times$256 canvas shows either the current z-slice or a maximum intensity projection (MIP). Controls include z-navigation buttons (+z, -z), MIP toggle, Undo, and Done. The status bar shows current z-position and points placed.

\paragraph{Annotation Behavior.}
The annotator begins by toggling MIP to see the full trajectory, then navigates slice-by-slice to place markers in order. Before each placement, they verify by checking adjacent slices. With 10\% probability, a point is placed incorrectly and must be undone. The sequence ends with MIP on and Done clicked.

\paragraph{Evaluation Metrics.}
Predicted and ground-truth markers are matched using the Hungarian algorithm for optimal bipartite assignment. A match is valid only if both spatial tolerances are satisfied: XY distance $\leq 0.02$ in normalized coordinates (2\% of the 256-pixel canvas) and Z distance $\leq 1.5$ slices (out of 16 total slices). Ground-truth Z coordinates are converted from normalized [0, 1] to slice units via $z_{\text{slice}} = z \cdot 16 - 0.5$. These tolerances approximate human annotation precision. We report recall (fraction of ground-truth markers matched), precision (fraction of predicted markers that are correct), and their harmonic mean F1. For matched pairs, we additionally report RMSE$_{xy}$ (in normalized coordinates) and MAE$_z$ (in slice units) to quantify localization accuracy. All metrics are computed per sample and then averaged (arithmetic mean) across samples; F1 is the primary metric used for comparison plots.

\begin{figure}[h]
\centering
\vspace{1.5em}
\begin{tabular}{ccc}
\includegraphics[width=0.18\textwidth]{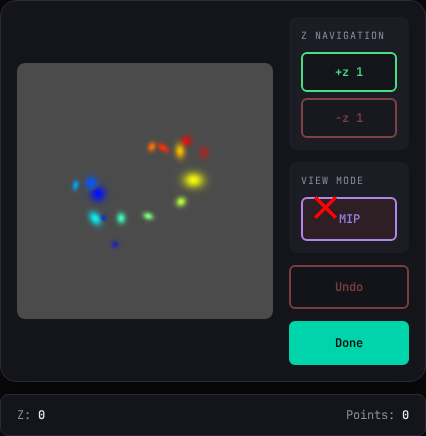} &
\includegraphics[width=0.18\textwidth]{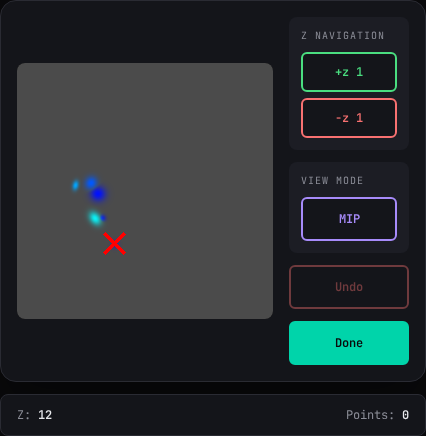} &
\includegraphics[width=0.18\textwidth]{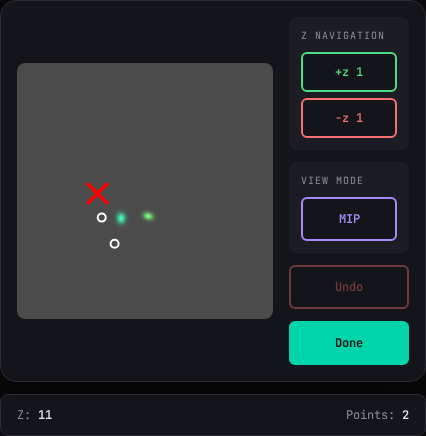} \\
(a) & (b) & (c) \\[0.5em]
\includegraphics[width=0.18\textwidth]{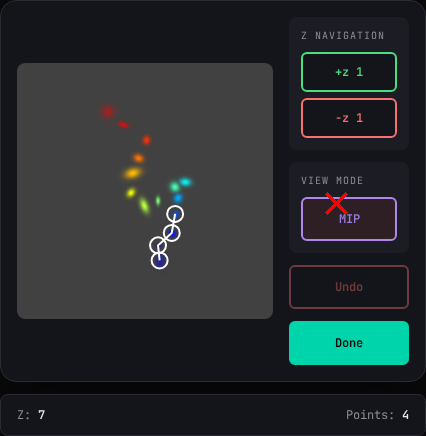} &
\includegraphics[width=0.18\textwidth]{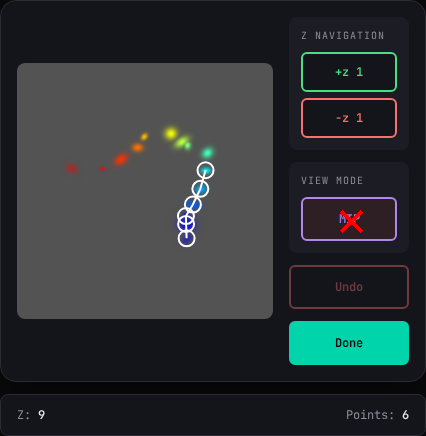} &
\includegraphics[width=0.18\textwidth]{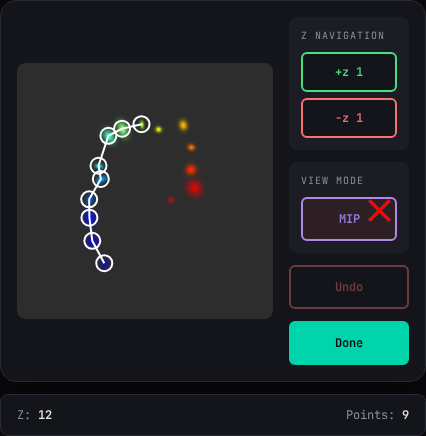} \\
\multicolumn{3}{c}{(d)} \\
\end{tabular}
\vspace{1.5em}
\caption{\textbf{\taskCDT.} Top: annotation sequence for modified hiragana from Omniglot: (a) initial MIP view showing full trajectory, (b) navigating to place first marker, (c) incorrect placement before undo. Bottom: diverse patterns (Greek, Georgian alphabets) in MIP view showing annotation progress.}
\label{fig:colored_dot_tracking}
\end{figure}

\subsubsection{\taskNT}
\label{app:tasks:neuron_tracking}

The annotator tracks 6--10 identical green neurons across 10 frames as the network undergoes elastic deformation, translation, and rotation. All neurons appear as identical Gaussian blobs, and identity comes purely from spatial continuity. This models calcium imaging analysis where neurons must be tracked through tissue deformation.

\paragraph{GUI.}
A 256$\times$256 canvas shows the current frame with placed markers as labeled circles. A sidebar contains neuron ID buttons (01--10), navigation buttons (prev/next), and control buttons (cancel/undo/done).

\paragraph{Annotation Behavior.}
The annotator works through neurons in groups of 4, placing each across all frames before moving to the next group. For each neuron: select ID button, place marker, advance frame, repeat. With 10\% probability, a placement is off-target and must be undone.

\paragraph{Evaluation Metrics.}
We evaluate tracking quality with three complementary metrics. A marker is considered correctly matched if its Euclidean distance to the ground-truth position is within a tolerance of 0.05 in normalized coordinates (5\% of canvas size, approximately 13 pixels on the 256$\times$256 canvas). \emph{Track accuracy rate} measures the fraction of neurons correctly tracked across all frames (matched neurons / total neurons), computed per sample and then averaged across samples. \emph{Position RMSE} quantifies localization error for markers placed outside the tolerance threshold, computed as $\sqrt{\frac{1}{N}\sum_i d_i^2}$ where $d_i$ is the Euclidean distance (in normalized coordinates) between placed and ground-truth positions, pooled across all samples. \emph{Detection rate} captures completeness: the fraction of neuron-frame pairs where any marker was placed (regardless of accuracy), averaged across samples. For comparison plots, we report the sample-averaged track accuracy rate as the primary metric.

\begin{figure}[h]
\centering
\vspace{1.5em}
\begin{tabular}{ccc}
\includegraphics[width=0.18\textwidth]{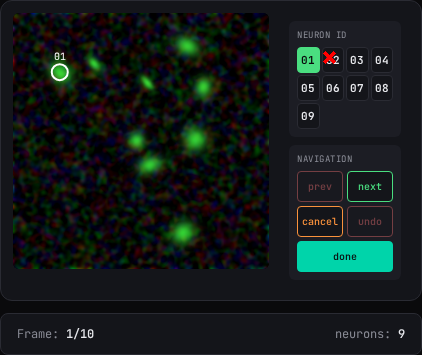} &
\includegraphics[width=0.18\textwidth]{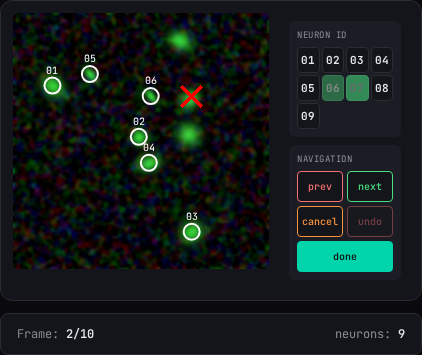} &
\includegraphics[width=0.18\textwidth]{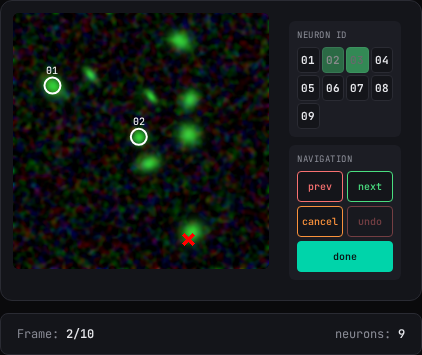} \\
(a) & (b) & (c) \\[0.5em]
\includegraphics[width=0.18\textwidth]{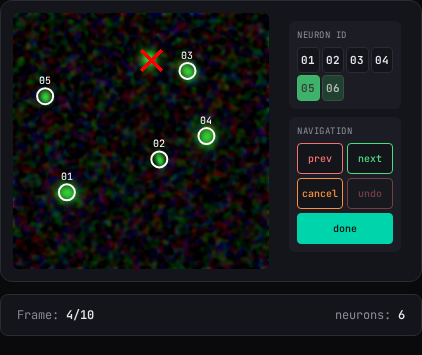} &
\includegraphics[width=0.18\textwidth]{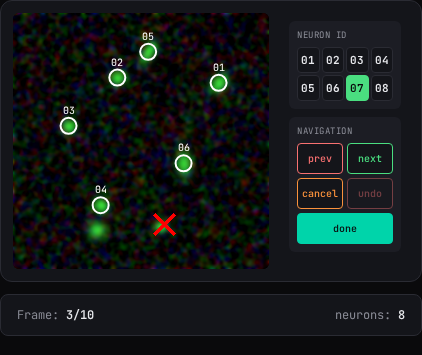} &
\includegraphics[width=0.18\textwidth]{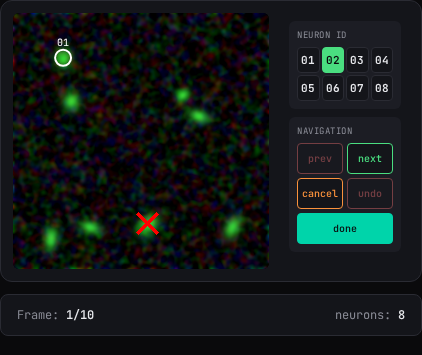} \\
\multicolumn{3}{c}{(d)} \\
\end{tabular}
\vspace{1.5em}
\caption{\textbf{\taskNT.} Top: annotation sequence: (a) neuron 01 placed, selecting 02, (b) mid-group with 6 neurons placed, placing 07, (c) misclick error on background. Bottom: diversity in neuron count (6--8) and annotation progress.}
\label{fig:neuron_tracking}
\end{figure}

\subsubsection{\taskCLT}
\label{app:tasks:cell_lineage_tracking}

The annotator tracks cell divisions in a developing embryo rendered as Voronoi-tessellated cells within an oval boundary. Starting from a single root cell, each division requires selecting the parent marker and placing markers on both daughter cells. This models developmental biology lineage tracing where cell genealogy must be reconstructed.

\paragraph{GUI.}
A 256$\times$256 canvas shows cells as colored Voronoi regions. Green circles mark cells in the current frame; white circles show propagated markers from previous frames. Navigation buttons (prev/next/undo/done) control the annotation.

\paragraph{Annotation Behavior.}
At frame 0, the annotator places a root marker. For each division event: navigate to the division frame, select the parent marker, then place markers on both children. Non-dividing cells auto-propagate. With 15\% probability, a marker is misplaced and must be undone.

\paragraph{Evaluation Metrics.}
We represent each lineage tree as a set of directed edges (parent $\to$ child relationships) and compare the user tree $E_u$ against ground truth $E_g$ using exact edge matching. Edge recall measures completeness (how many true lineage relationships were captured, $|E_u \cap E_g| / |E_g|$), while edge precision measures correctness (how many annotated relationships are valid, $|E_u \cap E_g| / |E_u|$). We combine these via F1 score. Tree similarity quantifies overall structural agreement: $1 - d/(|E_u| + |E_g|)$, where $d$ is the symmetric difference (edges that must be added or removed to match). Division accuracy specifically evaluates cell division events, measuring the fraction of true divisions (parent splitting into exactly two children) that were correctly identified with the correct parent-child assignments. All metrics are computed per sample and averaged across samples; we report the mean edge F1 score as the primary metric for comparisons.

\begin{figure}[h]
\centering
\vspace{1.5em}
\begin{tabular}{ccc}
\includegraphics[width=0.18\textwidth]{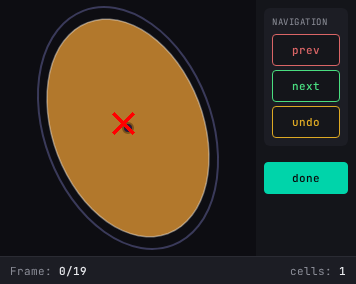} &
\includegraphics[width=0.18\textwidth]{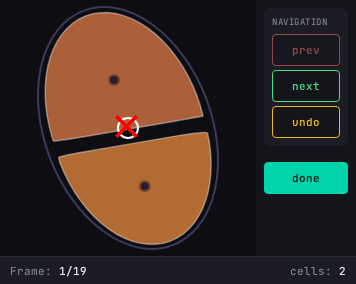} &
\includegraphics[width=0.18\textwidth]{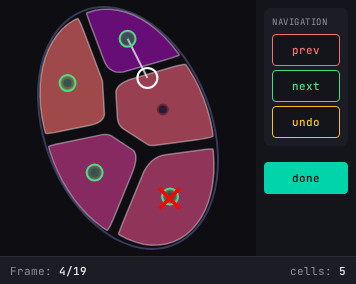} \\
(a) & (b) & (c) \\[0.5em]
\includegraphics[width=0.18\textwidth]{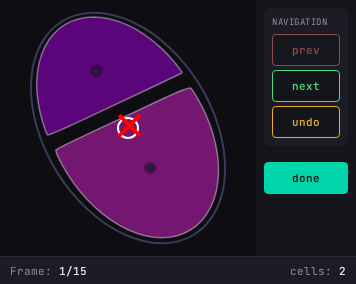} &
\includegraphics[width=0.18\textwidth]{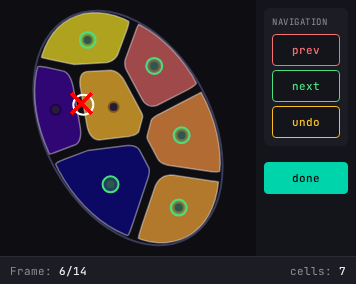} &
\includegraphics[width=0.18\textwidth]{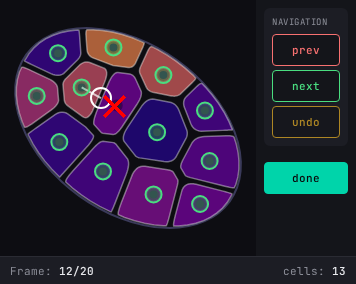} \\
\multicolumn{3}{c}{(d)} \\
\end{tabular}
\vspace{1.5em}
\caption{\textbf{\taskCLT.} Top: annotation sequence: (a) placing root marker on single cell, (b) selecting parent after first division, (c) misclick on wrong cell. Bottom (d): progression from 2 cells to 13 cells showing increasing lineage complexity.}
\label{fig:cell_lineage_tracking}
\end{figure}

\subsubsection{\taskABT}
\label{app:tasks:animal_behavioral_tracking}

The annotator tracks 4--8 animals across 10 video frames by placing front (head) and back (tail) markers on each. Animals have variable morphology (body segments, limbs, head shapes) but consistent appearance within a sequence. This models behavioral tracking studies where position and orientation must be annotated.

\paragraph{GUI.}
A 384$\times$384 canvas shows the current frame. A sidebar contains animal selection buttons (01--08), marker type buttons (Front/Back), and navigation controls. Markers appear as labeled circles (e.g., ``01F'', ``02B'').

\paragraph{Annotation Behavior.}
The annotator works through animals in order: select animal, place front marker on head, place back marker on tail, repeat for all animals, then advance frame. With 10\% probability each for near errors (slight offset) and far errors (random misclick), followed by undo.

\paragraph{Evaluation Metrics.}
We evaluate three aspects of annotation quality. \emph{Animal match rate} measures tracking accuracy: for each sample, we compute the fraction of correctly placed marker pairs (front and back within 5\% of the canvas, i.e., ${\sim}19$ pixels for the 384-pixel canvas) out of all expected pairs (animals $\times$ frames), then average across samples. \emph{Position RMSE} quantifies localization precision by computing root-mean-square error of position offsets for incorrectly placed markers, on a normalized $[0,1]$ scale. \emph{Detection rate} captures completeness: the fraction of samples where all required markers were placed (i.e., no missing annotations).

\begin{figure}[h]
\centering
\vspace{1.5em}
\begin{tabular}{ccc}
\includegraphics[width=0.18\textwidth]{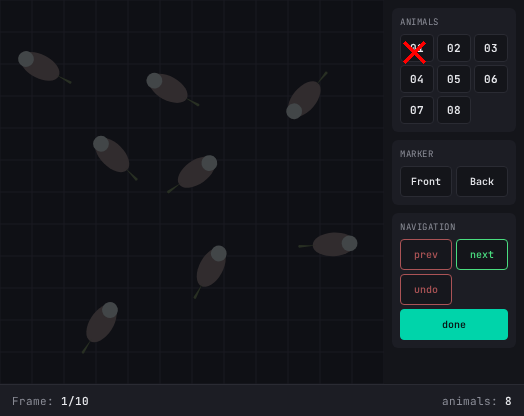} &
\includegraphics[width=0.18\textwidth]{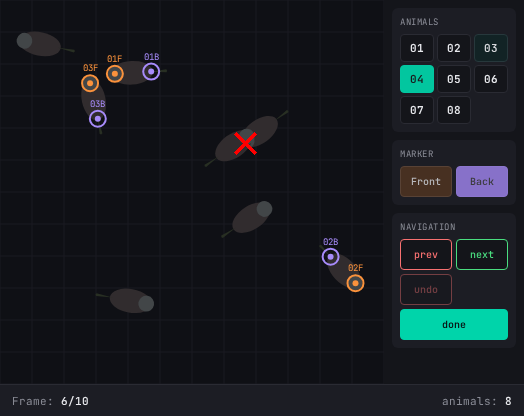} &
\includegraphics[width=0.18\textwidth]{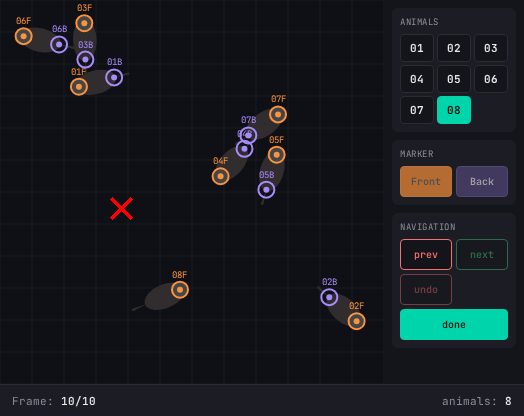} \\
(a) & (b) & (c) \\[0.5em]
\includegraphics[width=0.18\textwidth]{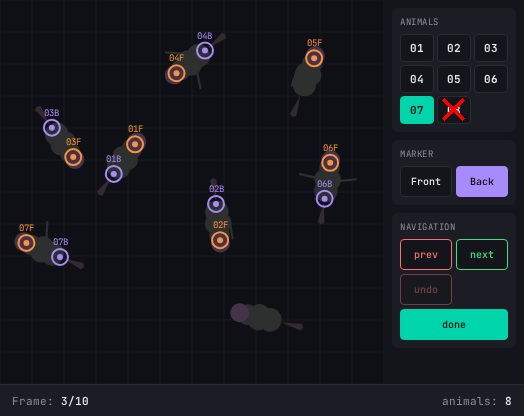} &
\includegraphics[width=0.18\textwidth]{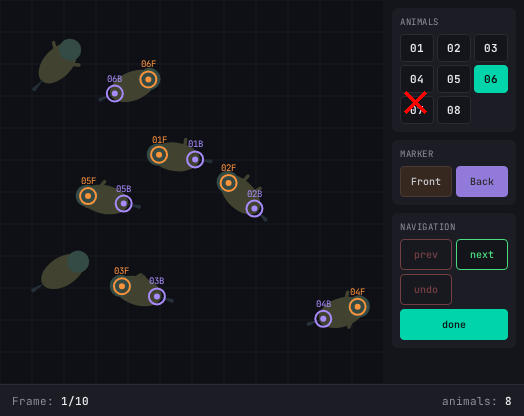} &
\includegraphics[width=0.18\textwidth]{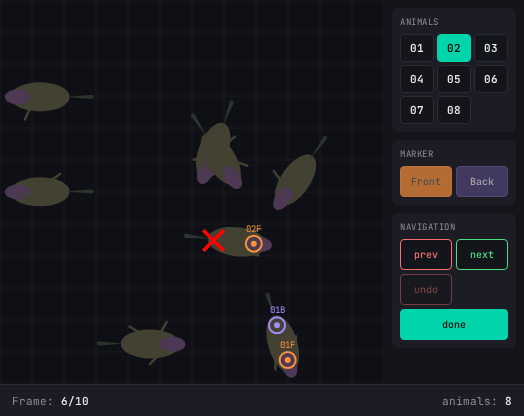} \\
\multicolumn{3}{c}{(d)} \\
\end{tabular}
\vspace{1.5em}
\caption{\textbf{\taskABT.} Top: annotation sequence: (a) selecting first animal, (b) mid-annotation with 3 animals marked on frame 6, (c) far error misclick near completion. Bottom: morphological diversity: insects with antennae, fish-like bodies, elongated forms.}
\label{fig:animal_behavioral_tracking}
\end{figure}

\subsubsection{\taskALT}
\label{app:tasks:animal_limb_tracking}

The annotator places keypoint markers on animal body parts across video frames. Animals (birds, insects, spiders, snakes) are rendered as shapes without visible keypoint markers, so positions must be inferred from morphology. This models pose estimation annotation where keypoints are implicit rather than marked.

\paragraph{GUI.}
A 256$\times$256 canvas shows the current frame. A tabbed sidebar organizes keypoints by body part (Body, Legs, Wings, etc.), with buttons for each keypoint. Navigation buttons (prev/next) switch frames. The status bar shows frame number and keypoints placed.

\paragraph{Annotation Behavior.}
The annotator works through tabs in order, clicking each keypoint button then placing it on the rendered animal. Limbs articulate $\sim$70° between frames, requiring re-inference of positions. With 10\% probability, a keypoint is placed slightly off (near error) and must be undone.

\paragraph{Evaluation Metrics.}
We evaluate keypoint localization using a Percentage of Correct Keypoints (PCK) proxy: the fraction of placed keypoints that fall within 3 pixels of their ground-truth positions. For each sample, we compute per-sample keypoint accuracy as $\text{correct\_keypoints} / \text{total\_keypoints}$, then report the arithmetic mean across all samples as the final keypoint accuracy metric used in plots. This measures spatial precision, specifically whether the model correctly infers body part locations from rendered morphology. We also track completion rate (whether the model signals task completion by clicking ``done'' after placing keypoints) to assess task understanding. The partial credit score combines both aspects: completed tasks receive full PCK credit, while incomplete tasks receive half credit ($0.5 \times \text{PCK}$), penalizing models that place accurate keypoints but fail to recognize when annotation is finished.

\begin{figure}[h]
\centering
\vspace{1.5em}
\begin{tabular}{ccc}
\includegraphics[width=0.18\textwidth]{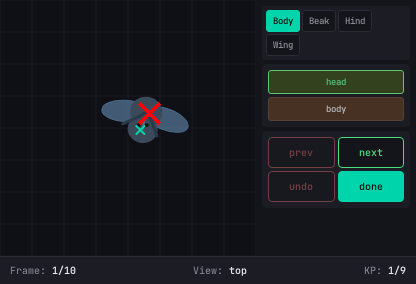} &
\includegraphics[width=0.18\textwidth]{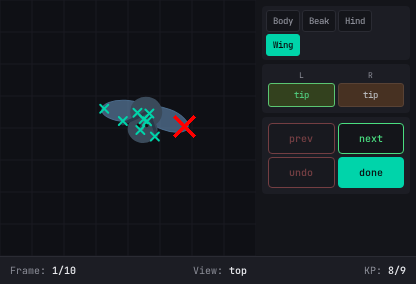} &
\includegraphics[width=0.18\textwidth]{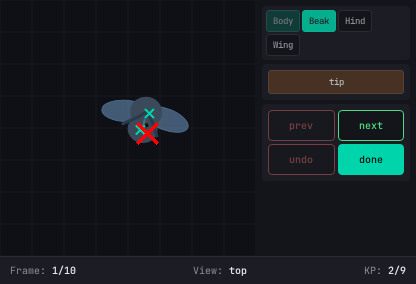} \\
(a) & (b) & (c) \\[0.5em]
\includegraphics[width=0.18\textwidth]{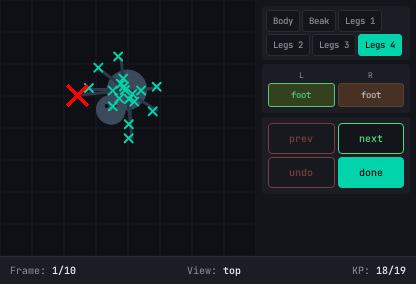} &
\includegraphics[width=0.18\textwidth]{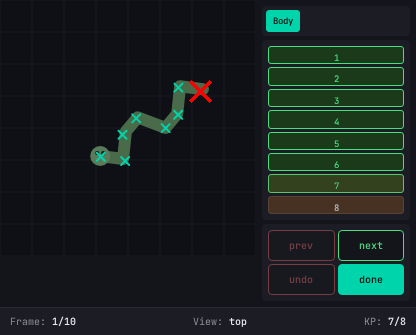} &
\includegraphics[width=0.18\textwidth]{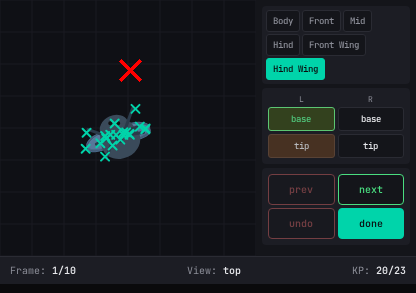} \\
\multicolumn{3}{c}{(d)} \\
\end{tabular}
\vspace{1.5em}
\caption{\textbf{\taskALT.} Top: bird annotation sequence: (a) early annotation with Body tab, (b) later with Wing tab and 8/9 keypoints placed, (c) near-error on beak before undo. Bottom: morphological diversity: spider with 8 legs, snake with spine-only keypoints, flying insect with wings.}
\label{fig:animal_limb_tracking}
\end{figure}

\subsubsection{\taskMIA}
\label{app:tasks:multichannel_image_alignment}

The annotator places corresponding landmark points across multiple image channels that have spatial misalignment. Each channel shows the same underlying scene with different appearance and geometric distortion.

\paragraph{Data Sources.}
50\% synthetic images (5 DGPs: Gaussian blobs, Perlin noise, geometric shapes, Voronoi cells, gradient fields) and 50\% natural images (ImageNette with multichannel projection). Each instance has 3--6 channels with 3--9 landmark points.

\paragraph{GUI.}
The interface displays a 256$\times$256 image (current channel) with channel selector dots and point buttons. Placed markers show as numbered circles. Status bar indicates current channel, point index, and completion count.

\paragraph{Annotation Behavior.}
For each landmark: click to place on current channel, switch to next channel, repeat across all channels. First 3 points are positioned for optimal affine estimation. 15\% error rate with undo correction; 1px click noise.

\paragraph{Evaluation Metrics.}
We evaluate alignment quality via the match rate, which measures the fraction of annotated landmarks whose positions fall within 2 pixels of the ground truth across all channels ($n_{\text{matched}} / n_{\text{annotated}}$). To assess task progress, we report the completion rate $\min(1, n_{\text{landmarks}} / 3)$, reflecting that at least 3 corresponding landmark pairs are needed to estimate an affine transformation (6 degrees of freedom). We also track the total annotation count as a measure of annotator engagement. For plotting and model comparison, we report the mean match rate averaged across all evaluation samples.

\begin{figure}[h]
\centering
\vspace{1.5em}
\begin{tabular}{ccc}
\includegraphics[width=0.18\textwidth]{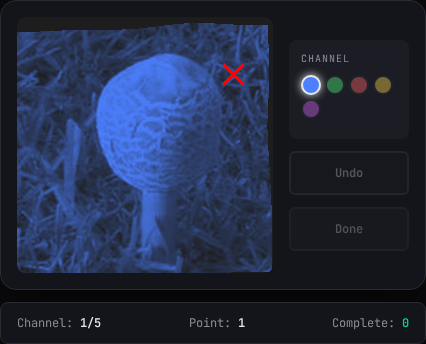} &
\includegraphics[width=0.18\textwidth]{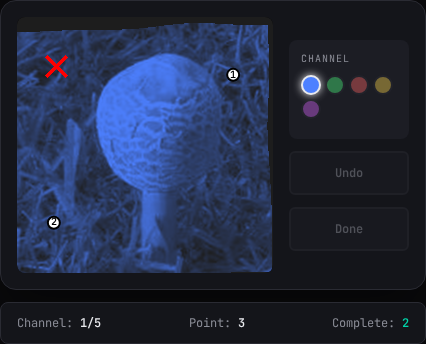} &
\includegraphics[width=0.18\textwidth]{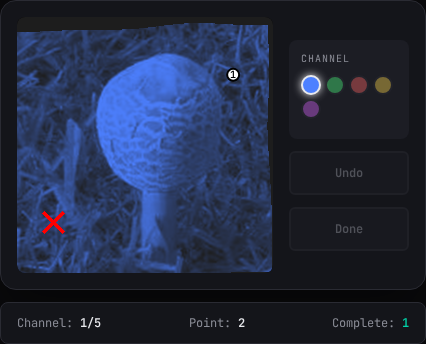} \\
(a) & (b) & (c) \\[0.5em]
\includegraphics[width=0.18\textwidth]{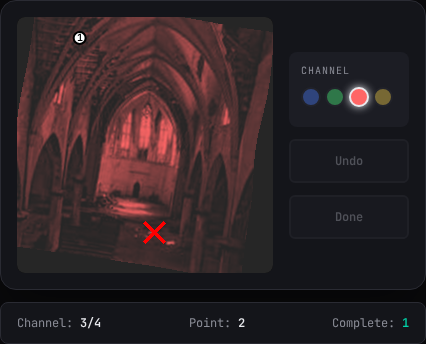} &
\includegraphics[width=0.18\textwidth]{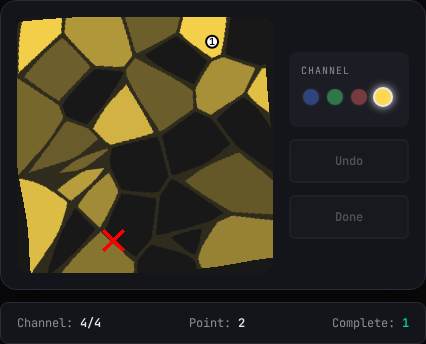} &
\includegraphics[width=0.18\textwidth]{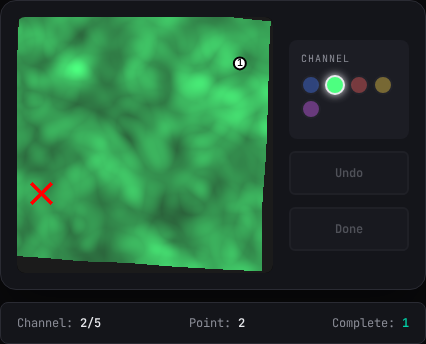} \\
\multicolumn{3}{c}{(d)} \\
\end{tabular}
\vspace{1.5em}
\caption{\textbf{\taskMIA.} Top: annotation sequence on a natural image (mushroom): (a) initial landmark placement, (b) mid-progress with 2 points complete, (c) wrong click location. Bottom: data diversity: church interior (natural), Voronoi cells (synthetic), Perlin noise (synthetic).}
\label{fig:multichannel_image_alignment}
\end{figure}

\subsubsection{\taskSPF}
\label{app:tasks:spectral_plume_finding}

The annotator identifies real plumes among confounders by checking their presence across multiple spectral bands, then draws their boundaries. Real plumes appear in exactly 3 of 5 bands; confounders appear in 1, 2, or all 5 bands.

\paragraph{Data.}
Synthetic scenes with 1--3 real plumes and 3--6 confounders per 256$\times$256 image. Objects are organic blob shapes with soft edges. Background uses aerial/satellite imagery textures.

\paragraph{GUI.}
The interface shows the current spectral band with toggle buttons (b1--b5) to switch views. Markers can be placed to track objects across bands. A Draw button enters polygon mode for boundary annotation; Cancel removes ongoing polygons.

\paragraph{Annotation Behavior.}
(1) Explore: toggle through bands to identify objects and count their band appearances. (2) Place markers on candidate plumes. (3) Draw: click polygon vertices around each verified plume; click near first point to close. (4) Done when all plumes are outlined.

\paragraph{Evaluation Metrics.}
We evaluate both detection quality and segmentation accuracy. Predicted polygons are matched to ground-truth plume boundaries using greedy matching with an IoU threshold of 0.3: we compute an IoU matrix between all predicted and ground-truth polygons, sort pairs by IoU descending, and greedily assign matches above the threshold. Detection recall measures the fraction of true plumes that were found, capped at 1 to avoid inflating scores when excess predictions match the same target. Detection precision measures the fraction of predicted plumes that correspond to real plumes (number of matched polygons / number of predictions). Count accuracy penalizes over- or under-counting: $\max(0, 1 - |n_{\text{found}} - n_{\text{expected}}| / \max(n_{\text{expected}}, 3))$. Mean IoU averages the polygon overlap scores for all matched plumes, capturing segmentation quality. All metrics are computed per sample and averaged across samples for reporting. The combined score balances detection and segmentation: $0.5 \cdot F_1 + 0.5 \cdot \text{mean IoU}$, where $F_1$ is the harmonic mean of recall and precision. This combined score is the primary metric used for comparisons.

\begin{figure}[h]
\centering
\vspace{1.5em}
\begin{tabular}{ccc}
\includegraphics[width=0.18\textwidth]{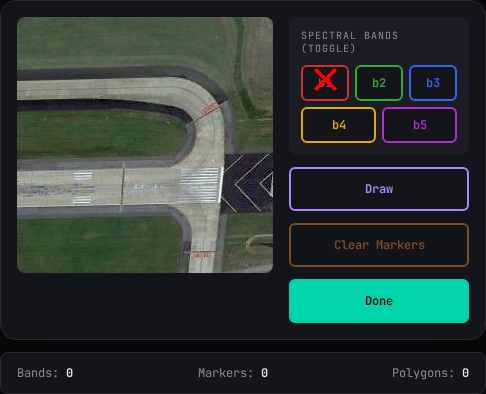} &
\includegraphics[width=0.18\textwidth]{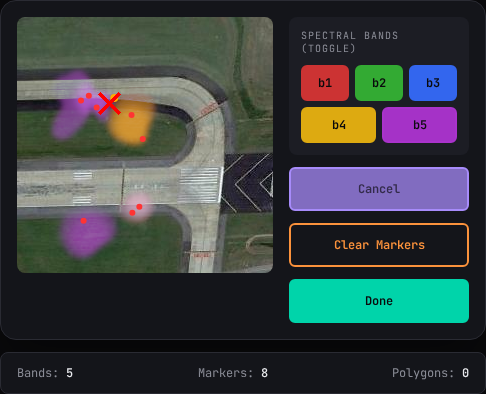} &
\includegraphics[width=0.18\textwidth]{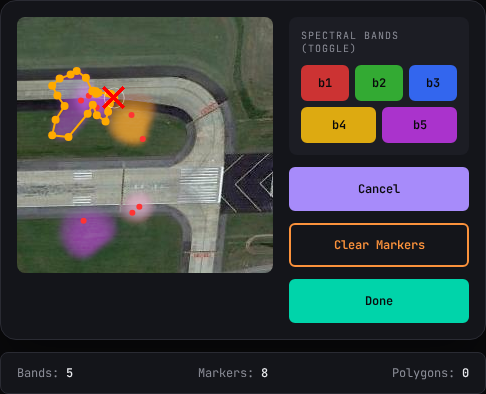} \\
(a) & (b) & (c) \\[0.5em]
\includegraphics[width=0.18\textwidth]{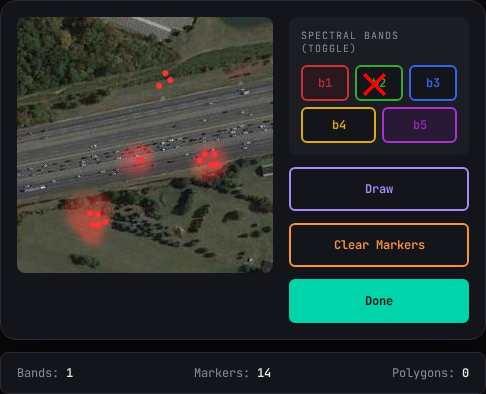} &
\includegraphics[width=0.18\textwidth]{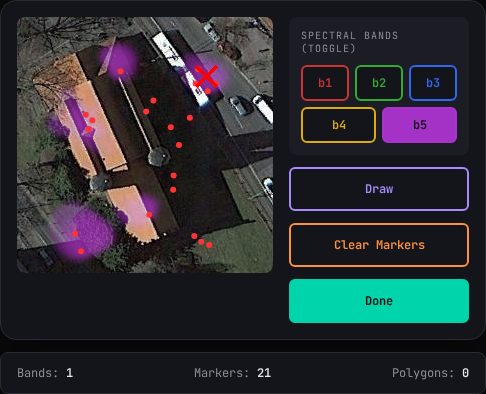} &
\includegraphics[width=0.18\textwidth]{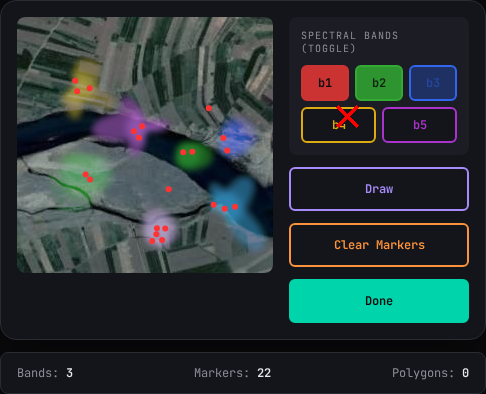} \\
\multicolumn{3}{c}{(d)} \\
\end{tabular}
\vspace{1.5em}
\caption{\textbf{\taskSPF.} Top: annotation sequence: (a) initial band toggle, (b) drawing mode with 5 bands toggled and 8 markers, (c) polygon outline nearly complete. Bottom: scene diversity: highway intersection, building/parking area, agricultural fields.}
\label{fig:spectral_plume_finding}
\end{figure}

\subsubsection{\taskRNC}
\label{app:tasks:road_network_construction}

The annotator traces road networks by placing nodes at intersections and connecting them with edges. The goal is to reconstruct the graph topology of visible road segments in aerial imagery.

\paragraph{Data.}
Aerial/satellite images showing road networks with varying complexity. Each instance has ground truth node positions (intersections, endpoints) and edge connectivity. Images are 256$\times$256 with roads highlighted by lane markings.

\paragraph{GUI.}
The interface overlays the road image with interactive node/edge creation. Clicking places a node; clicking an existing node in ``Connect to...'' mode creates an edge. Undo removes the last action; Remove deletes selected elements. Status shows node and edge counts.

\paragraph{Annotation Behavior.}
The annotator places nodes at road intersections and endpoints, then connects them with edges following visible road segments. Invalid placements (off-road, duplicate nodes) trigger immediate undo. Edges are created by selecting source and target nodes.

\paragraph{Evaluation Metrics.}
We evaluate both node placement and edge connectivity. Predicted nodes are matched to ground truth nodes using the Hungarian algorithm with a 6-pixel tolerance, yielding node recall (fraction of GT nodes found) and precision (fraction of predictions that are correct). For edges, we map each predicted edge through the node matching: an edge counts as correct only if both endpoints matched valid GT nodes and the corresponding GT edge exists. This captures whether the model correctly identified the road network's topological structure, not just node locations. We compute F1 scores for both nodes and edges. For each sample, Graph F1 = (Node F1 + Edge F1) / 2; we then average Graph F1 across samples for plotting.

\begin{figure}[h]
\centering
\vspace{1.5em}
\begin{tabular}{ccc}
\includegraphics[width=0.18\textwidth]{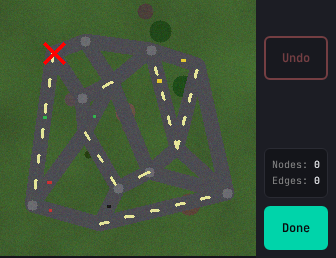} &
\includegraphics[width=0.18\textwidth]{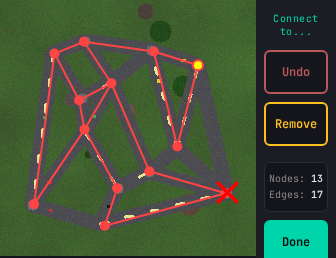} &
\includegraphics[width=0.18\textwidth]{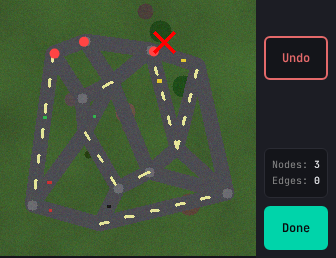} \\
(a) & (b) & (c) \\[0.5em]
\includegraphics[width=0.18\textwidth]{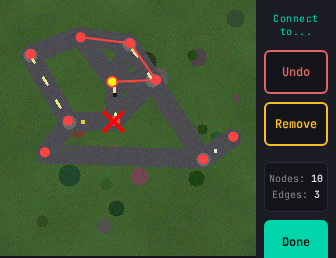} &
\includegraphics[width=0.18\textwidth]{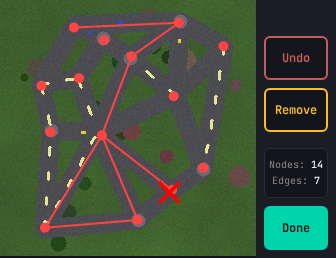} &
\includegraphics[width=0.18\textwidth]{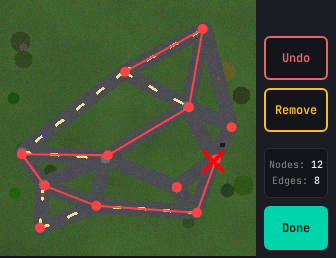} \\
\multicolumn{3}{c}{(d)} \\
\end{tabular}
\vspace{1.5em}
\caption{\textbf{\taskRNC.} Top: annotation sequence: (a) placing first node, (b) near completion with 13 nodes and 17 edges, (c) invalid node placement. Bottom: diverse road layouts with varying complexity (10--14 nodes, 3--8 edges).}
\label{fig:road_network_construction}
\end{figure}

\subsubsection{\taskTDE}
\label{app:tasks:3d_exploration_classification}

The annotator rotates a 3D object to explore it from multiple viewpoints, then classifies it into one of 9 classes. All objects consist of 5 colored spheres (red, blue, green, orange, purple) arranged in distinct configurations. From certain viewing angles, different arrangements appear similar due to occlusion, so rotation is required to reveal the true 3D structure.

\paragraph{Object Classes.}
The 9 classes divide into 5 planar and 4 truly three-dimensional arrangements:

\begin{table}[h]
\centering
\small
\begin{tabular}{clll}
\toprule
\textbf{\#} & \textbf{Class} & \textbf{Description} & \textbf{3D?} \\
\midrule
1 & Line & 5 collinear spheres & No \\
2 & Plus & 4 arms + 1 center (cross) & No \\
3 & Pentagon & 5 in regular ring & No \\
4 & T-shape & 3 across + 2 stem & No \\
5 & L-shape & 3 horizontal + 2 vertical & No \\
6 & Methane & 1 center + 4 tetrahedral & Yes \\
7 & Square pyramid & 4 base + 1 apex & Yes \\
8 & Trigonal bipyramid & 3 equatorial + 2 poles & Yes \\
9 & Bowtie & 2 triangles sharing vertex & Yes \\
\bottomrule
\end{tabular}
\vspace{1.5em}
\caption{\textbf{3D Object Classes.} The 9 object classes for \taskTDE.}
\label{tab:3d_classes}
\end{table}

Each class has a distinct topology (neighbor relationships), making them distinguishable given sufficient viewpoints. The 4 truly 3D classes require rotation to classify correctly.

\paragraph{GUI.}
The interface displays a 256$\times$256 Three.js canvas showing the 3D object. A control panel provides rotation buttons (+X, -X, +Y, -Y, each rotating 30°) and 9 class buttons. Objects start at a random orientation (uniform over SO(3)). A confirm button finalizes the selection.

\paragraph{Annotation Behavior.}
The virtual annotator performs 3--10 rotations (mean 5, Gaussian distribution) to explore the object, avoiding consecutive presses of the same button. After exploration, they select a class button. With 30\% probability, a misclick occurs (wrong class selected) followed by correction. Finally, confirm is clicked.

\paragraph{Evaluation Metrics.}
We evaluate four aspects of performance. \emph{Flat vs.\ 3D accuracy} measures whether the model correctly identifies if a shape is planar (classes I, +, P, T, L) or truly three-dimensional (classes M, A, Y, X), a coarse but fundamental distinction. \emph{Class similarity} awards partial credit for near-misses: 1.0 for exact matches, 0.3 for confusions within the same group (e.g., mistaking one flat shape for another flat shape, or one 3D shape for another 3D shape), and 0 for cross-group errors. \emph{Exploration quality} captures how thoroughly the model examines the object before classifying, computed as $\min(r/5, 1)$ where $r$ is the number of rotations performed (saturating at 5 rotations). Finally, \emph{exploration-adjusted accuracy} rewards correct classifications in proportion to their exploration quality: for correct predictions it equals $\min(r/5, 1)$, for incorrect predictions it equals 0. All metrics are computed per sample and then averaged (simple mean) across all samples; we report the mean exploration-adjusted accuracy in plots.

\begin{figure}[h]
\centering
\vspace{1.5em}
\begin{tabular}{ccc}
\includegraphics[width=0.18\textwidth]{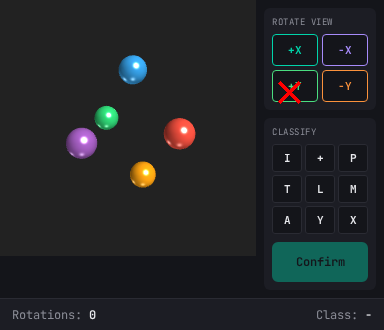} &
\includegraphics[width=0.18\textwidth]{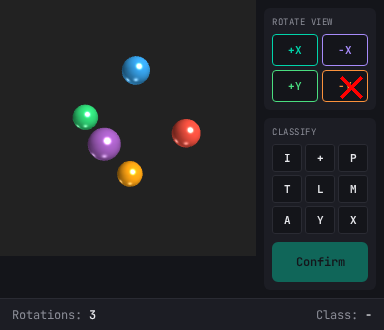} &
\includegraphics[width=0.18\textwidth]{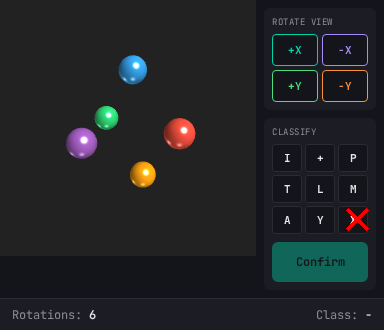} \\
(a) & (b) & (c) \\[0.5em]
\includegraphics[width=0.18\textwidth]{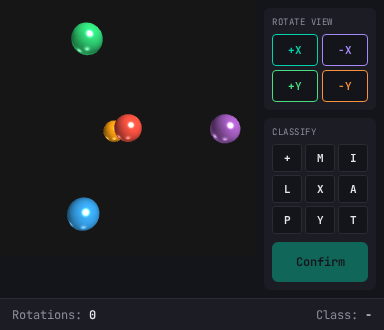} &
\includegraphics[width=0.18\textwidth]{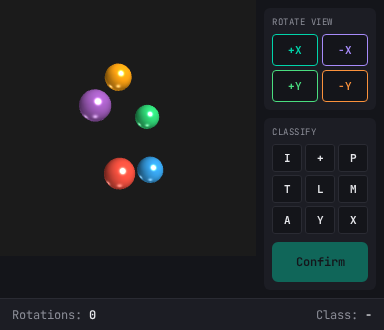} &
\includegraphics[width=0.18\textwidth]{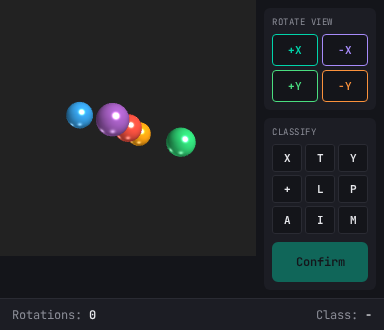} \\
\multicolumn{3}{c}{(d)} \\
\end{tabular}
\vspace{1.5em}
\caption{\textbf{\taskTDE.} Top: annotation sequence for square pyramid: (a) initial random orientation, (b) after several rotations exploring structure, (c) misclick on wrong class button. Bottom: class diversity: methane (truly 3D), pentagon (flat), plus/cross (flat).}
\label{fig:3d_exploration}
\end{figure}

\subsection{OOD Task: \taskSM}
\label{app:tasks:ood}

The annotator clicks all objects matching a template shape shown in the sidebar. This task tests generalization to an unseen task format not included in training.

\paragraph{Data.}
Each instance contains 12 randomly placed objects from 4 shape classes (circle, triangle, square, star) in various colors. A template shape is shown in the sidebar; 2--5 objects match it. Objects are placed on a dark background without overlap.

\paragraph{GUI.}
The interface shows a ``MATCH THIS'' sidebar with the template shape and a ``FOUND: X/N'' counter tracking progress. Clicked matching objects display a green checkmark. A Done button completes the annotation.

\paragraph{Annotation Behavior.}
The annotator scans the scene and clicks each object matching the template. The counter increments with each correct click. After finding all targets, Done is clicked.

\paragraph{Evaluation Metrics.}
We evaluate \taskSM performance using recall and precision-style metrics. A click registers as hitting an object if it falls within 17 pixels of the object center (object radius of 12 pixels plus 5-pixel tolerance). The all-targets rate measures the fraction of instances where every matching object was successfully clicked, assessing completeness of visual search. The no-extras rate measures the fraction of instances where no non-matching objects were clicked, assessing discrimination accuracy. Full correctness requires both: all targets clicked and no extras, meaning the model must both find all matching shapes and avoid false positives. We also track completion rate (whether Done was clicked) to assess understanding of task termination. For plots, we report the all-targets rate averaged across all test instances.

\begin{figure}[h]
\centering
\vspace{1.5em}
\begin{tabular}{ccc}
\includegraphics[width=0.18\textwidth]{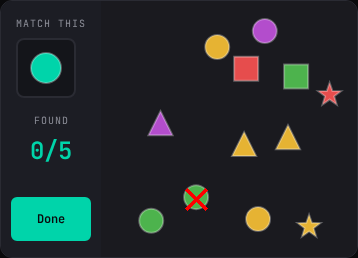} &
\includegraphics[width=0.18\textwidth]{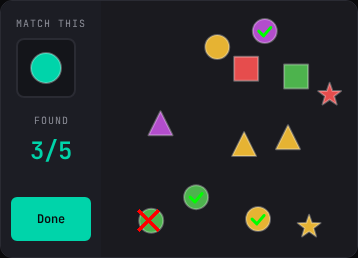} &
\includegraphics[width=0.18\textwidth]{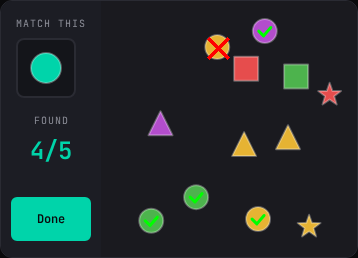} \\
(a) & (b) & (c) \\[0.5em]
\includegraphics[width=0.18\textwidth]{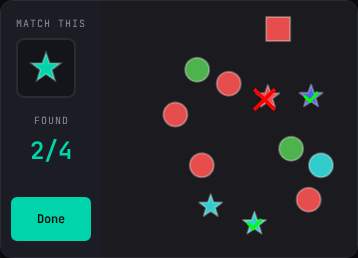} &
\includegraphics[width=0.18\textwidth]{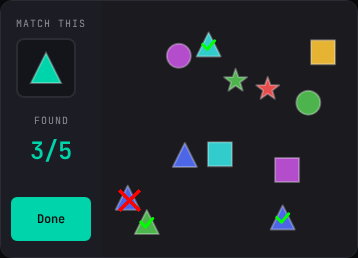} &
\includegraphics[width=0.18\textwidth]{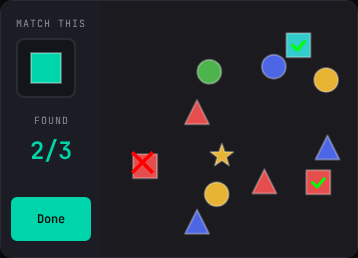} \\
\multicolumn{3}{c}{(d)} \\
\end{tabular}
\vspace{1.5em}
\caption{\textbf{\taskSM(OOD).} Top: annotation sequence for circle template: (a) initial state (0/5), (b) mid-progress (3/5), (c) near completion (4/5). Bottom: template diversity: star (2/4), triangle (3/5), square (2/3).}
\label{fig:shape_matching}
\end{figure}

\section{Methodology}
\label{app:methodology}

\subsection{Model}
\label{app:methodology:model}

We use a Vision-Language Model (VLM) that processes interleaved sequences of GUI screenshots and click coordinates. The architecture consists of a DINOv2 vision encoder and a transformer head.

\paragraph{Vision Encoder.}
We use DINOv2 with registers~\citep{darcet2024visiontransformersneedregisters} as the vision backbone. All encoder weights are unfrozen during training. Images are processed at native resolution with patch size 14, then spatially pooled to a fixed grid of $12 \times 9 = 108$ tokens per image via adaptive average pooling. Learnable position embeddings are added to the pooled features.

\begin{figure}[h]
\centering
\includegraphics[width=0.45\textwidth]{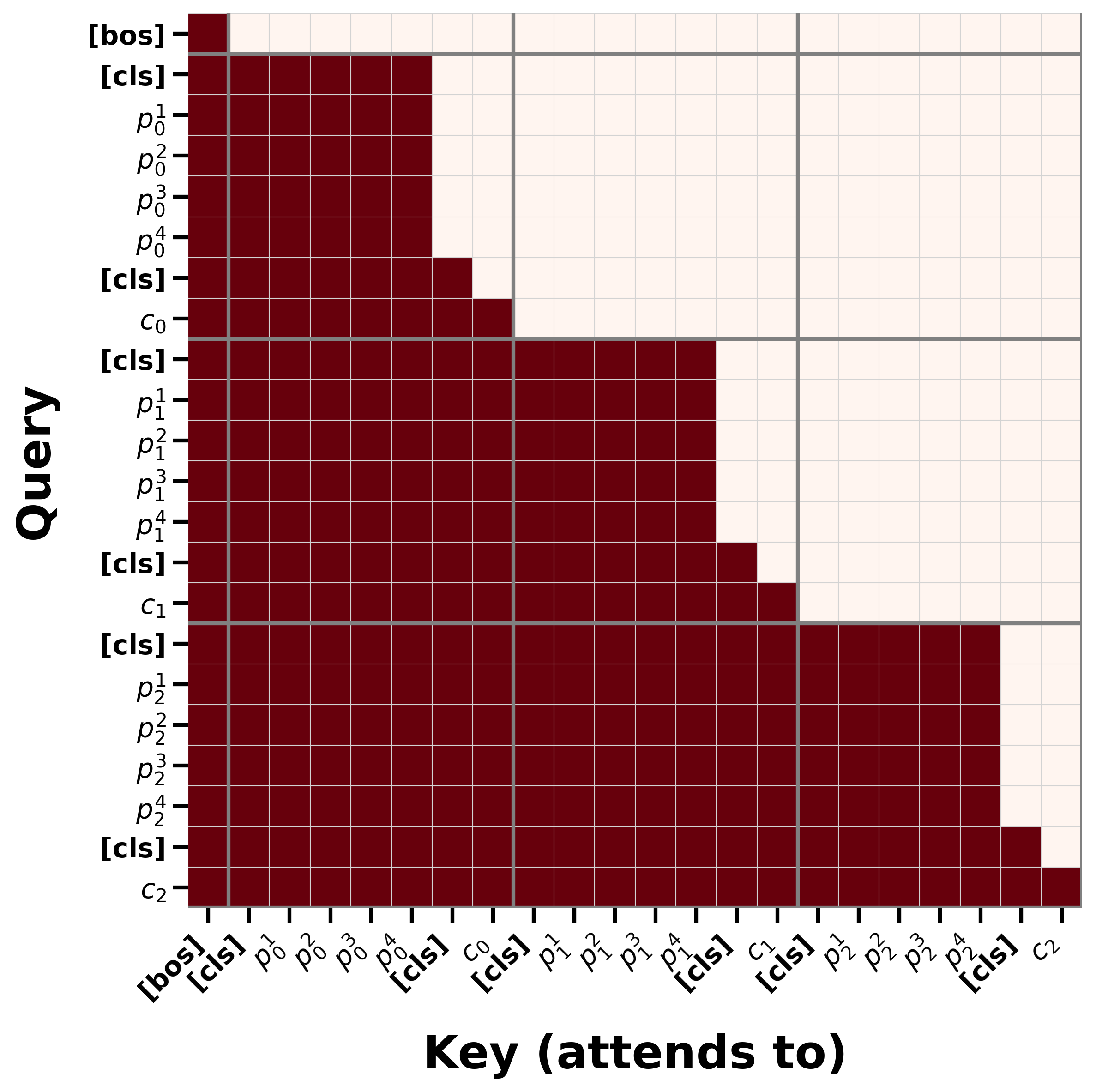}
\vspace{1.5em}
\caption{\textbf{Block-Causal Attention Mask.} Illustration with 3 patches per frame; the actual model uses 108). Blue indicates allowed attention. Within each frame, all tokens attend bidirectionally. Across frames, attention is causal: Frame 1 attends to Frame 0, but not vice versa.}
\label{fig:attention_mask}
\end{figure}

\paragraph{Transformer Head.}
The transformer head processes the interleaved sequence of image tokens and click tokens. Each block uses multi-head self-attention with RoPE positional encoding (base frequency 10000), layer normalization (pre-norm), MLP with expansion ratio 4.0 and GELU activation, and dropout 0.1. See Table~\ref{tab:model_sizes} for size configurations. Attention is \emph{block-causal}: bidirectional within each image's patch tokens, but causal across the sequence (each frame can only attend to previous frames). See Figure~\ref{fig:attention_mask} for an illustration.

\paragraph{Sequence Format.}
The input sequence has the following structure:
\begin{center}
\texttt{[bos] [cls] $\text{img}_0$ [cls] $\text{click}_0$ [cls] $\text{img}_1$ [cls] $\text{click}_1$ $\cdots$ [cls] $\text{img}_t$ [cls]}
\end{center}
where $\text{img}_i$ denotes the 108 patch tokens for frame $i$, and $\text{click}_i$ is a single token encoding the $(x, y)$ coordinate. Modality embeddings distinguish token types: \texttt{cls}(0), \texttt{image}(1), \texttt{click}(2), \texttt{bos}(3). The context length is 20 frames.

\paragraph{Coordinate Prediction.}
Tasks have different native image sizes (ranging from $336 \times 258$ to $524 \times 436$ pixels). For multi-task training, all images are resized to the maximum dimensions ($524 \times 436$). Coordinates are stored as relative values in $[0, 1]$ and discretized into bins matching the maximum pixel dimensions: 524 bins for $x$ and 436 bins for $y$. This provides pixel-accurate prediction on the resized images. Input coordinates use separate learned embeddings for $x$ and $y$, which are concatenated and projected to the model dimension. Output coordinates are predicted via separate linear heads:
\begin{align}
    \hat{x} &= \text{Linear}_{x}(h) \in \mathbb{R}^{524} \\
    \hat{y} &= \text{Linear}_{y}(h) \in \mathbb{R}^{436}
\end{align}
where $h$ is the hidden state at the \texttt{[cls]} token preceding the target click. The loss is the sum of cross-entropy losses over $x$ and $y$.

\paragraph{Model Sizes.}
We train four model sizes with log-linear parameter progression. Table~\ref{tab:model_sizes} shows the configurations.

\begin{table}[h]
\centering
\small
\begin{subtable}[t]{0.48\textwidth}
\centering
\begin{tabular}{lccccc}
\toprule
\textbf{Backbone} & \textbf{Dim} & \textbf{Heads} & \textbf{Layers} & \textbf{Params} \\
\midrule
ViT-S/14 & 384 & 6 & 12 & 22M \\
ViT-B/14 & 768 & 12 & 12 & 86M \\
ViT-L/14 & 1024 & 16 & 24 & 304M \\
\bottomrule
\end{tabular}
\vspace{1.5em}
\caption{\textbf{DINOv2 Vision Encoder.} With registers. Patch size 14, MLP ratio 4.0.}
\label{tab:dino_configs}
\end{subtable}
\hfill
\begin{subtable}[t]{0.48\textwidth}
\centering
\begin{tabular}{lcccccc}
\toprule
\textbf{Size} & \textbf{Backbone} & \textbf{Dim} & \textbf{Layers} & \textbf{Heads} & \textbf{Total} \\
\midrule
Very Small & ViT-S/14 & 256 & 4 & 4 & 25M \\
Small & ViT-S/14 & 384 & 6 & 6 & 28M \\
Base & ViT-B/14 & 512 & 8 & 8 & 95M \\
Large & ViT-L/14 & 768 & 12 & 12 & 320M \\
\bottomrule
\end{tabular}
\vspace{1.5em}
\caption{\textbf{Transformer Head.} MLP ratio 4.0, dropout 0.1.}
\label{tab:head_configs}
\end{subtable}
\caption{\textbf{Model Configurations.} (a) DINOv2 backbone specifications. (b) Transformer head and total parameter counts.}
\label{tab:model_sizes}
\end{table}

\subsection{Training}
\label{app:methodology:training}

\begin{table}[h]
\centering
\small
\begin{tabular}{ll}
\toprule
\textbf{Hyperparameter} & \textbf{Value} \\
\midrule
Optimizer & AdamW \\
Weight decay & 0.01 \\
Betas & (0.9, 0.999) \\
Learning rate (head) & $10^{-4}$ \\
Learning rate (backbone) & $10^{-5}$ (10$\times$ reduction) \\
Gradient clipping & max norm 1.0 \\
\midrule
LR schedule & Linear warmup + cosine decay to 10\% \\
Warmup steps & 4000 (very small, small), 2000 (base), 1000 (large) \\
\midrule
Batch size per GPU & 4 \\
Gradient accumulation & 2 steps \\
GPUs & 4$\times$ NVIDIA A100 80GB \\
Effective batch size & 32 \\
\bottomrule
\end{tabular}
\vspace{1.5em}
\caption{\textbf{Training Hyperparameters.}}
\label{tab:training_hparams}
\end{table}

\paragraph{Training Steps.}
To compare models at equal compute~\citep{kaplan2020scaling,hoffmann2022training}, we scale training steps inversely with model size:

\begin{table}[h]
\centering
\small
\begin{tabular}{lccc}
\toprule
\textbf{Size} & \textbf{Params} & \textbf{Steps} & \textbf{Checkpoints} \\
\midrule
Very Small & 25M & 380,000 & every 2,000 \\
Small & 28M & 300,000 & every 2,000 \\
Base & 95M & 100,000 & every 1,000 \\
Large & 320M & 30,000 & every 500 \\
\bottomrule
\end{tabular}
\vspace{1.5em}
\caption{\textbf{FLOP-Matched Training Configurations.} Steps are scaled so that total compute (params $\times$ steps) is approximately constant.}
\label{tab:training_steps}
\end{table}

We also compare models at equal loss to test whether larger models are more data-efficient (Section~\ref{sec:multi-task}).

\paragraph{Single-Task vs.\ Multi-Task.}
For single-task experiments, we train the base model on \taskCDT for one epoch (109,355 steps), processing approximately 3.5M context windows (70M images at 20 frames per window). For multi-task experiments, we train on all 9 tasks jointly with uniform sampling. The combined multi-task dataset contains 21.8M context windows; approximately 2.4\% of samples use padding tokens for shorter sequences (primarily \taskTDE). Multi-task models are trained for varying steps depending on model size (Table~\ref{tab:training_steps}), ranging from 960K samples (Large) to 12.2M samples (Very Small). The training procedure is otherwise identical.

\subsection{Evaluation}
\label{app:methodology:evaluation}

We evaluate models in two modes: \emph{teacher-forced} and \emph{autoregressive}. Task-specific evaluation metrics are described alongside each task in Appendix~\ref{app:tasks:main}.

\paragraph{Teacher-Forced Evaluation.}
Given ground-truth action history, the model predicts the next action. We measure per-action-type accuracy (e.g., button clicks vs.\ canvas placements) and canvas placement error at multiple thresholds (1px, 3px, 5px, 10px). This provides fine-grained learning dynamics without compounding errors.

\paragraph{Autoregressive Evaluation.}
The model interacts with the GUISimulator in a closed loop, predicting clicks from screenshots until it clicks ``Done'' or reaches a maximum step limit ($2\times$ ground-truth length). The simulator executes each click through the GUI's JavaScript interface, determining action type from coordinates alone, so the model cannot specify action types directly.

\paragraph{Inference Parameters.}
\begin{itemize}[leftmargin=*,nosep]
    \item Temperature: 0.4
    \item No beam search
    \item Context window: 20 frames (same as training)
\end{itemize}

\paragraph{Strict vs.\ Generous Metrics.}
Binary ``is\_correct'' evaluation is harsh: partial progress receives zero credit. We additionally compute \emph{generous metrics} using Hungarian matching to optimally align predicted and ground-truth outputs, then compute precision, recall, F1, and position errors on matched pairs. This enables finer-grained analysis of model capability. See each task's Evaluation Metrics paragraph in Appendix~\ref{app:tasks:main} for task-specific definitions.

\paragraph{Evaluation Sets.}
For teacher-forced evaluation, we sample 1024 random 20-frame windows from test sequences. For autoregressive evaluation, we run 50--300 complete episodes per checkpoint (50 for training dynamics, 300 for final evaluation). Confidence intervals are computed via binomial standard deviation for accuracy metrics.

\subsection{ICL \& Fine-Tuning}
\label{app:methodology:icl}

We evaluate whether multi-task pretraining enables adaptation to a held-out task (\taskSM, Appendix~\ref{app:tasks:ood}) through in-context learning or fine-tuning. All experiments use the base model checkpoint at 100,000 steps (95M parameters).

\paragraph{In-Context Learning Protocols.}
We test three ICL conditions, all evaluated on 256 test instances with temperature 0.4 and maximum 50 generation steps:

\begin{itemize}[leftmargin=*,nosep]
    \item \textbf{Zero-shot (ZS)}: The model receives only the first frame of a test instance and generates subsequent actions autoregressively. This tests whether the model can infer the task from the visual template alone.

    \item \textbf{Prefix}: The model receives the first 2 ground-truth frame-action pairs from a test sequence, then generates the remaining actions. This tests within-sequence continuation.

    \item \textbf{Few-shot (FS)}: Three complete demonstration sequences from the training set are prepended to the context before the test instance's first frame. Since \taskSM sequences are short (5--8 steps), three demos fit within the 20-frame context window. This tests true in-context learning from examples.
\end{itemize}

All three conditions achieve negligible accuracy ($<$2\%), indicating that multi-task pretraining does not induce in-context learning capabilities for novel annotation tasks.

\paragraph{Fine-Tuning Protocol.}
We fine-tune from the pretrained checkpoint using full fine-tuning with differential learning rates: $10^{-4}$ for the transformer head and $10^{-5}$ for the DINOv2 backbone (0.1$\times$ scaling). Other hyperparameters: 500 warmup steps, cosine decay, weight decay 0.01, gradient clipping at 1.0. We train for 1,000 steps on subsets of varying size (50, 100, 150, 250, 350, 500 sequences) and 5,000 steps on the full dataset (7,800 sequences).

Fine-tuning achieves 76.6\% accuracy with just 500 sequences, saturating near this level; 7,800 sequences yields similar performance (77.7\%). See Figures~\ref{fig:ft_variants} and~\ref{fig:ft_losses} for training dynamics and data efficiency curves.

\paragraph{From-Scratch Baseline.}
To isolate the contribution of pretraining, we train the identical architecture from random initialization on the \taskSM task. We test two learning rate configurations: an aggressive setting ($3 \times 10^{-4}$, 2000 warmup steps) and a matched setting (same as fine-tuning). Both are trained for 5,000 steps on 7,800 sequences.

Training from scratch fails entirely (0\% accuracy) regardless of hyperparameters, demonstrating that multi-task pretraining provides essential inductive biases for GUI annotation that cannot be recovered through extended training on the target task alone.

\subsection{Model Internal Analysis}
\label{app:methodology:internals}

We probe the internal representations of the trained multi-task model to understand what information is encoded during annotation. All experiments use the base model checkpoint at 100,000 steps (95M parameters), extracting activations from layer 6 of 8 (75\% depth, 512 dimensions) at the [cls] token position preceding each action prediction.

\paragraph{Activation Extraction.}
For each task, we sample 100 test sequences and perform a forward pass through the model, hooking the transformer layer to extract hidden states. Each action yields a 512-dimensional activation vector paired with metadata (frame index, action type, mistake labels, task-specific state). Activations are saved in HDF5 format for efficient downstream analysis.

\paragraph{Linear Probe Training.}
All probes use the same methodology: activations are standardized (zero mean, unit variance), split 80/20 into train/test sets (stratified for classification), and fit with either logistic regression (classification; $C=1.0$, balanced class weights) or ridge regression (continuous labels; $\alpha=1.0$). We report test-set metrics and 5-fold cross-validation scores.

\paragraph{Mistake and Correction Probes (Figure~\ref{fig:internals}a).}
We define ``mistake'' as an action immediately preceding an undo, and ``correction'' as the undo action itself. For the pooled analysis across all 9 tasks, we train a single logistic regression probe on 130k actions ($\sim$4.4\% mistakes). The mistake probe achieves ROC AUC = 0.87 (CV: $0.83 \pm 0.07$), while the correction probe achieves ROC AUC = 0.99 (CV: $0.96 \pm 0.04$). The higher correction accuracy reflects that corrections have distinctive activation patterns (reversal actions with consistent structure), while mistakes are more heterogeneous.

\paragraph{Diverse Probes (Figure~\ref{fig:internals}b).}
We train probes for multiple latent concepts across tasks:
\begin{itemize}[leftmargin=*,nosep]
    \item \textbf{Task phase} (road\_network): Binary classification of node placement vs.\ edge construction phases achieves ROC AUC = 1.00, indicating the model strongly encodes workflow stage.
    \item \textbf{Object class} (3d\_exploration): 9-way classification of 3D object type achieves ROC AUC = 0.93, suggesting the model identifies object geometry during exploration.
    \item \textbf{Task progress}: Regression on normalized progress (fraction of annotation completed) achieves $R^2 = 0.65$--$0.92$ depending on task, with colored\_dot\_tracking showing strongest signal ($R^2 = 0.92$ for color progress).
    \item \textbf{Error states}: Per-task mistake probes achieve ROC AUC = 0.83--0.91 across tasks.
    \item \textbf{Temporal position}: Frame index classification achieves ROC AUC = 0.70--0.89 across tracking tasks, with cell\_lineage\_tracking showing best frame encoding (0.89).
\end{itemize}

\paragraph{Cross-Task PCA (Figure~\ref{fig:internals}c).}
We pool activations from all 9 tasks, standardize, and compute 2-component PCA. Mistakes (actions before undo) form a partially separable cluster from correct actions, suggesting a shared ``something is wrong'' representation. However, task identity also clusters in activation space, indicating both task-specific and task-general structure.

\paragraph{Leave-One-Task-Out Transfer (Figure~\ref{fig:internals}d).}
To quantify generalization of mistake representations, we train probes on 8 tasks and test on the held-out task. Mean transfer ROC AUC = 0.71 ($\pm 0.16$), with 8/9 tasks showing above-chance transfer. The outlier is 3d\_exploration\_classification (ROC AUC = 0.29), which has fundamentally different error structure (classification mistakes vs.\ spatial placement errors). This suggests mistake detection is partially universal but retains task-specific components.

\paragraph{Probe Direction Analysis.}
We analyze cosine similarity between per-task probe coefficient vectors. Off-diagonal similarity ranges from 0.35--0.50, indicating moderate alignment of mistake directions across tasks. Some task pairs show stronger alignment (animal\_behavioral $\leftrightarrow$ animal\_limb $\leftrightarrow$ road\_network: similarity 0.6--0.7), while others are more isolated (colored\_dot\_tracking, 3d\_exploration).

\section{Additional Results}
\label{app:results}

\subsection{Single-Task Training}

\begin{figure*}[ht]
\centering
\includegraphics[width=\textwidth]{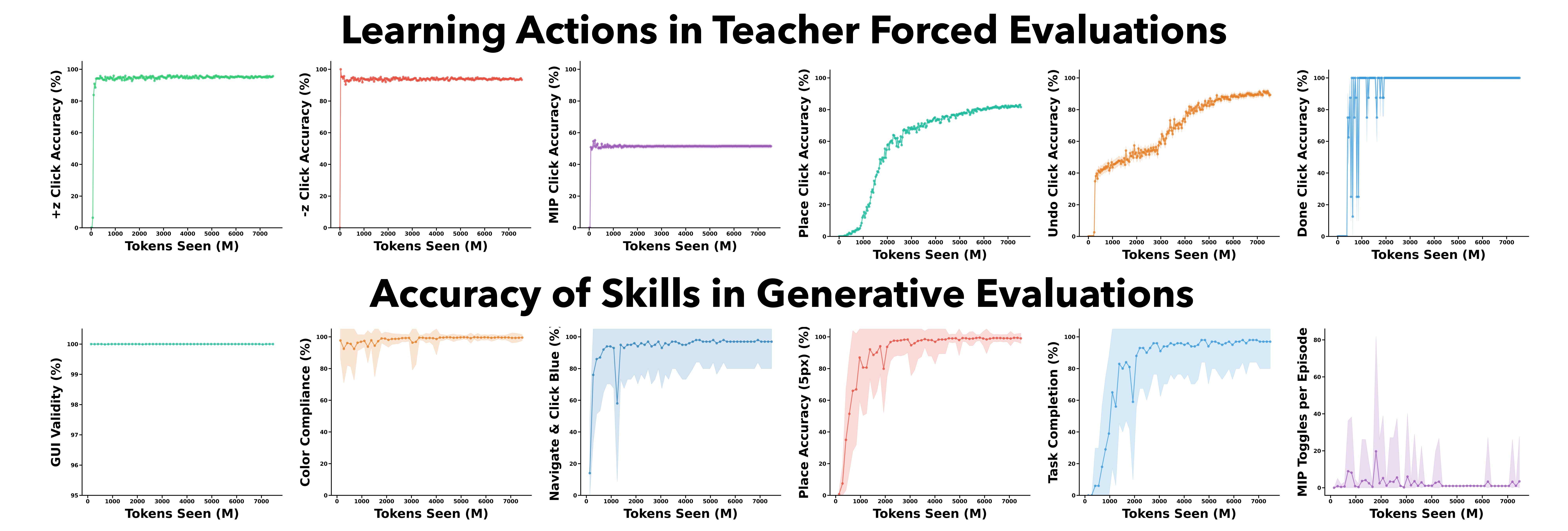}
\caption{\textbf{Single-Task Training Dynamics.} Top: Accuracy of various actions (as assessed by teacher-forced evaluation) in single-task \taskCDT. Shaded area denotes 1 standard deviation. Bottom: Accuracy of various skills (as assessed by generative evaluation) in single-task \taskCDT. The shaded area denotes 1 standard deviation.}
\label{fig:single_task_cdt_1}
\end{figure*}

\begin{figure*}[ht]
\centering
\includegraphics[width=\textwidth]{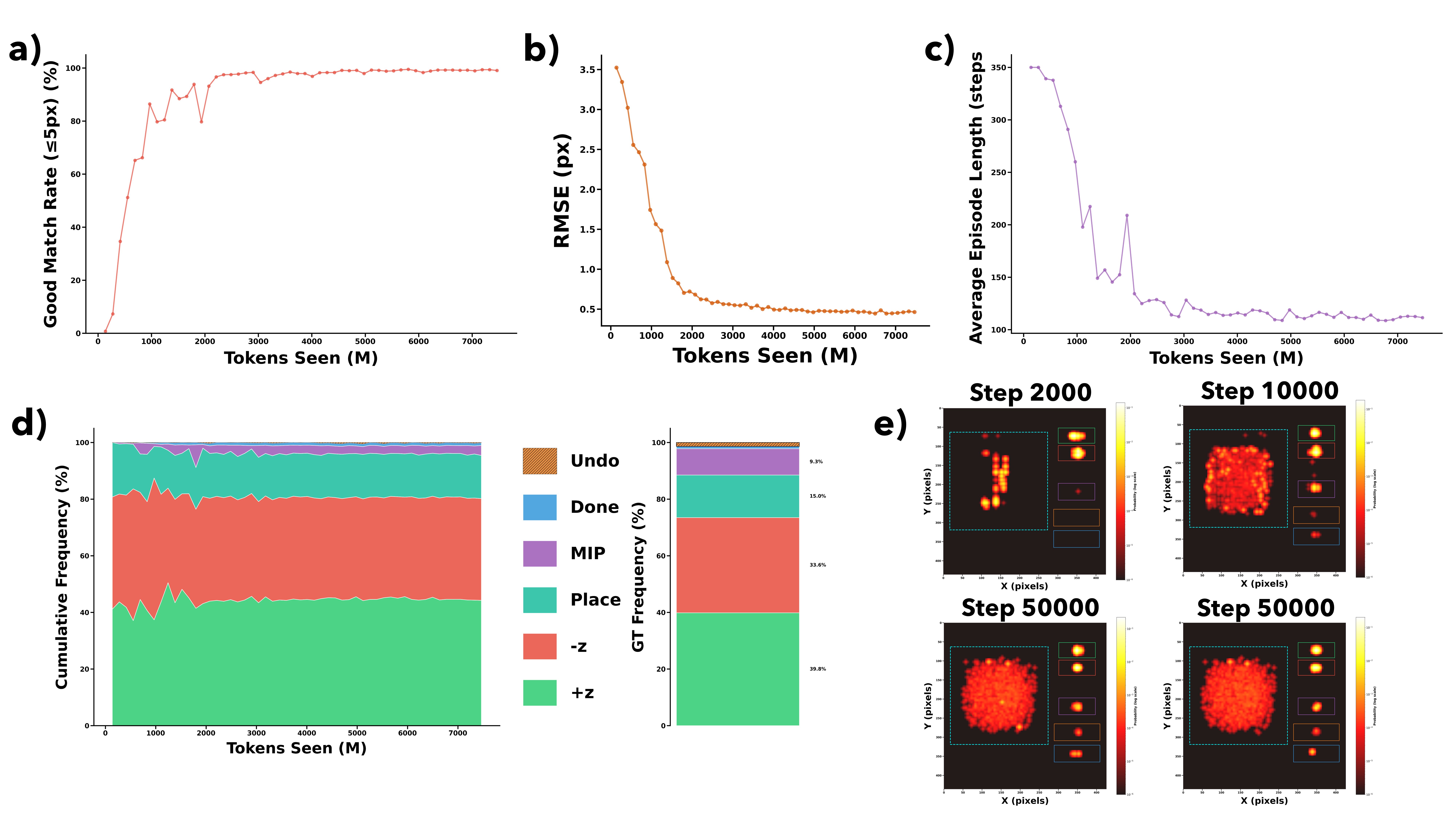}
\caption{\textbf{Additional Generative Evaluation Results.} Results for single-task \taskCDT. (a) Matches are determined via a Hungarian matching procedure between model annotations and the ground truth annotations. A good match is such that it is within 5 px of the ground truth annotation. (b) RMSE of good matches. (c) Episode length decreases over training. (d) Frequency of various actions during generative evaluation compared to ground truth distribution. (e) Heatmap of model clicks throughout training.}
\label{fig:single_task_cdt_2}
\end{figure*}

We analyze training dynamics on a single task (\taskCDT) to understand skill acquisition in isolation. Figure~\ref{fig:single_task_cdt_1} shows how action accuracy evolves during training under teacher-forced evaluation, alongside skill accuracy under generative evaluation. Figure~\ref{fig:single_task_cdt_2} provides additional generative metrics including match quality, RMSE, episode length, action distributions, and click heatmaps.

\clearpage

\subsection{Multi-Task Scaling}

\begin{figure*}[ht]
\centering
\includegraphics[width=0.6\textwidth]{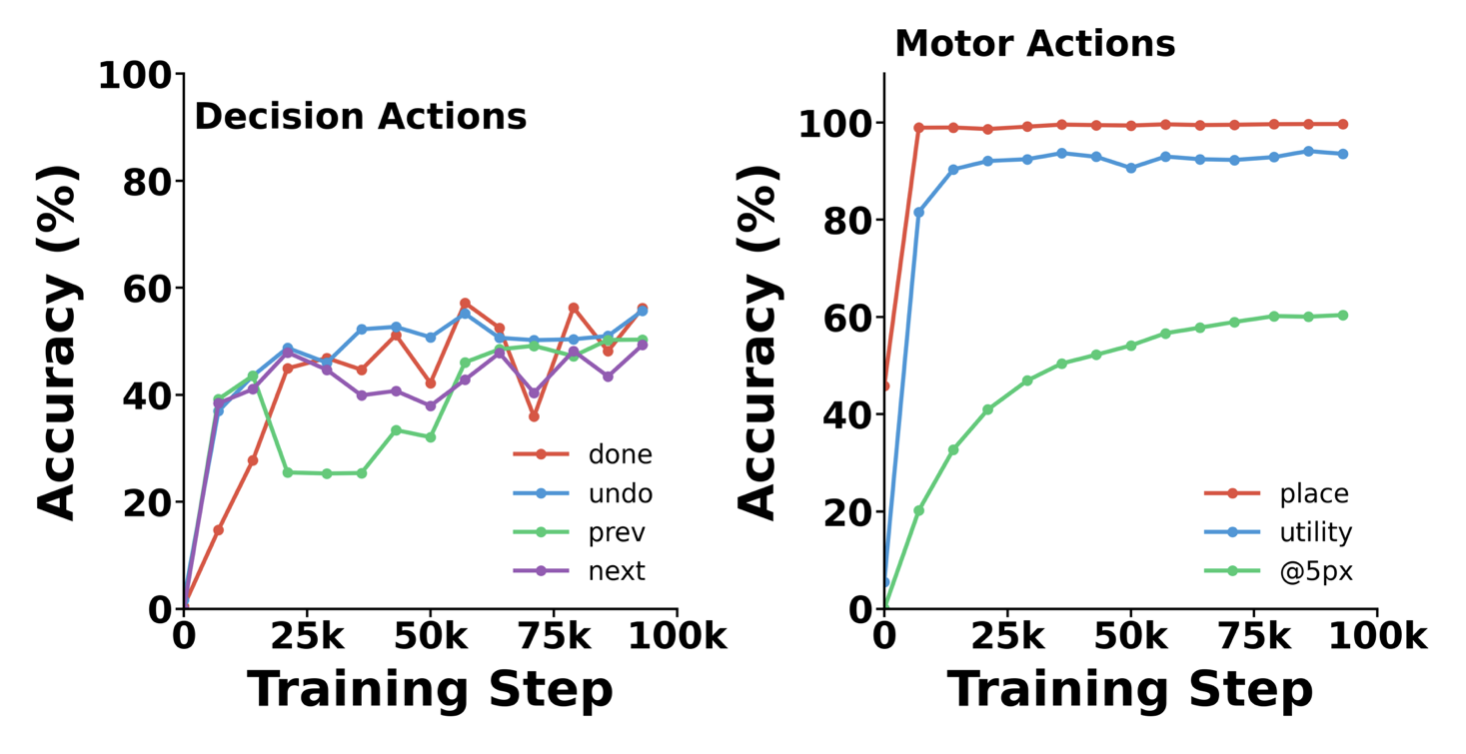}
\caption{\textbf{Decision vs Motor Metrics.} Training dynamics of downstream task metrics. (a) Decision action classification accuracy for done, undo, prev, and next actions. (b) Motor metrics including place action accuracy, utility action accuracy, and placement precision within 5 pixels.}
\label{fig:decision_vs_motor}
\end{figure*}

\begin{figure*}[ht]
\centering
\includegraphics[width=0.5\textwidth]{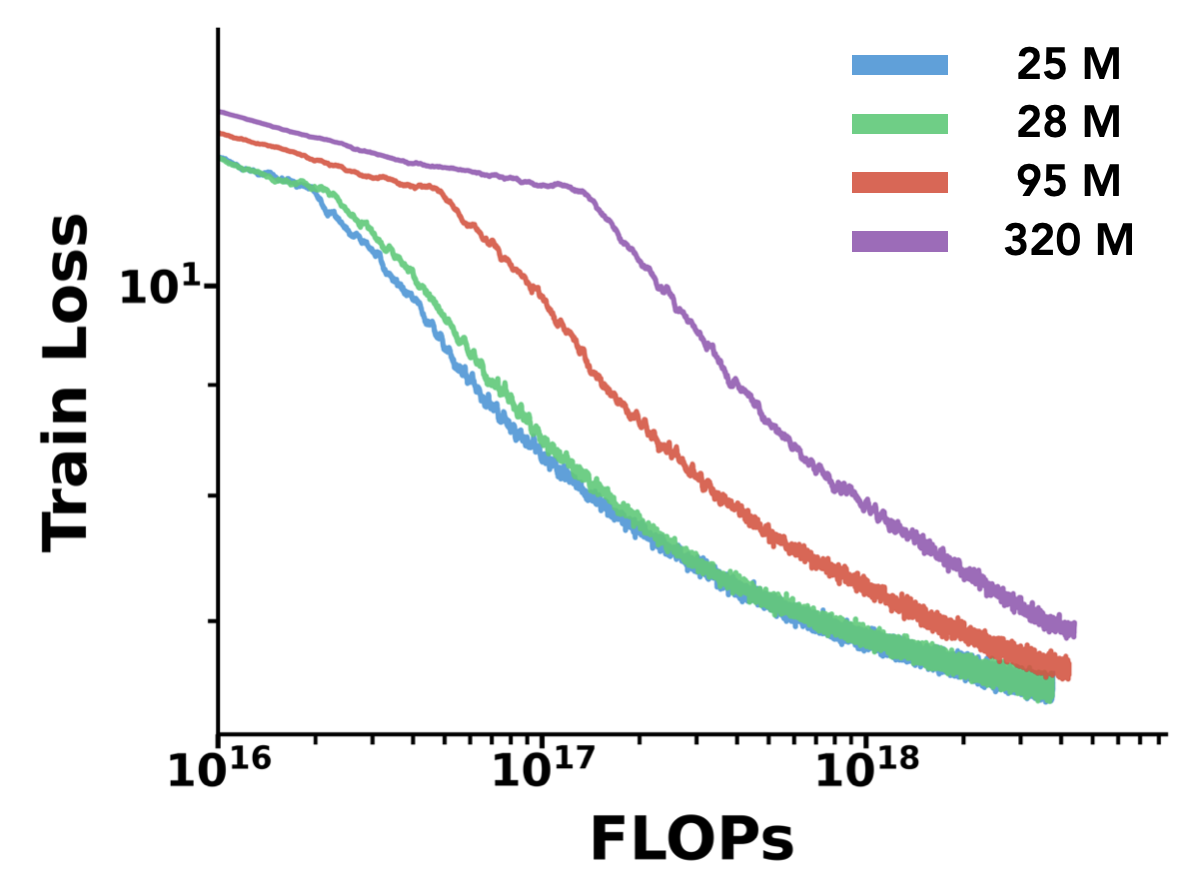}
\caption{\textbf{Scaling with Compute.} Training loss as a function of compute across model scales. Loss (log scale) versus cumulative training FLOPs (log scale) for four model sizes: Very Small (25M), Small (28M), Base (95M), and Large (320M parameters). The plot is cropped to FLOPs $\geq 10^{16}$ to focus on the converged training regime. All models follow similar loss trajectories when plotted against compute, with larger models achieving marginally lower loss at equivalent FLOPs. The overlapping curves suggest that within this model size range, scaling up parameters provides diminishing returns relative to simply training smaller models longer, consistent with the finding that downstream task performance does not improve monotonically with model size.}
\label{fig:app_scaling_flops}
\end{figure*}

\begin{figure*}[ht]
\centering
\includegraphics[width=0.6\textwidth]{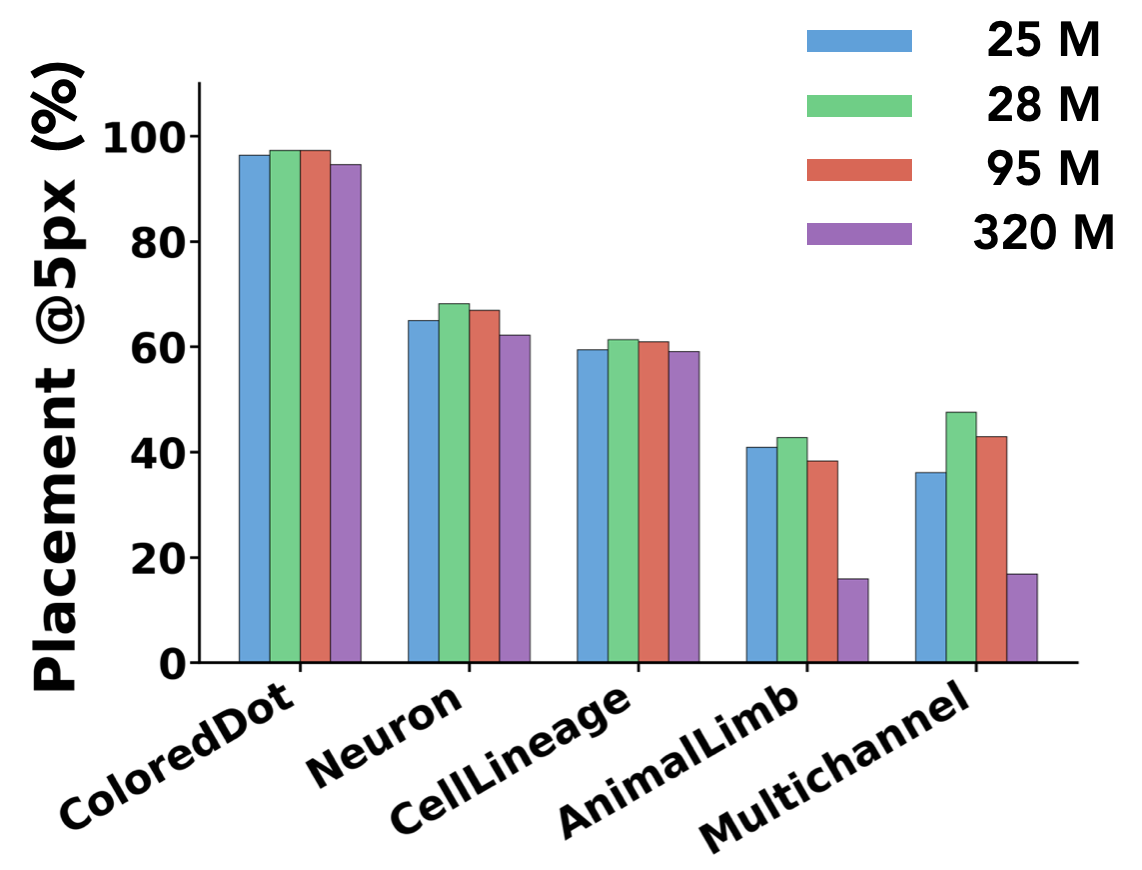}
\caption{\textbf{Placement Precision Across Model Scales.} Fraction of placement actions landing within 5 pixels of the ground truth target under teacher-forced evaluation, grouped by task and model size. This metric isolates spatial precision from action selection by evaluating only placement actions. \taskCDT achieves consistently high precision (94--97\%) across all scales, while \taskNT and \taskCLT show moderate precision (59--68\%). Notably, placement accuracy degrades substantially for the Large model on \taskALT (40.9\% $\to$ 16.0\%) and \taskMIA (47.6\% $\to$ 16.9\%), resulting in overall placement precision dropping from 63.5\% (Small) to 49.8\% (Large). This inverse scaling suggests that larger models may overfit to action-type prediction at the expense of fine-grained spatial localization.}
\label{fig:place_precision_as_model_scales}
\end{figure*}

\begin{figure*}[ht]
\centering
\includegraphics[width=0.6\textwidth]{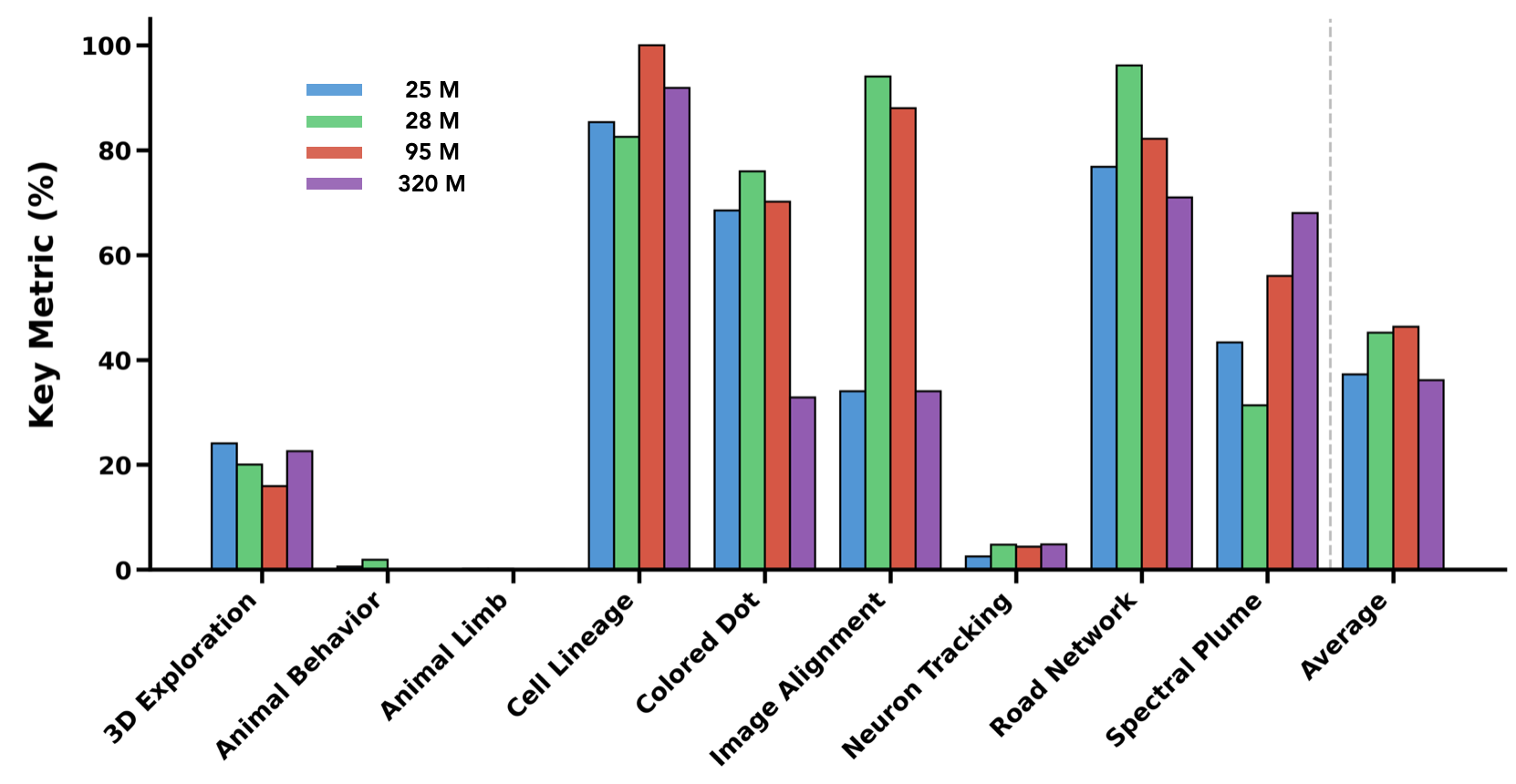}
\caption{\textbf{Task-Specific Performance Across Model Scales.} Key metrics for each of the 9 tasks under autoregressive evaluation on models of increasing size (Very Small: 25M, Small: 28M, Base: 95M, Large: 320M parameters). Each task uses its most informative metric: F1 scores for tracking tasks, accuracy for classification, and completion rate for alignment. Performance varies substantially across tasks, with \taskCLT achieving 82--100\% and \taskRNC reaching 71--96\%, while \taskABT and \taskALT remain below 2\% across all scales. Notably, scaling does not consistently improve performance: the Large model underperforms smaller models on several tasks (\taskCDT, \taskMIA, \taskRNC), and average performance peaks at the Base model (46.3\%) before declining at Large (36.1\%).}
\label{fig:task_specific_performance_multitask}
\end{figure*}

We study the training dynamics of downstream decision and motor task metrics (Figure~\ref{fig:decision_vs_motor}). We examine how performance scales with model size and compute in the multi-task setting. Figure~\ref{fig:app_scaling_flops} shows training loss as a function of FLOPs across model scales, revealing diminishing returns from increased parameters. Figure~\ref{fig:place_precision_as_model_scales} isolates spatial precision by measuring placement accuracy within 5 pixels, showing inverse scaling for some tasks. Figure~\ref{fig:task_specific_performance_multitask} reports task-specific metrics across all 9 tasks, demonstrating that larger models do not consistently improve performance.

\clearpage

\subsection{Downstream Adaptation}

\begin{figure*}[ht]
\centering
\includegraphics[width=0.5\textwidth]{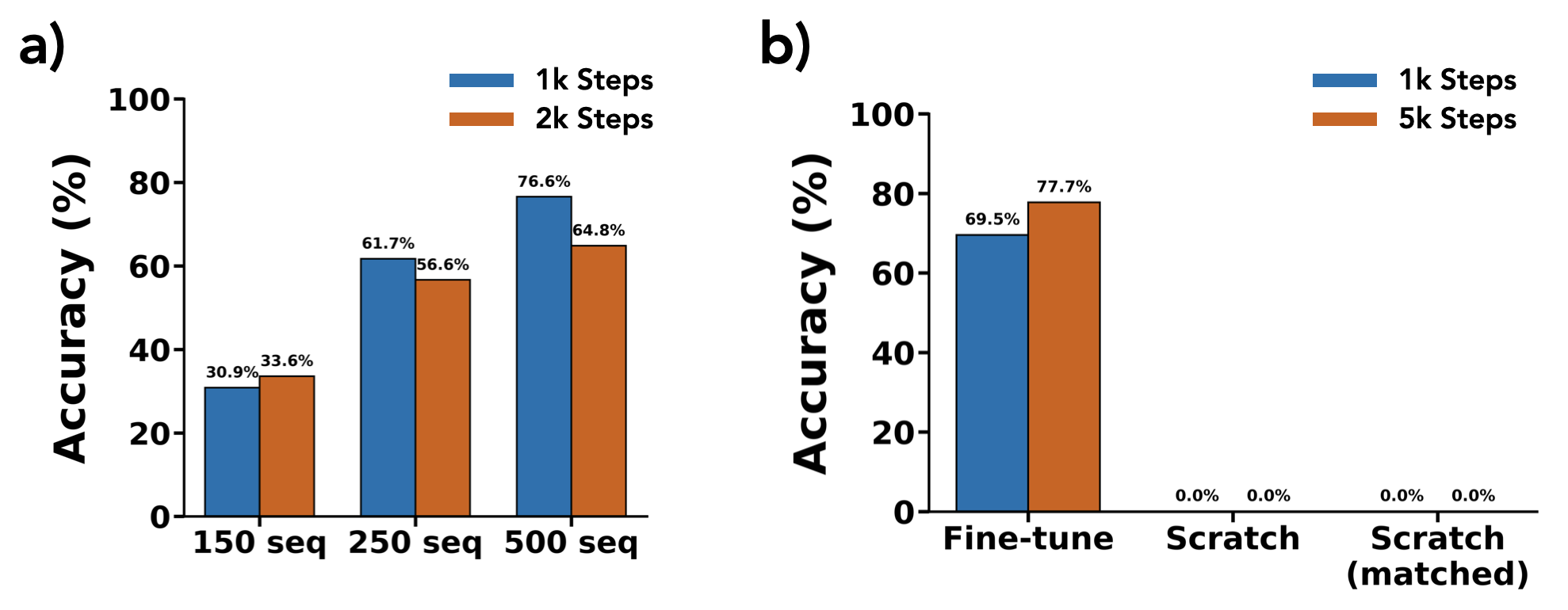}
\caption{\textbf{Fine-Tuning Variants.} (a) Comparison of 1,000 vs 2,000 training steps across different dataset sizes. Longer training improves performance with limited data (150 sequences) but causes overfitting with larger datasets, with accuracy dropping from 76.6\% to 64.8\% for 500 sequences. (b) Fine-tuning from a pretrained checkpoint versus training from scratch, evaluated at 1,000 and 5,000 steps. Fine-tuning achieves 69.5\% accuracy after just 1,000 steps and 77.7\% after 5,000 steps, while training from scratch fails completely (0\% accuracy) regardless of training duration, demonstrating that multitask pretraining provides essential inductive biases that cannot be recovered through extended training alone.}
\label{fig:ft_variants}
\end{figure*}

\begin{figure*}[ht]
\centering
\includegraphics[width=0.5\textwidth]{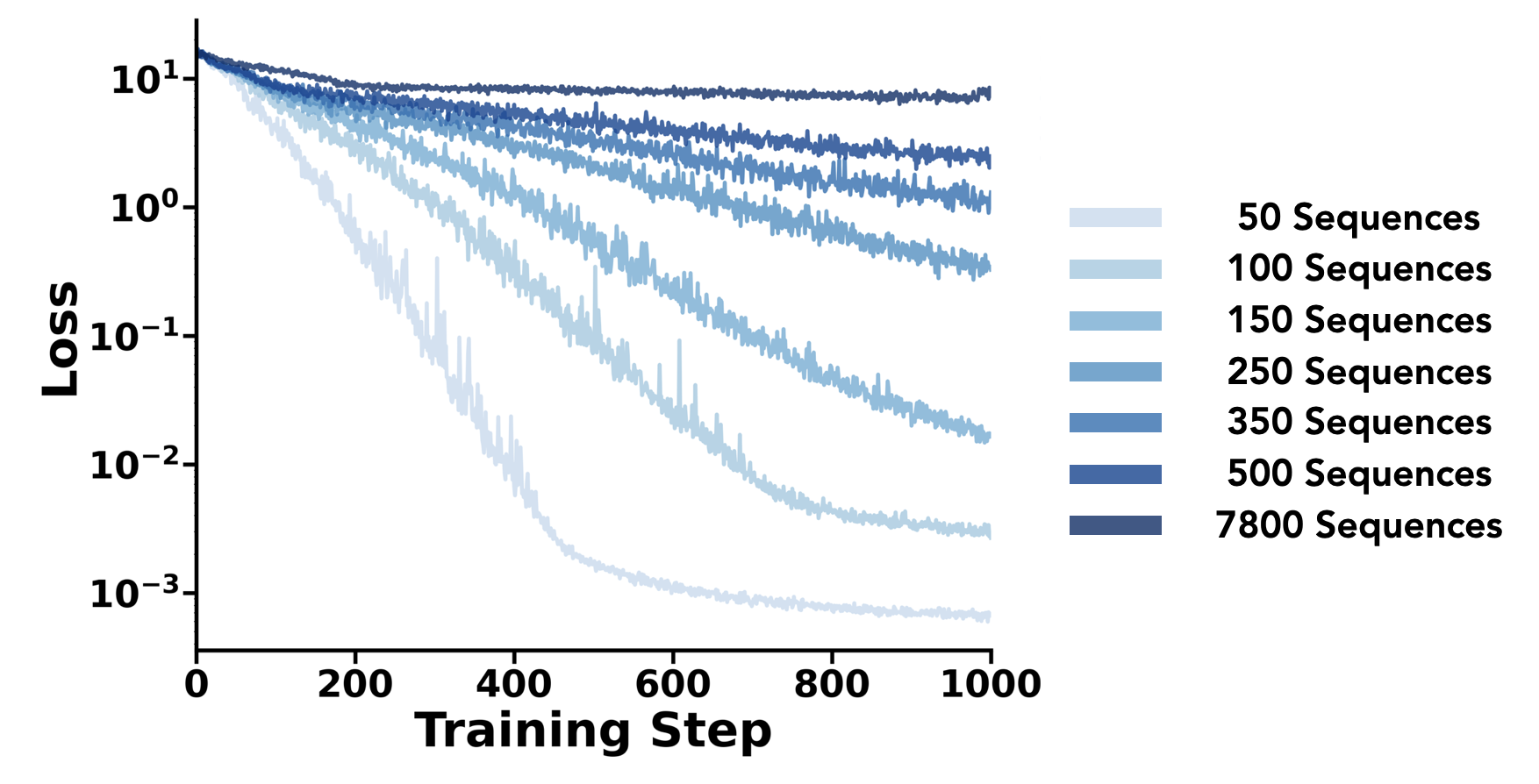}
\caption{\textbf{Fine-Tuning Loss Curves.} Training loss (log scale) over the first 1,000 steps for models fine-tuned on varying amounts of data, from 50 to 7,800 sequences. Smaller datasets (lighter colors) exhibit faster loss reduction and reach lower final loss values, indicating rapid overfitting to the limited training data. Larger datasets (darker colors) show more gradual loss decay, consistent with better generalization. The divergence between training loss and downstream accuracy (where 500 sequences outperforms smaller datasets despite higher loss) highlights that lower training loss does not necessarily correspond to better task performance.}
\label{fig:ft_losses}
\end{figure*}

We provide additional analysis of fine-tuning behavior on the out-of-distribution \taskSM task. Figure~\ref{fig:ft_variants} compares training duration and dataset size effects, and demonstrates that training from scratch fails entirely while fine-tuning succeeds. Figure~\ref{fig:ft_losses} shows loss curves across dataset sizes, illustrating the divergence between training loss and downstream accuracy.

\clearpage

\subsection{Training Ablations}

\begin{figure*}[ht]
\centering
\includegraphics[width=0.4\textwidth]{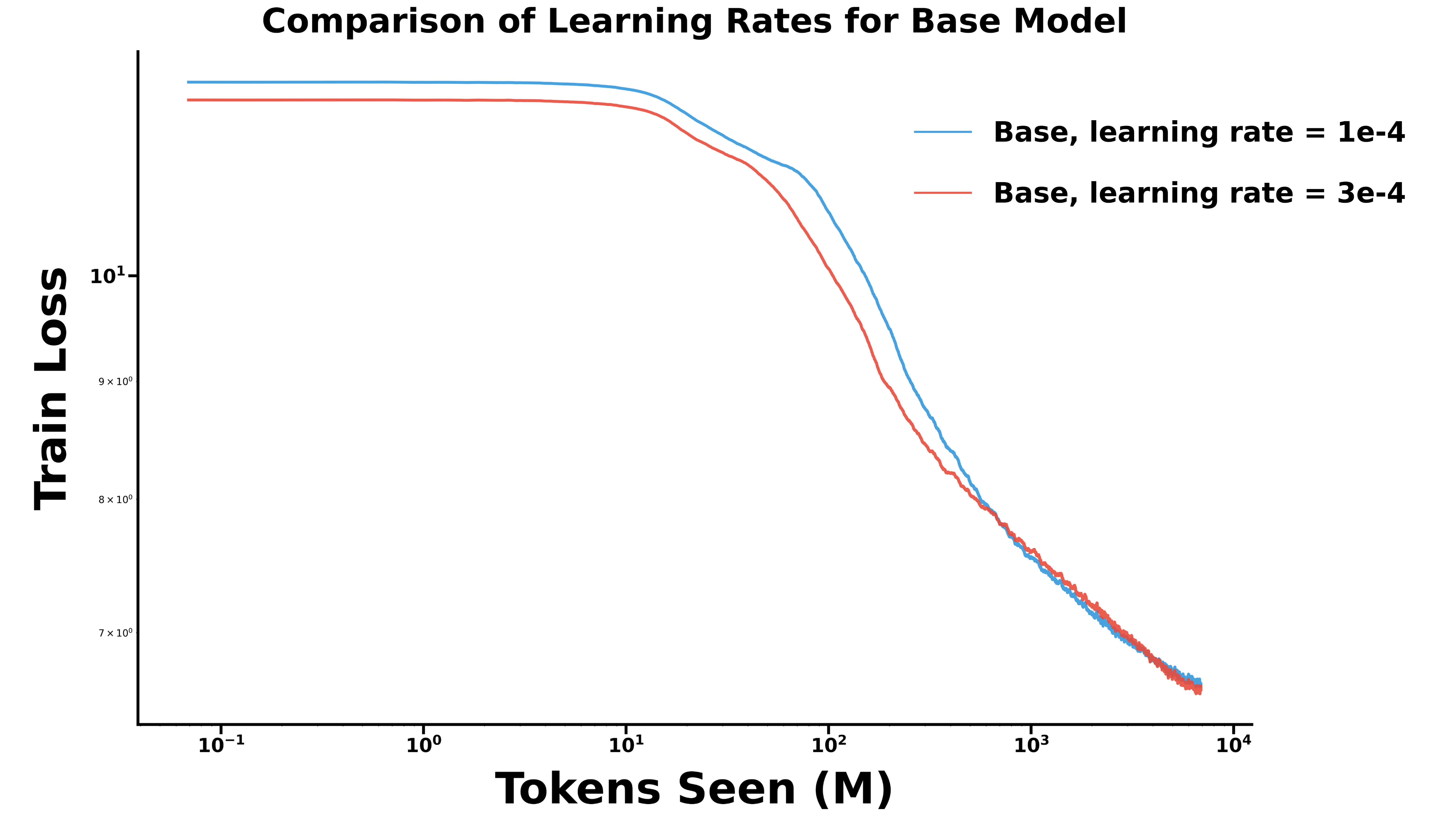}
\caption{\textbf{Learning Rate Comparison.} Train loss for the base model variant across learning rates.}
\label{fig:base_model_multitask_lr}
\end{figure*}

\begin{figure*}[ht]
\centering
\includegraphics[width=0.6\textwidth]{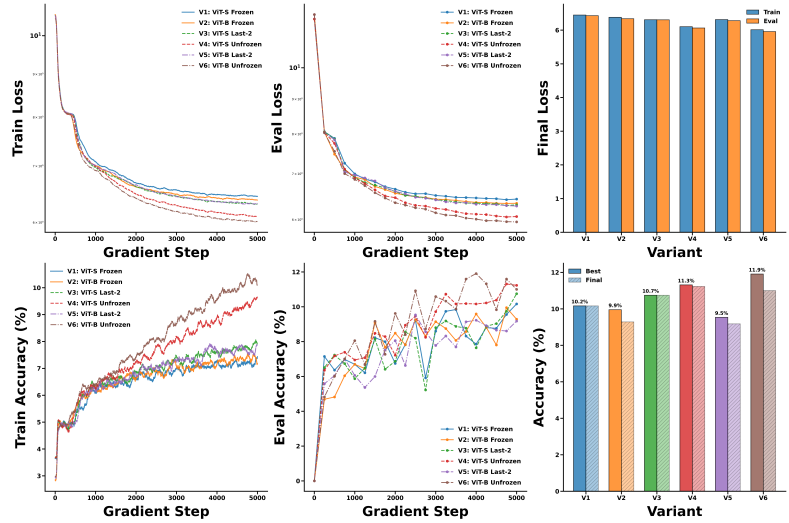}
\caption{\textbf{DINOv2 Training Strategies.} Comparing model sizes and training approaches on \taskCDT.}
\label{fig:dinov2_appendix}
\end{figure*}

\begin{figure*}[ht]
\centering
\includegraphics[width=0.4\textwidth]{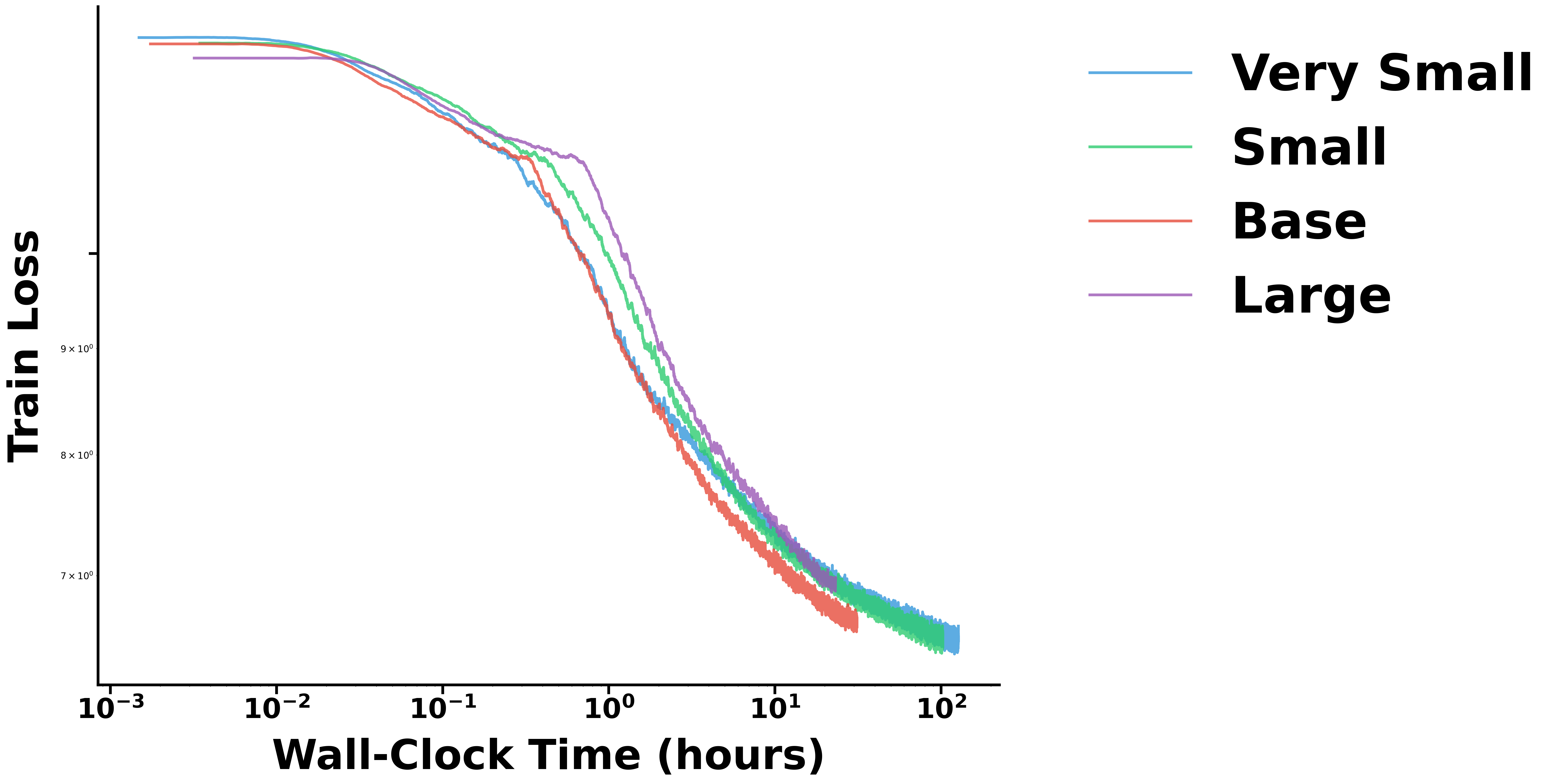}
\caption{\textbf{Training Time Analysis.} Train loss as a function of wall-clock time across four model sizes.}
\label{fig:loss_vs_time}
\end{figure*}

\begin{figure*}[ht]
\centering
\includegraphics[width=0.8\textwidth]{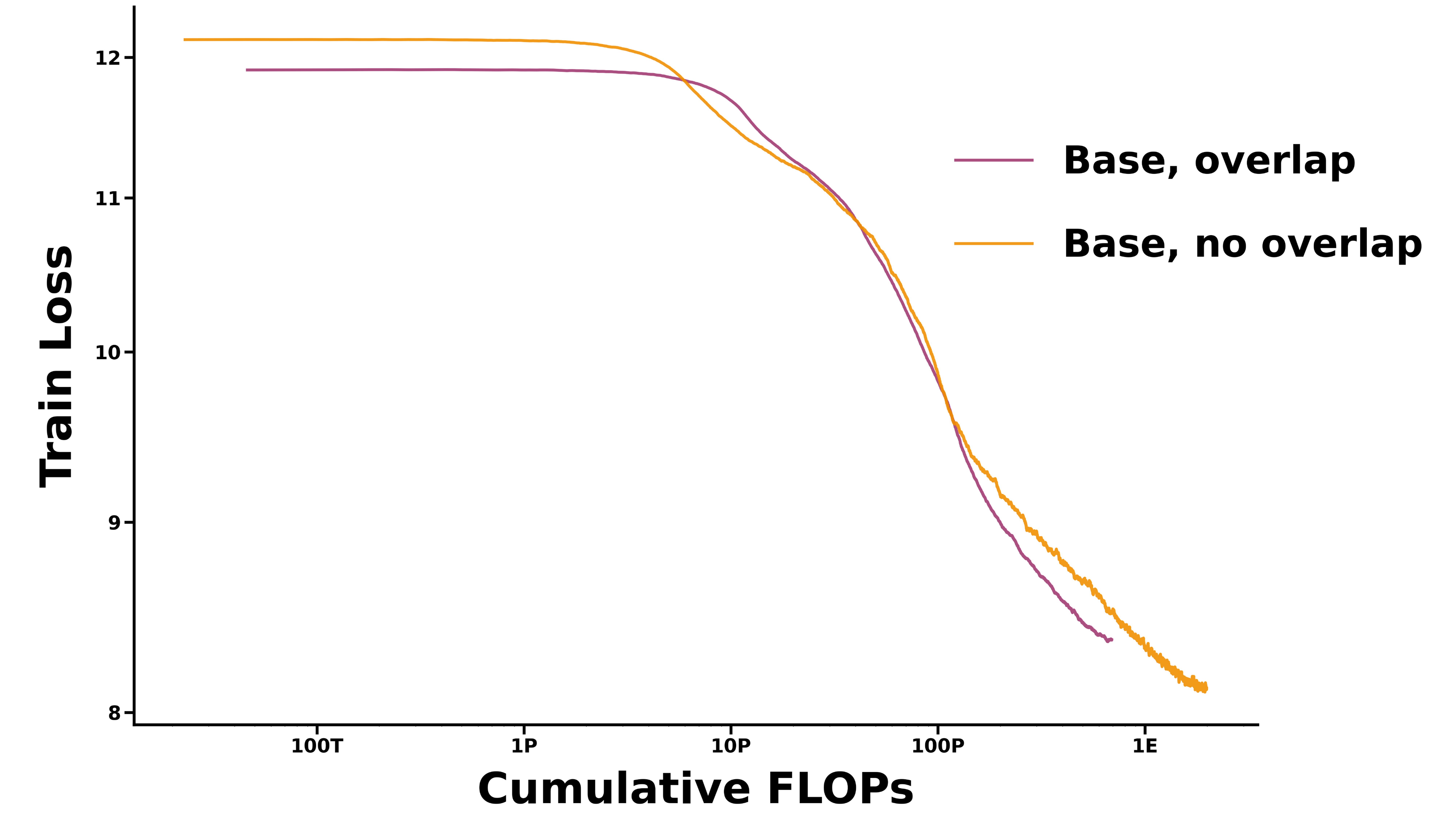}
\caption{\textbf{Data Augmentation Comparison.} In overlap, episodes in the train dataset are converted into context for the model via a sliding-window approach. In no overlap, episodes are split into nearly non-intersecting context windows.}
\label{fig:overlap_vs_no_overlap}
\end{figure*}

We report ablation studies on key training hyperparameters and design choices. Figure~\ref{fig:base_model_multitask_lr} compares learning rates for the base model. Figure~\ref{fig:dinov2_appendix} evaluates DINOv2 variants and training strategies on single-task performance. Figure~\ref{fig:loss_vs_time} shows training loss versus wall-clock time across model sizes. Figure~\ref{fig:overlap_vs_no_overlap} compares sliding-window versus non-overlapping context window construction for data augmentation.

\subsection{VLM Baselines}
\label{app:vlm_baselines}

We compare our 95M behavioral cloning model (step 100K of multi-task training) against two frontier vision--language models on all 9 synthetic tasks: \textbf{Gemini 3 Flash Preview} and \textbf{Qwen3-VL-32B-Instruct}, both accessed via OpenRouter. We provide each VLM with a fully descriptive scaffold: a system prompt describing the task and GUI affordances, the current text state (z-position, placed markers, etc.), and the 3 most recent screenshots per step. Coordinate predictions are calibrated against each VLM's native pixel-action grounding format. Despite this scaffold and the orders-of-magnitude scale advantage, BC outperforms both VLMs on every quantitative metric we report.

\begin{figure}[h]
\centering
\includegraphics[width=0.95\linewidth]{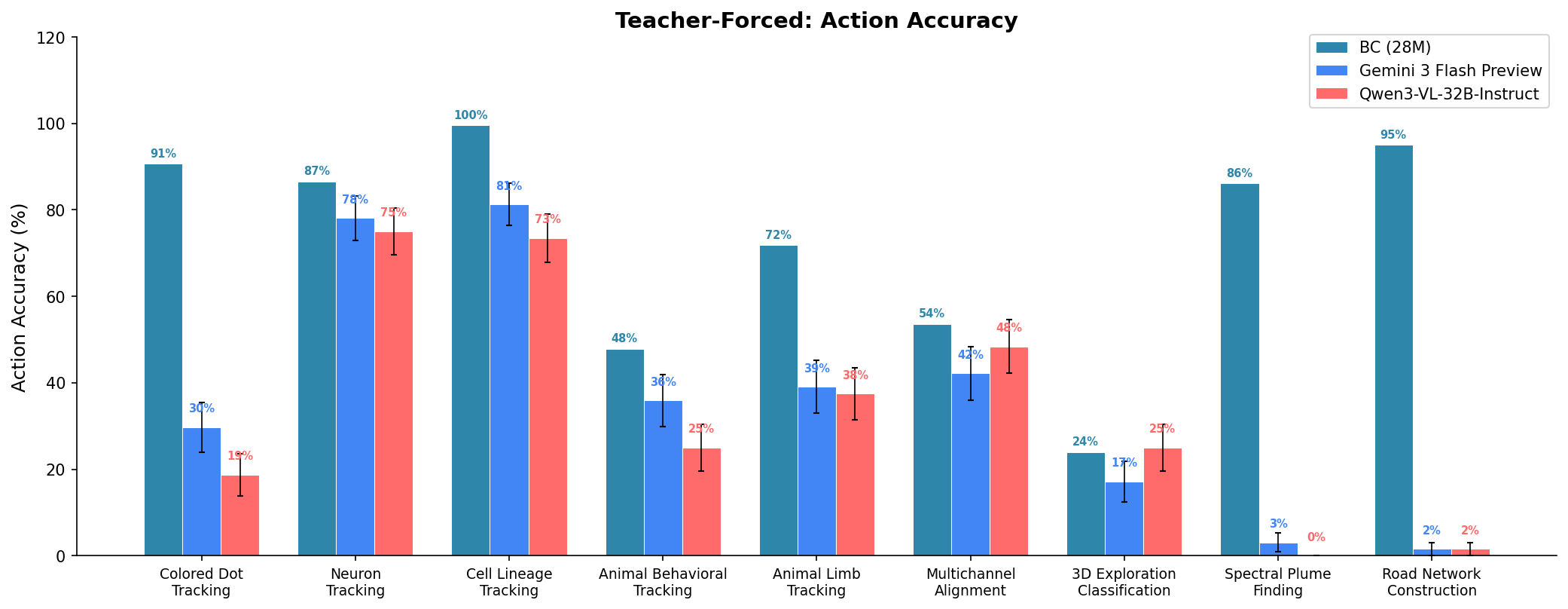}
\caption{\textbf{Teacher-forced action accuracy} across 9 tasks for Gemini 3 Flash Preview and Qwen3-VL-32B-Instruct.}
\label{fig:app_vlm_action_acc}
\end{figure}

\paragraph{Teacher-forced action accuracy.}
With ground-truth screenshots and no compounding errors, action accuracy varies dramatically by task type (Table~\ref{tab:app_vlm_action_acc}, Figure~\ref{fig:app_vlm_action_acc}). Both VLMs do well on tasks that effectively reduce to clicking dot-like targets (\taskCLT 81\%/73\%, \taskNT 78\%/75\%), but degrade sharply on tasks requiring graph topology (\taskRNC: 1.6\%/1.6\%) or multi-band reasoning (\taskSPF: 3.1\%/0\%). Sample size $n=64$ per task; standard errors are bootstrapped with 10K resamples.

\begin{table}[h]
\centering
\small
\caption{\textbf{Teacher-forced action accuracy} on 9 tasks, $n=64$ per task. Standard errors are bootstrapped with 10K resamples.}
\label{tab:app_vlm_action_acc}
\begin{tabular}{lcc}
\toprule
\textbf{Task} & \textbf{Gemini 3 Flash} & \textbf{Qwen3-VL-32B} \\
\midrule
\taskCLT          & \textbf{81.2\%} $\pm$4.9 & 73.4\% $\pm$5.6 \\
\taskNT                & \textbf{78.1\%} $\pm$5.1 & 75.0\% $\pm$5.4 \\
\taskMIA   & 42.2\% $\pm$6.1 & \textbf{48.4\%} $\pm$6.3 \\
\taskALT           & \textbf{39.1\%} $\pm$6.1 & 37.5\% $\pm$6.0 \\
\taskABT     & \textbf{35.9\%} $\pm$6.0 & 25.0\% $\pm$5.4 \\
\taskCDT           & \textbf{29.7\%} $\pm$5.7 & 18.8\% $\pm$4.9 \\
\taskTDE                 & 17.2\% $\pm$4.7 & \textbf{25.0\%} $\pm$5.4 \\
\taskSPF         & \textbf{3.1\%} $\pm$2.2  & 0.0\% \\
\taskRNC      & 1.6\% $\pm$1.5  & 1.6\% $\pm$1.6 \\
\bottomrule
\end{tabular}
\end{table}

\begin{figure}[h]
\centering
\includegraphics[width=0.95\linewidth]{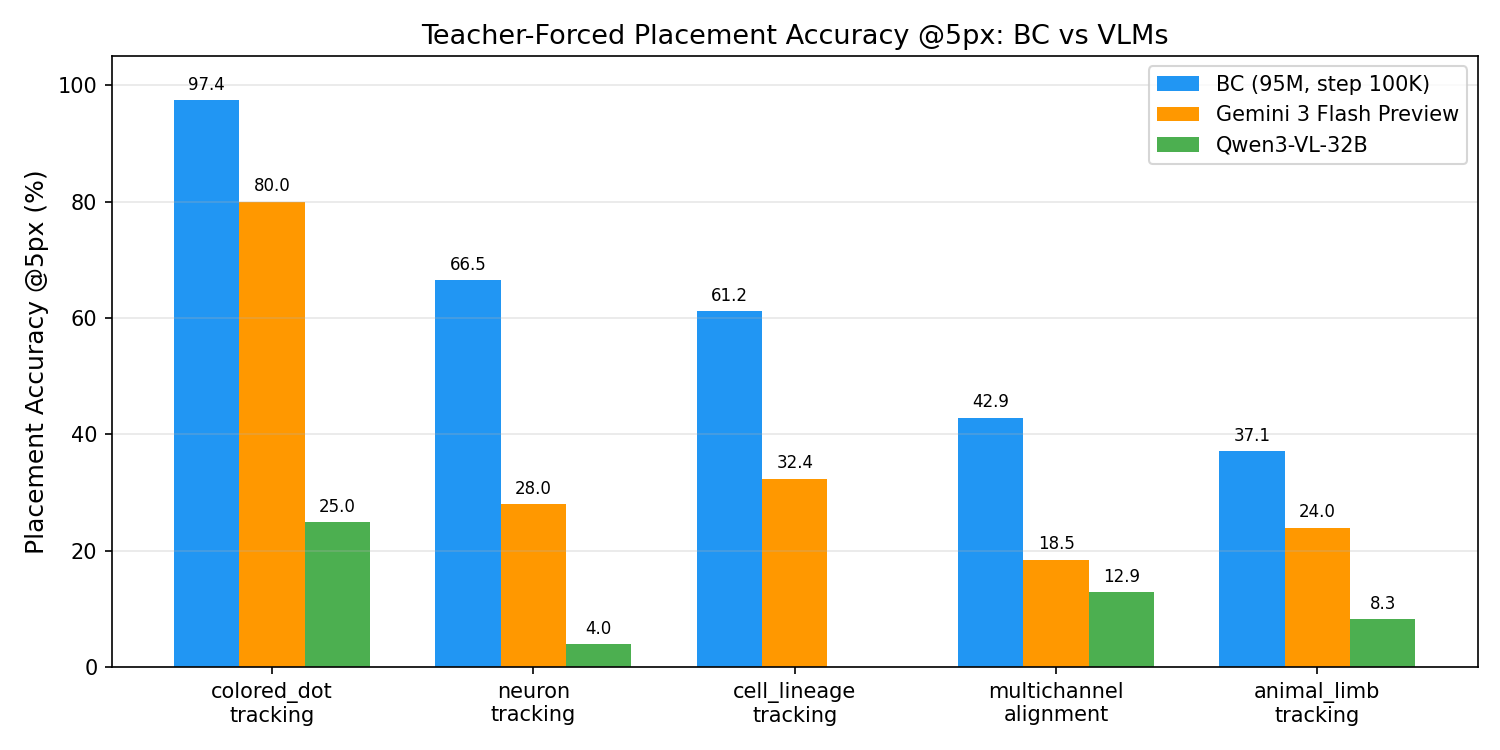}
\caption{\textbf{Teacher-forced placement accuracy @5px:} BC (95M, step 100K) vs.\ Gemini 3 Flash Preview vs.\ Qwen3-VL-32B-Instruct on the 5 tasks with canvas placement actions.}
\label{fig:app_vlm_placement}
\end{figure}

\paragraph{Teacher-forced placement accuracy.}
On the 5 tasks with canvas placement actions, BC outperforms both VLMs (Table~\ref{tab:app_vlm_placement}, Figure~\ref{fig:app_vlm_placement}). The gap is largest on tasks requiring fine pixel-level localization (\taskCDT: BC 97.4\% vs.\ Gemini 80.0\% vs.\ Qwen 25.0\%). VLM placement sample counts are small (5--34 per task) because many of their predicted actions target buttons rather than the canvas; we restrict the comparison to predictions that target the canvas.

\begin{table}[h]
\centering
\small
\caption{\textbf{Teacher-forced placement accuracy @5px.} BC is the 95M model at step 100K of multi-task training.}
\label{tab:app_vlm_placement}
\begin{tabular}{lccc}
\toprule
\textbf{Task} & \textbf{BC (95M)} & \textbf{Gemini} & \textbf{Qwen} \\
\midrule
\taskCDT           & \textbf{97.4\%} & 80.0\% & 25.0\% \\
\taskNT                & \textbf{66.5\%} & 28.0\% & 4.0\% \\
\taskCLT          & \textbf{61.2\%} & 32.4\% & 0.0\% \\
\taskMIA   & \textbf{42.9\%} & 18.5\% & 12.9\% \\
\taskALT           & \textbf{37.1\%} & 24.0\% & 8.3\% \\
\bottomrule
\end{tabular}
\end{table}

\begin{figure}[h]
\centering
\includegraphics[width=0.95\linewidth]{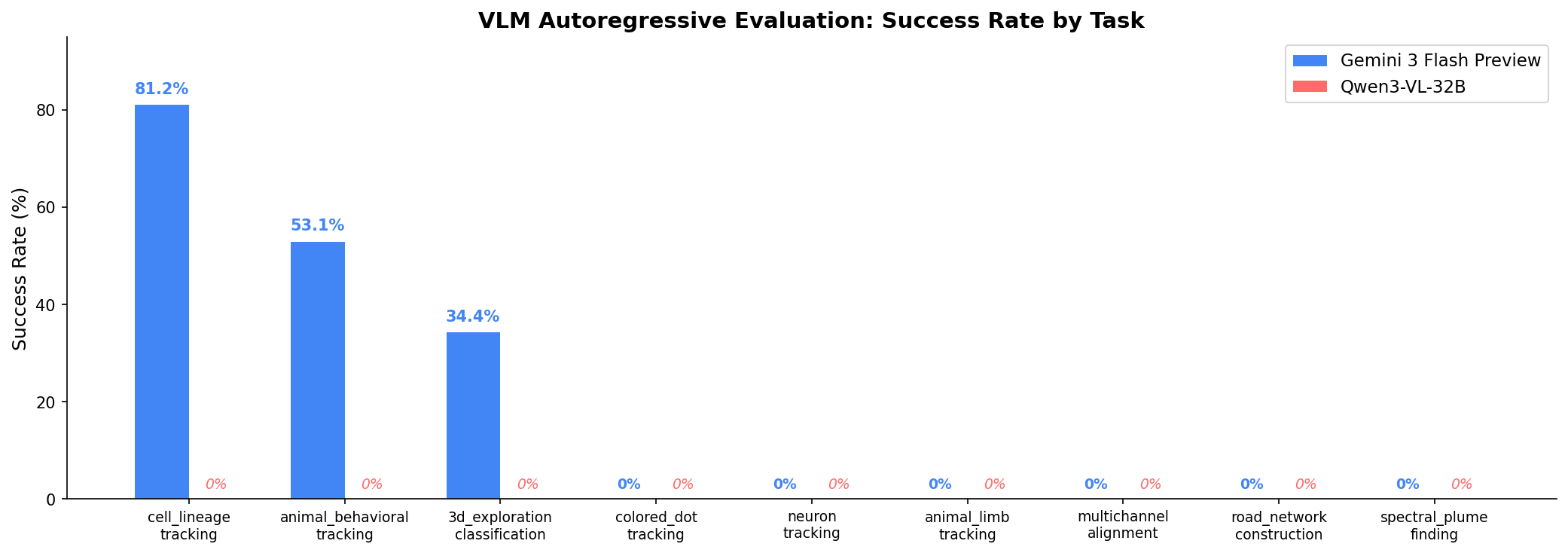}
\caption{\textbf{Autoregressive success rate} for Gemini and Qwen across 9 tasks, 32 instances per task.}
\label{fig:app_vlm_autoreg}
\end{figure}

\paragraph{Autoregressive success rate.}
In autoregressive evaluation (32 instances per task, with the same scaffold as above), Gemini achieves non-zero success on 3/9 tasks (\taskCLT 81.2\%, \taskABT 53.1\%, \taskTDE 34.4\%); Qwen achieves 0\% on all 9 tasks, getting stuck in repetition loops or API failures (Table~\ref{tab:app_vlm_autoreg}, Figure~\ref{fig:app_vlm_autoreg}). The general failure under autoregressive conditions is expected: these are long-horizon tasks (50--600 steps) where errors compound rapidly without recovery.

\begin{table}[h]
\centering
\small
\caption{\textbf{Autoregressive success rate} on 9 tasks, 32 instances per task, with scaffold (text state + 3 screenshots per step).}
\label{tab:app_vlm_autoreg}
\begin{tabular}{lcc}
\toprule
\textbf{Task} & \textbf{Gemini 3 Flash} & \textbf{Qwen3-VL-32B} \\
\midrule
\taskCLT          & \textbf{81.2\%} & 0\% \\
\taskABT     & \textbf{53.1\%} & 0\% \\
\taskTDE                 & \textbf{34.4\%} & 0\% \\
\taskCDT           & 0\% & 0\% \\
\taskNT                & 0\% & 0\% \\
\taskALT           & 0\% & 0\% \\
\taskMIA   & 0\% & 0\% \\
\taskRNC      & 0\% & 0\% \\
\taskSPF         & 0\% & 0\% \\
\bottomrule
\end{tabular}
\end{table}

\FloatBarrier
\subsection{UI Layout Adaptation}
\label{app:ui_adaptation}

Tests whether the model memorizes pixel positions tied to a specific GUI layout or learns generalizable annotation behavior. The pretrained 95M multi-task model (step 100K) is fine-tuned on 8 diverse UI layout variants of \taskCDT and evaluated on those 8 in-distribution (ID) variants plus 3 held-out OOD variants combining visual axes in novel ways. Visual axes varied: panel position (left/right/top/bottom/split), theme (dark/light/retro), button style (rounded/pill/square), button size, canvas border, status position. Fine-tuning: 5K steps on $8 \times 500 = 4{,}000$ sequences (${\sim}599$K frames).

\begin{figure}[h]
\centering
\setlength{\tabcolsep}{1pt}
\renewcommand{\arraystretch}{0.9}
\small
\begin{tabular}{cccc}
\includegraphics[width=0.21\linewidth]{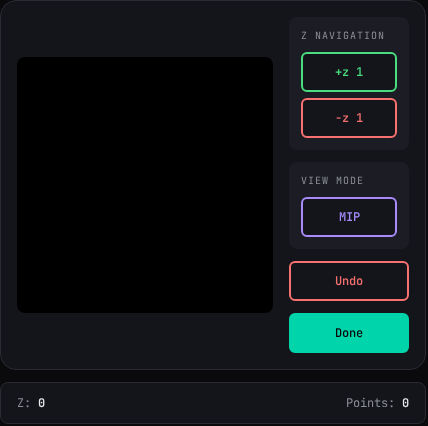} &
\includegraphics[width=0.21\linewidth]{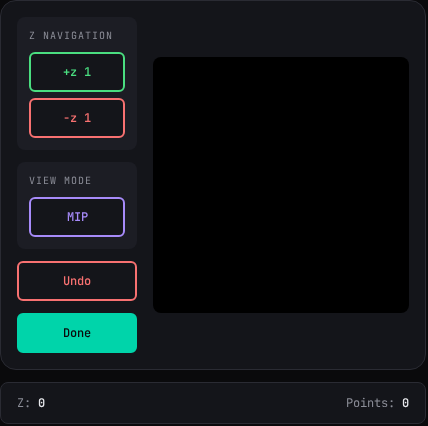} &
\includegraphics[width=0.21\linewidth]{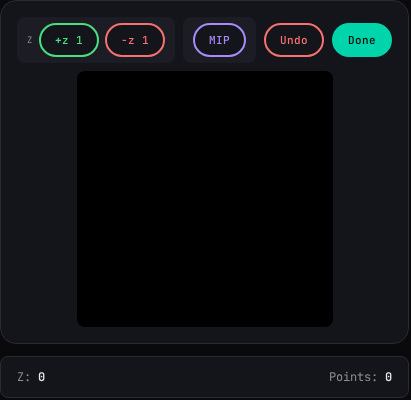} &
\includegraphics[width=0.21\linewidth]{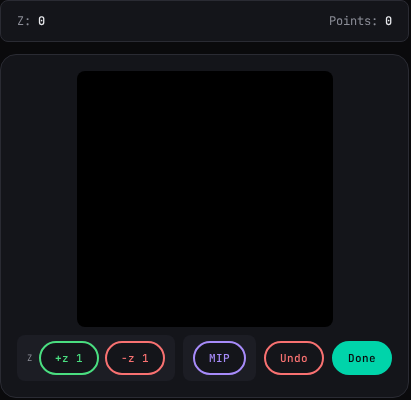} \\
original & left\_panel & top\_toolbar & bottom\_toolbar \\[0.3em]
\includegraphics[width=0.21\linewidth]{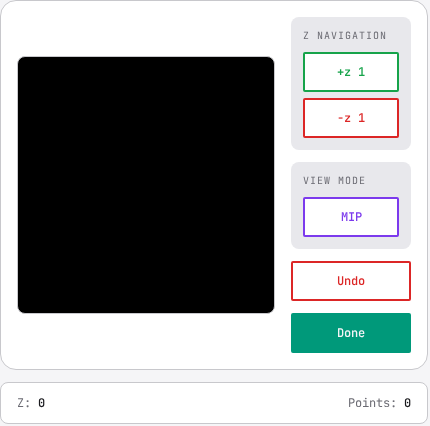} &
\includegraphics[width=0.21\linewidth]{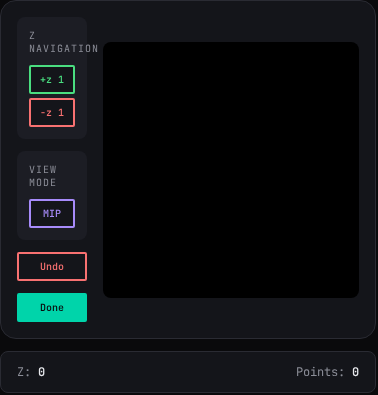} &
\includegraphics[width=0.21\linewidth]{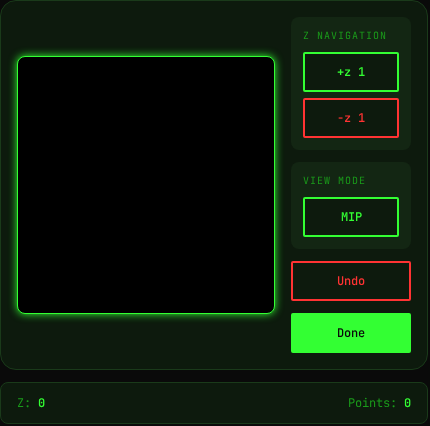} &
\includegraphics[width=0.21\linewidth]{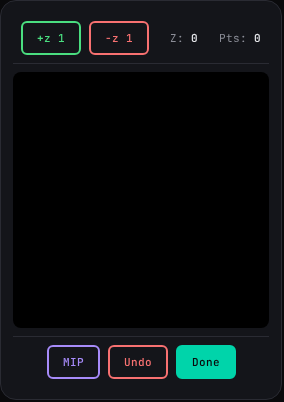} \\
light\_right & minimal\_left & retro & split \\[0.5em]
\includegraphics[width=0.21\linewidth]{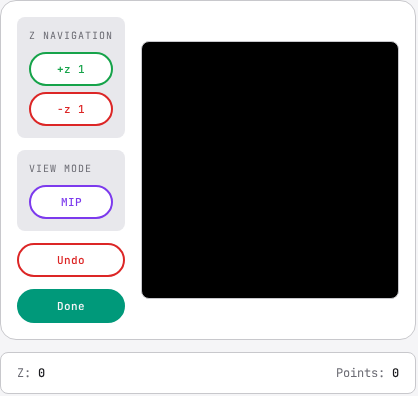} &
\includegraphics[width=0.21\linewidth]{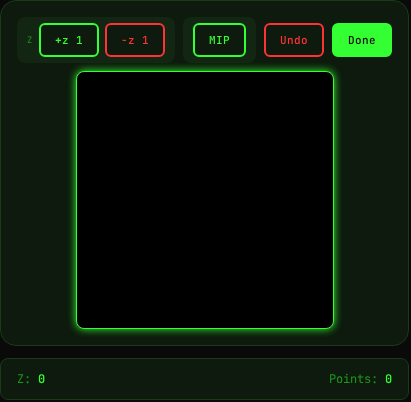} &
\includegraphics[width=0.21\linewidth]{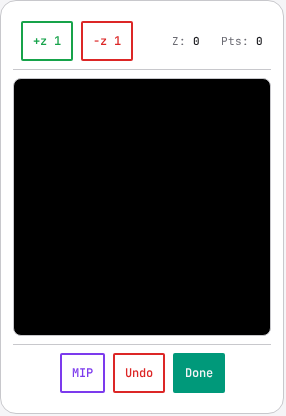} & \\
\textit{ood\_light\_left} & \textit{ood\_retro\_top} & \textit{ood\_light\_split} & \\
\end{tabular}
\caption{\textbf{UI variants.} Top two rows: 8 in-distribution variants used for fine-tuning. Bottom row: 3 out-of-distribution variants held out from fine-tuning; each combines visual axes in a way not present in the training set (e.g., \textit{ood\_light\_left} pairs the light theme with a left panel, while training only paired light with right panels).}
\label{fig:app_ui_variants}
\end{figure}

\paragraph{Teacher-forced evaluation.}
The base model's accuracy correlates with visual similarity to the original layout (22\% on original, 2--9\% on rearranged layouts). Fine-tuning brings all variants to 22--35\%, including OOD variants never seen during training (avg 26.6\%), confirming the model learns UI-invariant behavior. Performance on the original layout is preserved (22.4\% $\to$ 23.2\%).

\begin{figure}[h]
\centering
\includegraphics[width=0.95\linewidth]{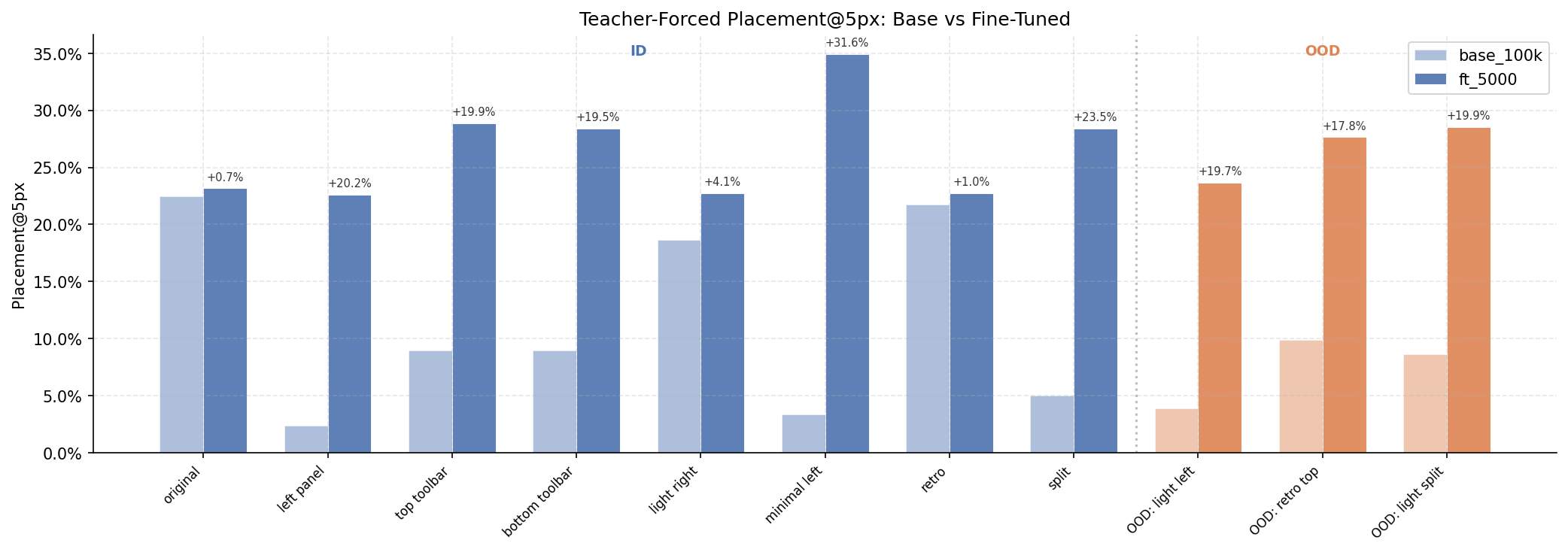}
\caption{\textbf{Teacher-forced placement accuracy @5px} across all 11 UI variants, grouped by base (pretrained on the original layout only) vs.\ fine-tuned (5K steps on the 8 ID variants).}
\label{fig:app_ui_tf_bar}
\end{figure}

\begin{table}[h]
\centering
\small
\caption{\textbf{Teacher-forced placement accuracy @5px} per UI variant. Base $=$ multi-task pretrained at step 100K; FT $=$ fine-tuned for 5K steps on the 8 ID variants.}
\label{tab:app_ui_tf}
\begin{tabular}{llccc}
\toprule
\textbf{Variant} & \textbf{Split} & \textbf{Base @5px} & \textbf{FT @5px} & $\Delta$ \\
\midrule
original        & ID  & 22.4\% & \textbf{23.2\%} & $+0.8$ \\
retro           & ID  & 21.7\% & \textbf{22.8\%} & $+1.1$ \\
light\_right    & ID  & 18.6\% & \textbf{22.7\%} & $+4.1$ \\
top\_toolbar    & ID  & 9.0\%  & \textbf{28.9\%} & $+19.9$ \\
bottom\_toolbar & ID  & 8.9\%  & \textbf{28.4\%} & $+19.5$ \\
split           & ID  & 5.0\%  & \textbf{28.4\%} & $+23.4$ \\
minimal\_left   & ID  & 3.3\%  & \textbf{34.9\%} & $+31.6$ \\
left\_panel     & ID  & 2.4\%  & \textbf{22.6\%} & $+20.2$ \\
\midrule
ood\_retro\_top   & OOD & 9.9\% & \textbf{27.7\%} & $+17.8$ \\
ood\_light\_split & OOD & 8.6\% & \textbf{28.5\%} & $+19.9$ \\
ood\_light\_left  & OOD & 3.9\% & \textbf{23.6\%} & $+19.7$ \\
\bottomrule
\end{tabular}
\end{table}

\begin{table}[h]
\centering
\small
\caption{\textbf{Teacher-forced placement accuracy @5px, averaged by split.}}
\label{tab:app_ui_summary}
\begin{tabular}{lcc}
\toprule
\textbf{Split} & \textbf{Base @5px (avg)} & \textbf{FT @5px (avg)} \\
\midrule
ID (8 variants)   & 11.4\% & \textbf{26.5\%} \\
OOD (3 variants)  & 7.5\%  & \textbf{26.6\%} \\
\bottomrule
\end{tabular}
\end{table}

\begin{figure}[h]
\centering
\includegraphics[width=0.95\linewidth]{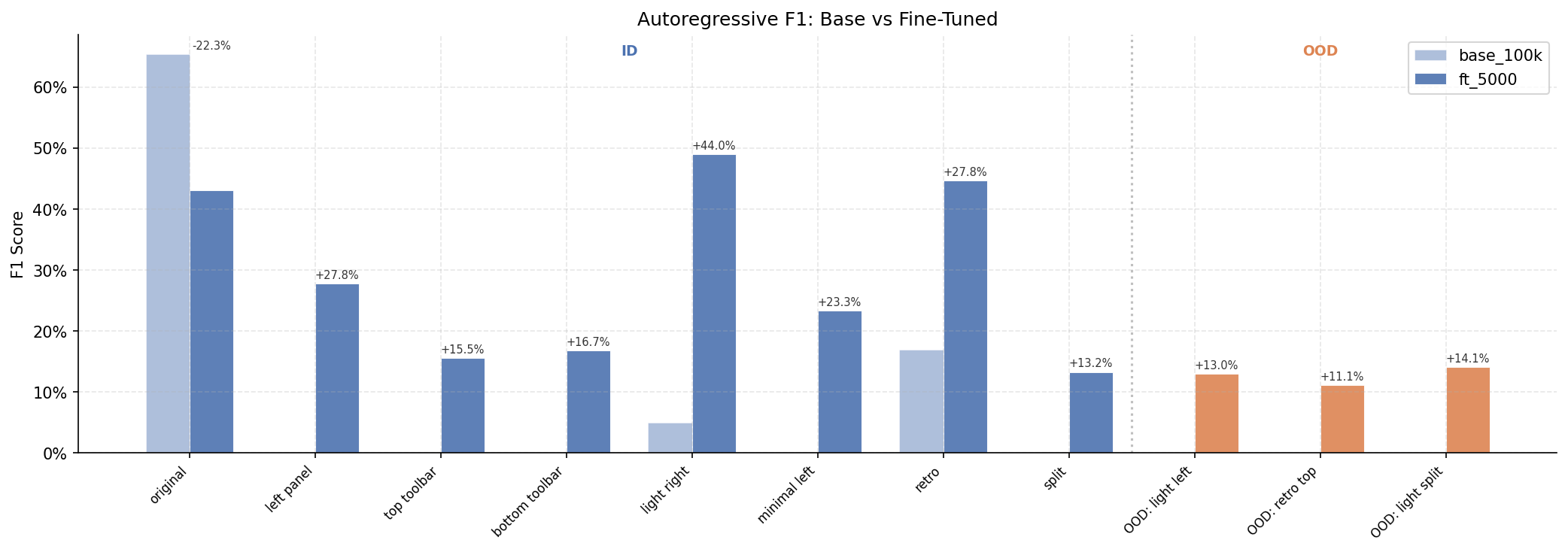}
\caption{\textbf{Autoregressive F1} across all 11 UI variants, base vs.\ fine-tuned (20 episodes per variant, max 300 steps).}
\label{fig:app_ui_ar_bar}
\end{figure}

\paragraph{Autoregressive evaluation.}
The base model only functions on the original layout (F1$=0.654$) and partially on retro (F1$=0.168$); all other layouts yield F1$=0$. Fine-tuning enables non-zero F1 on all 11 variants, including the 3 OOD layouts (Table~\ref{tab:app_ui_ar}, Figure~\ref{fig:app_ui_ar_bar}).

\begin{table}[h]
\centering
\small
\caption{\textbf{Autoregressive F1} per UI variant (20 samples per variant, max 300 steps).}
\label{tab:app_ui_ar}
\begin{tabular}{llcc}
\toprule
\textbf{Variant} & \textbf{Split} & \textbf{Base F1} & \textbf{FT F1} \\
\midrule
original        & ID  & \textbf{0.654} & 0.430 \\
retro           & ID  & 0.168 & \textbf{0.446} \\
light\_right    & ID  & 0.050 & \textbf{0.490} \\
left\_panel     & ID  & 0.000 & \textbf{0.278} \\
top\_toolbar    & ID  & 0.000 & \textbf{0.155} \\
bottom\_toolbar & ID  & 0.000 & \textbf{0.167} \\
minimal\_left   & ID  & 0.000 & \textbf{0.233} \\
split           & ID  & 0.000 & \textbf{0.133} \\
\midrule
ood\_light\_left  & OOD & 0.000 & \textbf{0.130} \\
ood\_retro\_top   & OOD & 0.000 & \textbf{0.111} \\
ood\_light\_split & OOD & 0.000 & \textbf{0.141} \\
\bottomrule
\end{tabular}
\end{table}

\FloatBarrier
\subsection{DAgger}
\label{app:dagger}

Tests whether DAgger~\citep{ross2011reduction} can improve autoregressive performance on the two tasks where BC achieves 0\% accuracy: animal\_behavioral\_tracking and animal\_limb\_tracking.

\textbf{Method.} $\beta$-DAgger ($\beta=0.1$). 50 instances per task collected with 90\% model policy, 10\% oracle corrections; state-aware greedy oracles provide ground-truth actions. DAgger data (upsampled $50\times$) mixed with original training data (${\sim}26\%$ of training mix). Fine-tuned for 5K steps.

\begin{figure}[h]
\centering
\includegraphics[width=0.95\linewidth]{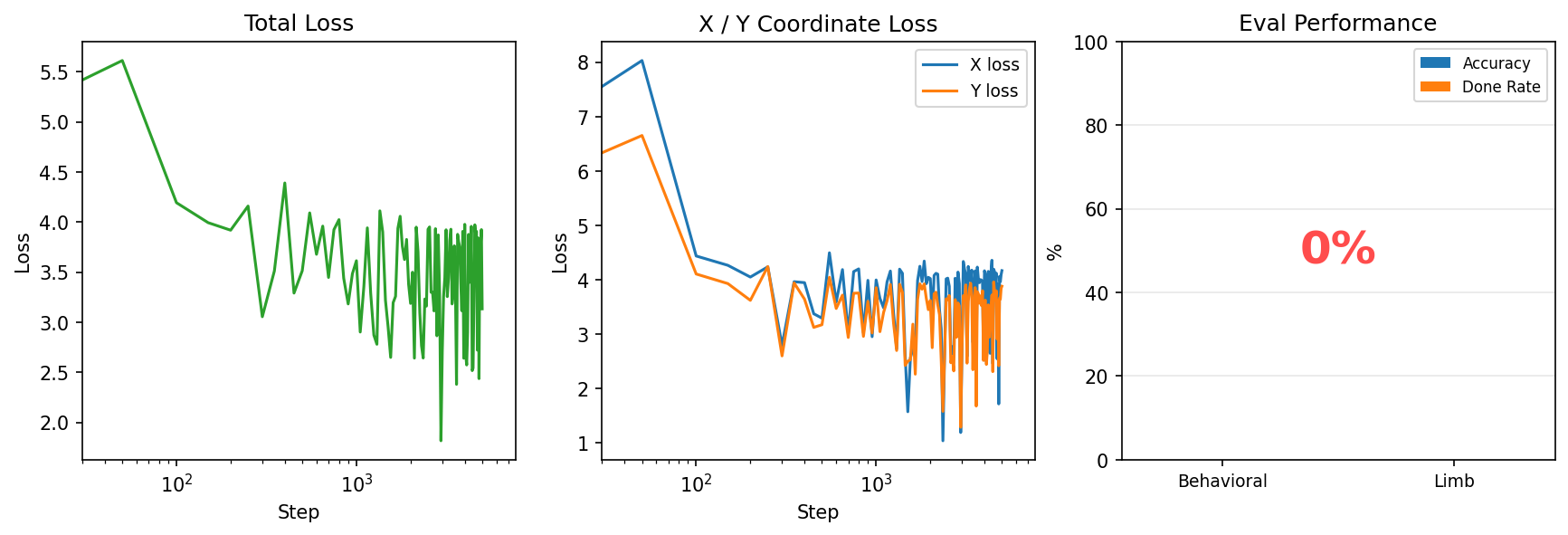}
\caption{\textbf{DAgger results.} Loss curves (left columns) and autoregressive evaluation (right columns) for animal\_behavioral\_tracking and animal\_limb\_tracking. Both tasks remain at 0\% accuracy and 0\% done rate after DAgger fine-tuning.}
\label{fig:app_dagger}
\end{figure}

\begin{table}[h]
\centering
\small
\caption{\textbf{DAgger autoregressive evaluation} (30 samples per task, max 500 steps).}
\label{tab:app_dagger}
\begin{tabular}{llccc}
\toprule
\textbf{Task} & \textbf{Condition} & \textbf{Accuracy} & \textbf{Done rate} & \textbf{Avg steps} \\
\midrule
animal\_behavioral\_tracking & Baseline & 0\% & 1.3\% & 494 \\
animal\_behavioral\_tracking & DAgger   & 0\% & 0\%   & 500 \\
animal\_limb\_tracking       & Baseline & 0\% & 0\%   & 491 \\
animal\_limb\_tracking       & DAgger   & 0\% & 0\%   & 492 \\
\bottomrule
\end{tabular}
\end{table}

DAgger does not improve performance: both tasks remain at 0\% accuracy and 0\% done rate. All 60 evaluation instances hit the max-step limit with zero markers placed. This confirms that failure on these tasks reflects fundamental task difficulty (complex hierarchical annotation workflows), not compounding errors from distribution shift.

\clearpage

\section{Real-Data Experiments}
\label{app:real_data}

\subsection{Connectomics Tracing}
\label{app:connectomics_tracing}

We fine-tune the multi-task pretrained 95M-parameter model (100K-step base checkpoint) on real EM neuron tracing from two volumes: the H01 human cortex dataset~\citep{shapson2024petavoxel} and the \textit{C. elegans} nerve ring~\citep{witvliet2021connectomes}. In both cases, the model sees only raw EM with no segmentation overlay, must identify the target neuron in the centered frame at $z{=}0$, and trace its cross-section through 50 z-slices using $\pm$z navigation, Undo to correct, and Done to terminate. We use the same architecture, action space, and GUI as the synthetic tasks (Sec.~\ref{sec:framework}); only the data and the fine-tuned weights differ. Training hyperparameters are shared across the two runs: effective batch size 32 (2 per GPU $\times$ 4 gradient accumulation $\times$ 4$\times$A100), constant learning rate $1\mathrm{e}{-4}$.

\subsubsection{H01 Human Cortex}
\label{app:connectomics_tracing:h01}

\textbf{Data.} EM volumes from the H01 dataset~\citep{shapson2024petavoxel} at 16~nm/px. Ground-truth tracing targets are derived from the dense segmentation: 1{,}013 myelinated axon skeletons (28 held out). Fine-tuning data: ${\sim}6.5$M training frames from 52{,}000 episodes (4$\times$ rotation augmentation). The model is trained for 117K gradient steps (${\sim}1$ epoch).

\begin{figure}[h]
\centering
\begin{subfigure}{0.48\linewidth}
\centering
\includegraphics[width=\linewidth]{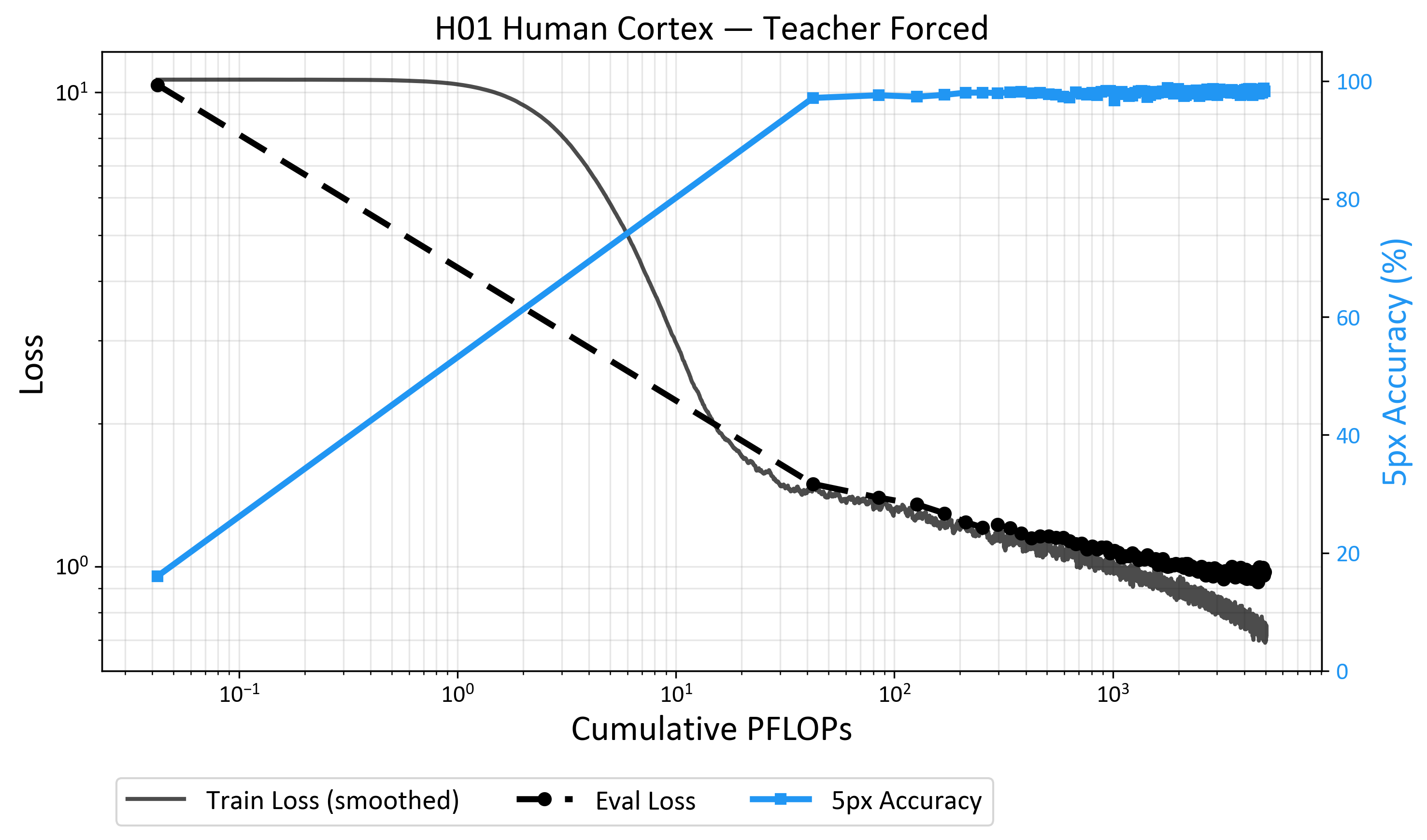}
\caption{Teacher-forced training loss and overall placement accuracy.}
\end{subfigure}\hfill
\begin{subfigure}{0.48\linewidth}
\centering
\includegraphics[width=\linewidth]{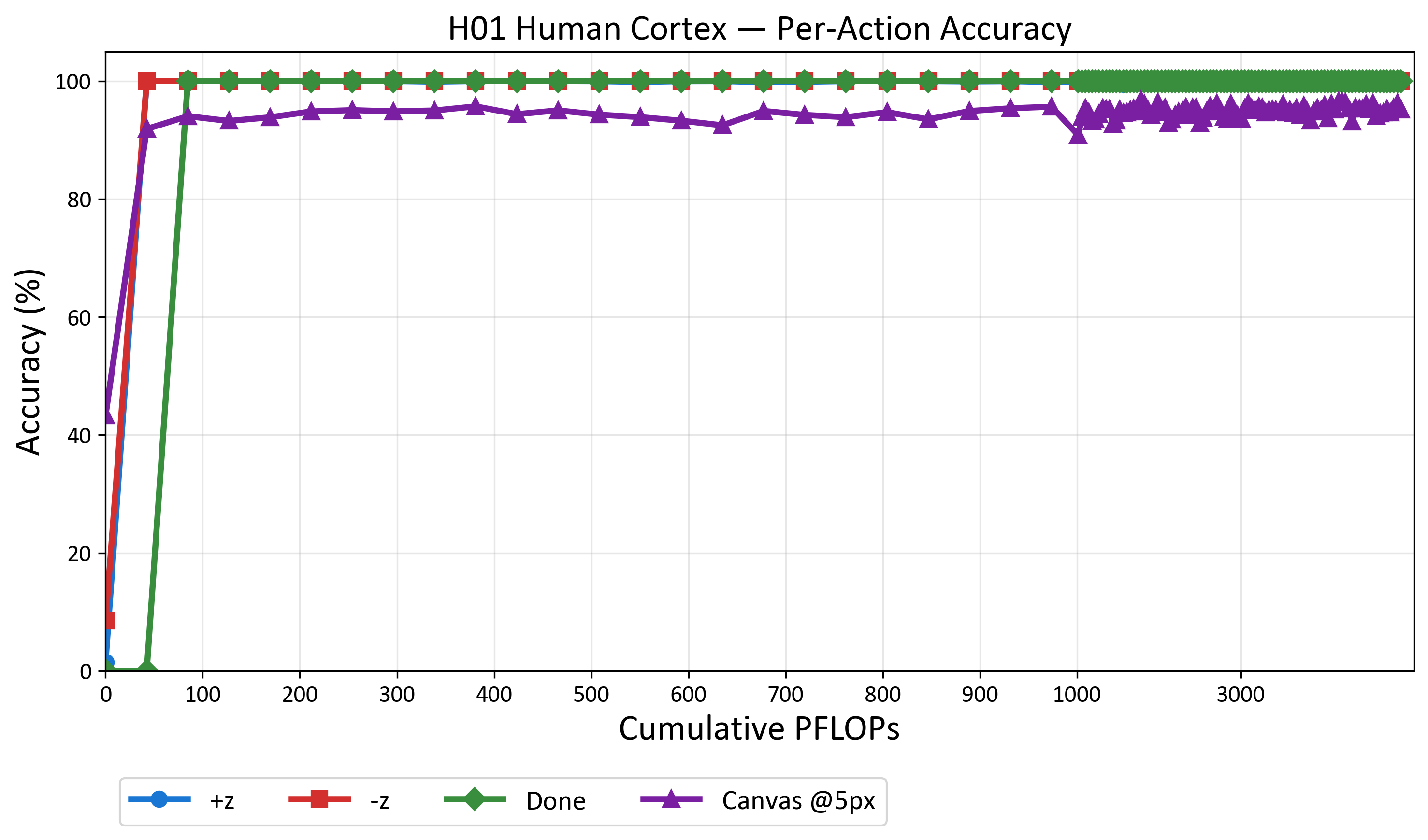}
\caption{Teacher-forced per-action accuracy.}
\end{subfigure}\\[0.5em]
\begin{subfigure}{0.85\linewidth}
\centering
\includegraphics[width=\linewidth]{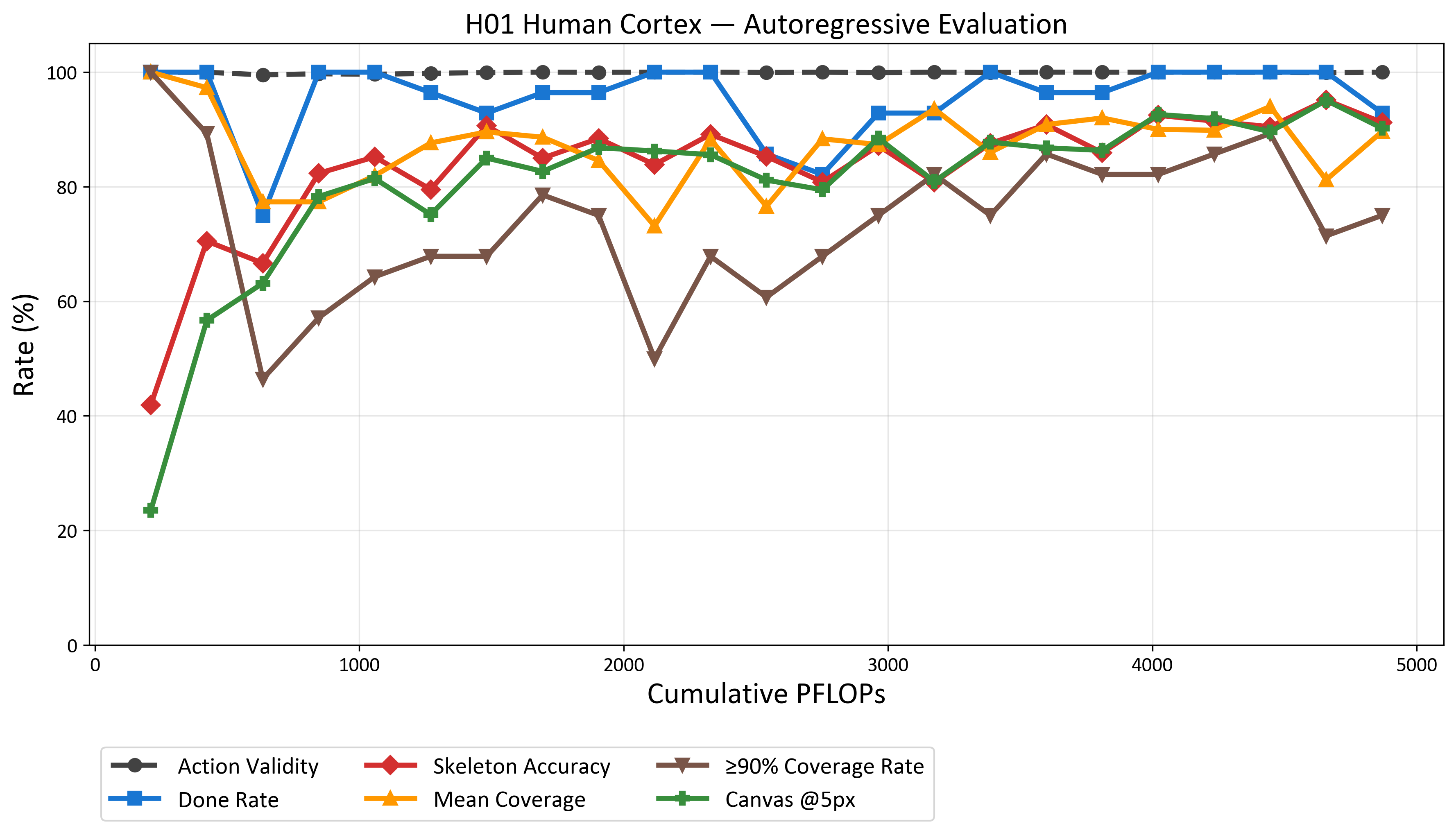}
\caption{Autoregressive metrics over training.}
\end{subfigure}
\caption{\textbf{H01 fine-tuning curves.} Top: teacher-forced. Loss decreases from 10.37 (step 1) to 0.95 (step 100K); action validity reaches 100\% by step 2K and button accuracy by step 5K, while canvas placement improves more gradually and peaks at 96.7\% @5px by step 42K---a fast-then-slow pattern consistent with the hierarchical skill emergence reported in \S\ref{sec:single-task}. Bottom: autoregressive (28 held-out neurons, 64 episodes per checkpoint, every 5K steps). Skeleton accuracy and canvas placement saturate near 95\% by step 110K; first-node accuracy and Done rate are at 100\% throughout.}
\label{fig:app_h01}
\end{figure}

\begin{table}[h]
\centering
\small
\caption{\textbf{H01: teacher-forced evaluation} on 28 held-out neurons across 118 checkpoints. ``Step 1'' is the pretrained checkpoint with no fine-tuning.}
\label{tab:app_h01_tf}
\begin{tabular}{lccc}
\toprule
\textbf{Metric} & \textbf{Step 1 (pretrained)} & \textbf{Best} & \textbf{Final (117K)} \\
\midrule
Overall @5px      & 16.1\% & \textbf{98.7\%} (step 116K) & 98.3\% \\
Canvas @5px       & 43.3\% & \textbf{96.7\%} (step 42K)  & 95.2\% \\
Button accuracy   & 4.7\%  & \textbf{100.0\%} (step 5K)  & 100.0\% \\
Action validity   & 99.0\% & \textbf{100.0\%} (step 2K)  & 100.0\% \\
Loss              & 10.37  & \textbf{0.95} (step 100K)   & 0.98 \\
\bottomrule
\end{tabular}
\end{table}

\begin{table}[h]
\centering
\small
\caption{\textbf{H01: autoregressive evaluation.} 28 held-out neurons, 64 episodes per checkpoint.}
\label{tab:app_h01_autoreg}
\begin{tabular}{lccc}
\toprule
\textbf{Metric} & \textbf{Step 5K} & \textbf{Step 95K} & \textbf{Best} \\
\midrule
Skeleton accuracy    & 41.9\% & 92.5\% & \textbf{95.1\%} (step 110K) \\
Canvas @5px          & 23.5\% & 92.6\% & \textbf{95.1\%} (step 110K) \\
Canvas @10px         & 37.4\% & 96.5\% & \textbf{97.5\%} (step 110K) \\
Done rate            & 100\%  & 100\%  & \textbf{100\%} \\
Mean coverage        & 100\%* & 90.0\% & 94.0\% (step 105K) \\
Coverage $\geq 90\%$ & 100\%* & 82.1\% & \textbf{89.3\%} (step 105K) \\
First-node accuracy  & 100\%  & 100\%  & \textbf{100\%} \\
\bottomrule
\end{tabular}
\\[0.3em]
{\footnotesize *Step 5K coverage is inflated because the model has not yet learned to terminate; it runs the full 500-step budget placing many extra markers.}
\end{table}

\subsubsection{\textit{C. elegans} Nerve Ring}
\label{app:connectomics_tracing:worm}

\textbf{Data.} EM volumes from the \textit{C. elegans} nerve ring connectome~\citep{witvliet2021connectomes} at 8~nm/px. This is a substantially harder visual setting than H01: \textit{C. elegans} neuropil is densely packed with very small neurites. The trace target follows the largest cross-section through z when a neuron bifurcates. Fine-tuning data: ${\sim}2.6$M training frames from 19{,}872 episodes (8$\times$ augmentation: 4 rotations $\times$ 2 flips). Trained for 66{,}397 gradient steps (${\sim}1$ epoch). Test split: 13 held-out neurons from a spatially separated region.

\begin{figure}[h]
\centering
\begin{subfigure}{0.48\linewidth}
\centering
\includegraphics[width=\linewidth]{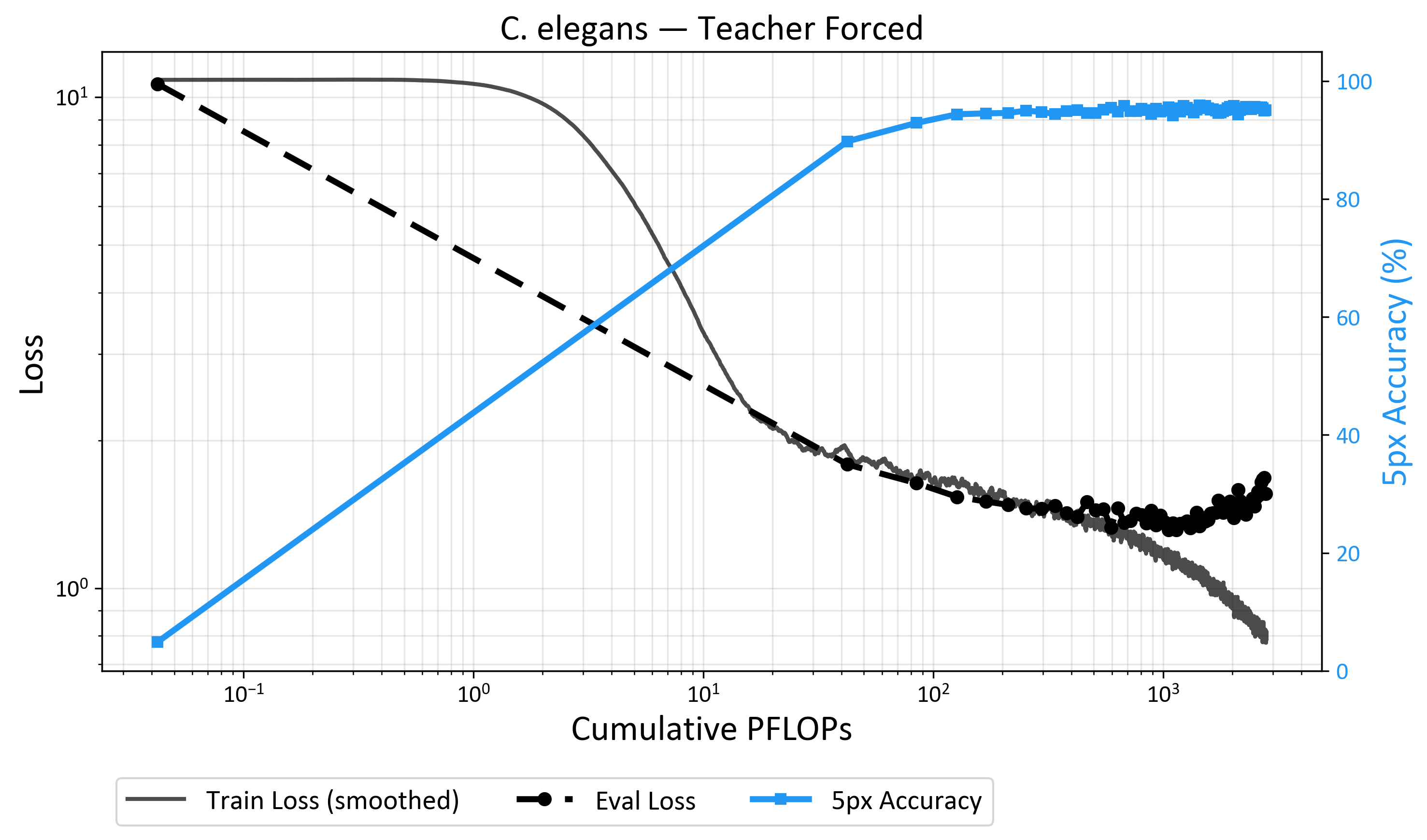}
\caption{Teacher-forced training loss and overall placement accuracy.}
\end{subfigure}\hfill
\begin{subfigure}{0.48\linewidth}
\centering
\includegraphics[width=\linewidth]{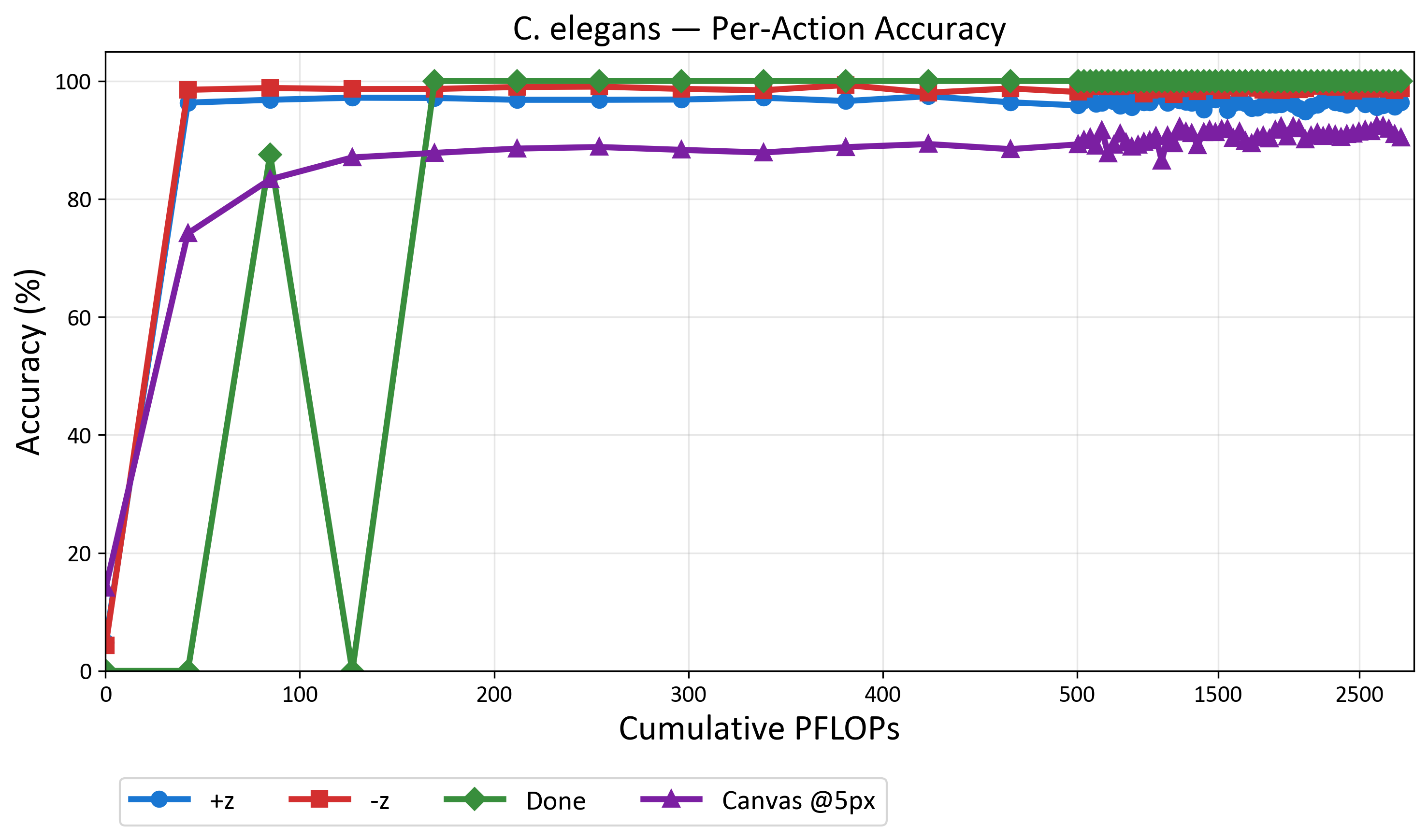}
\caption{Teacher-forced per-action accuracy.}
\end{subfigure}\\[0.5em]
\begin{subfigure}{0.85\linewidth}
\centering
\includegraphics[width=\linewidth]{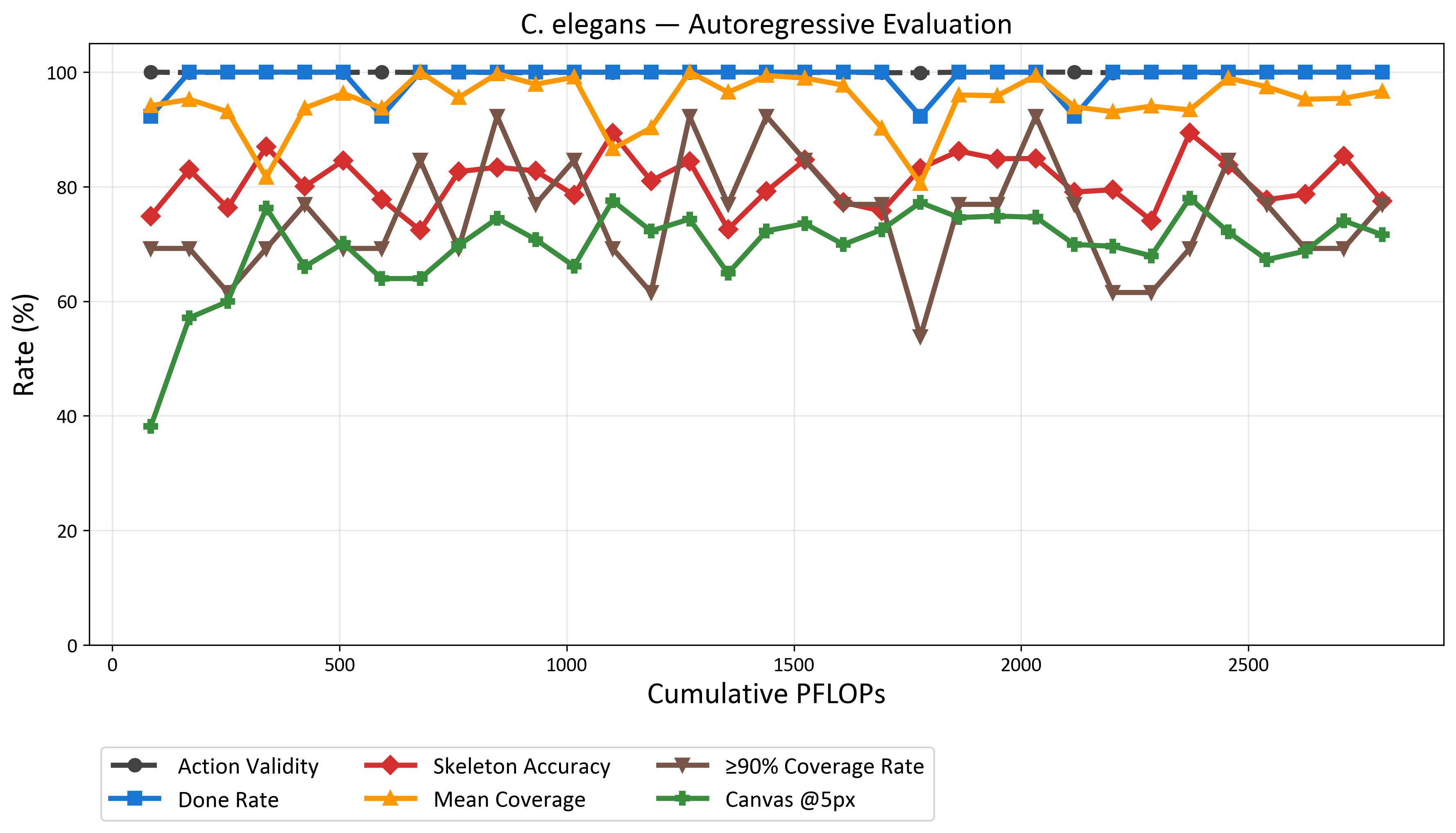}
\caption{Autoregressive (closed-loop) metrics over training.}
\end{subfigure}
\caption{\textbf{\textit{C. elegans} fine-tuning curves.} Top: teacher-forced. Canvas @5px jumps from 14.1\% to ${\sim}88\%$ in the first 5K steps and gradually saturates at 92.4\% by step 62K. Loss reaches its minimum (1.31) at step 27K and drifts slightly upward (1.56 by step 66K), suggesting mild overfitting on this smaller dataset, but downstream canvas accuracy stays stable. Bottom: autoregressive (13 held-out neurons, every 2K steps, max episode length 400). Skeleton accuracy peaks at 89.4\% (step 56K); coverage peaks separately at step 20K. Higher checkpoint-to-checkpoint variance than H01 reflects the smaller evaluation set.}
\label{fig:app_worm}
\end{figure}

\begin{table}[h]
\centering
\small
\caption{\textbf{\textit{C. elegans}: teacher-forced evaluation} on 13 held-out neurons across 67 checkpoints.}
\label{tab:app_worm_tf}
\begin{tabular}{lccc}
\toprule
\textbf{Metric} & \textbf{Step 1 (pretrained)} & \textbf{Best} & \textbf{Final (66K)} \\
\midrule
Overall @5px      & 4.9\%  & \textbf{95.8\%} (step 29K) & 95.1\% \\
Canvas @5px       & 14.1\% & \textbf{92.4\%} (step 62K) & 90.4\% \\
Button accuracy   & 4.6\%  & \textbf{98.2\%} (step 22K) & 97.5\% \\
Action validity   & 99.4\% & \textbf{100.0\%} (step 1K) & 100.0\% \\
Loss              & 10.65  & \textbf{1.31} (step 27K)   & 1.56 \\
\bottomrule
\end{tabular}
\end{table}

\begin{table}[h]
\centering
\small
\caption{\textbf{\textit{C. elegans}: autoregressive evaluation.} 13 held-out neurons.}
\label{tab:app_worm_autoreg}
\begin{tabular}{lccc}
\toprule
\textbf{Metric} & \textbf{Step 2K} & \textbf{Best (seg.)} & \textbf{Best (cov.)} \\
\midrule
Skeleton accuracy    & 74.9\% & \textbf{89.4\%} (step 56K) & 83.4\% (step 20K) \\
Canvas @5px          & 38.2\% & 78.0\% & 74.5\% \\
Canvas @10px         & 63.2\% & 83.9\% & 80.6\% \\
Done rate            & 92.3\% & 100\%  & 100\% \\
Mean coverage        & 94.2\% & 93.4\% & \textbf{99.7\%} (step 20K) \\
Coverage $\geq 90\%$ & 69.2\% & 69.2\% & \textbf{92.3\%} (step 20K) \\
First-node accuracy  & 100\%  & 100\%  & \textbf{100\%} \\
\bottomrule
\end{tabular}
\end{table}

\paragraph{Discussion.}
Across both datasets, fine-tuning peaks well before a full epoch through the data (H01 at step 42K of 117K; \textit{C. elegans} shows most progress within the first ${\sim}10$K of 66K steps), suggesting substantially fewer annotated traces would suffice in practice. The pretrained model already achieves 99\%+ action validity and non-trivial canvas @5px (43.3\% on H01, 14.1\% on \textit{C. elegans}) on these unseen real-data tasks at step 1, indicating that GUI mechanics learned on the synthetic suite transfer immediately and that fine-tuning is primarily adapting the visual encoder to EM imagery. H01 outperforms \textit{C. elegans} on every metric, which we attribute to (i) myelinated axons being visually more distinct than dense neuropil, (ii) 2.5$\times$ more training data, and (iii) a larger held-out set reducing variance.

\FloatBarrier
\subsection{Human Annotation Validation}
\label{app:human_annotation}

Validates the virtual annotator (Sec.~\ref{sec:framework}) against real humans. Four annotators performed 5 instances each of \taskCDT (20 annotations) with minimal instructions.

\paragraph{Annotator 2 exclusion.}
Excluded from quantitative comparisons: 93.8\% navigation actions (vs.\ 67--78\%), 392 average steps (vs.\ 71--118), and z-placement agreement of only 0--12\% with the other annotators.

\paragraph{Action distribution.}
The virtual annotator's action fractions fall within the human range (Table~\ref{tab:app_human_action_dist}). Its navigation-to-placement ratio (5.1) sits between Ann.\ 1 (5.3, cautious) and Ann.\ 3--4 (2.9--3.8, efficient), and its correction rate (6.6\%) is within the human range (2.4--7.6\%).

\begin{table}[h]
\centering
\small
\caption{\textbf{Action distribution} on \taskCDT: virtual annotator vs.\ three competent human annotators.}
\label{tab:app_human_action_dist}
\begin{tabular}{lcccc}
\toprule
 & \textbf{Virtual} & \textbf{Ann.\ 1} & \textbf{Ann.\ 3} & \textbf{Ann.\ 4} \\
\midrule
Navigation (\%)        & 73.8 & 78.1 & 70.0 & 66.6 \\
Placement (\%)         & 14.4 & 15.4 & 21.1 & 22.9 \\
MIP toggle (\%)        & 10.0 & 4.9  & 6.4  & 8.5 \\
Undo (\%)              & 1.0  & 0.8  & 1.3  & 0.6 \\
Nav/placement ratio    & 5.1  & 5.3  & 3.8  & 2.9 \\
Avg.\ total steps      & 118  & 111  & 88   & 71 \\
Correction rate (\%)   & 6.6  & 4.6  & 7.6  & 2.4 \\
\bottomrule
\end{tabular}
\end{table}

\paragraph{Model vs.\ human.}
The trained model (autoregressive, 100 episodes) closely matches Ann.\ 1: navigation 88 vs.\ 87, placements 15.6 vs.\ 16.6, total steps 110 vs.\ 111, with 97\% task completion (Table~\ref{tab:app_human_model_vs_human}).

\begin{table}[h]
\centering
\small
\caption{\textbf{Model vs.\ human} per-task statistics on \taskCDT.}
\label{tab:app_human_model_vs_human}
\begin{tabular}{lcccc}
\toprule
 & \textbf{Model} & \textbf{Ann.\ 1} & \textbf{Ann.\ 3} & \textbf{Ann.\ 4} \\
\midrule
Task completion (\%)     & 97   & 100  & 80   & 100 \\
Avg.\ navigation steps   & 88   & 87   & 63   & 47 \\
Avg.\ placements         & 15.6 & 16.6 & 17.0 & 16.2 \\
Avg.\ MIP toggles        & 5.4  & 5.2  & 5.2  & 6.0 \\
Avg.\ undos              & 0.1  & 0.8  & 1.4  & 0.4 \\
Avg.\ total steps        & 110  & 111  & 88   & 71 \\
Correction rate (\%)     & 0.8  & 4.6  & 7.6  & 2.4 \\
\bottomrule
\end{tabular}
\end{table}

\paragraph{Reaction times.}
Annotators take ${\sim}2.2\times$ longer before canvas placements (1289~ms median) than button presses (516--616~ms), reflecting precision demands (Table~\ref{tab:app_human_rt}).

\begin{table}[h]
\centering
\small
\caption{\textbf{Median inter-action intervals} across the three competent annotators (Ann.\ 1, 3, 4).}
\label{tab:app_human_rt}
\begin{tabular}{lcc}
\toprule
\textbf{Action type} & $N$ & \textbf{Median (ms)} \\
\midrule
$+$z (navigate forward)  & 527 & 516 \\
$-$z (navigate backward) & 461 & 616 \\
Place marker             & 249 & 1289 \\
MIP toggle               & 68  & 1118 \\
Done                     & 15  & 1950 \\
\bottomrule
\end{tabular}
\end{table}

Annotators span a cautious-to-efficient spectrum; both the virtual annotator and the trained model sit within it.

\clearpage

\section{Visualizations}
\label{app:visualizations}

Task execution videos are available at: \url{https://osf.io/qmhrx/} or in the supplementary materials.



\end{document}